\PassOptionsToPackage{numbers,sort}{natbib}

\documentclass[manuscript,screen]{acmart}
\renewcommand\footnotetextcopyrightpermission[1]{} 

\pdfoutput=1

\AtBeginDocument{%
  \providecommand\BibTeX{{%
    \normalfont B\kern-0.5em{\scshape i\kern-0.25em b}\kern-0.8em\TeX}}}

\setcopyright{rightsretained}
\acmJournal{TORS}
\acmYear{2024} \acmVolume{1} \acmNumber{1} \acmArticle{1} \acmMonth{1}\acmDOI{10.1145/3662738}

\usepackage[utf8]{inputenc}
\usepackage[ruled,vlined]{algorithm2e}
\usepackage{amsmath, bm}
\usepackage{xcolor}
\usepackage{soul}
\usepackage{geometry}
\usepackage{enumitem}
\usepackage{nicefrac}
\usepackage{caption}
\usepackage{subcaption}
\usepackage{xspace}
\usepackage{hyperref}
\usepackage{cleveref}
\usepackage{adjustbox}
\usepackage{multirow}
\usepackage{makecell}
\usepackage{booktabs}
\usepackage{tabularx}
\usepackage{natbib}
\usepackage{ragged2e}
\usepackage{comment}

\definecolor{gray_ablation}{rgb}{0.83, 0.83, 0.83}

\usepackage{pifont}
\newcommand{\cmark}{\ding{51}}%
\newcommand{\xmark}{\ding{55}}%

\PassOptionsToPackage{table,svgname,hyperref}{xcolor}

\definecolor{White}{gray}{0.995}
\usepackage{colortbl}%

\usepackage[table]{xcolor}
\newcommand{\CC}[1]{\cellcolor{gray!#1}}
\newcolumntype{M}{>{\RaggedRight\arraybackslash}X}

\DeclareMathOperator*{\argmax}{arg\,max}

\begin{document}

\title{Formalizing Multimedia Recommendation through Multimodal Deep Learning}

\author{Daniele Malitesta}
\affiliation{%
  \institution{Université Paris-Saclay, CentraleSupélec, Inria}
  \city{Gif-sur-Yvette}
  \country{France}}
  \authornote{Work done while at Politecnico di Bari as a PhD student.}
\email{daniele.malitesta@centralesupelec.fr}

\author{Giandomenico Cornacchia}
\affiliation{%
  \institution{IBM Research Europe}
  \city{Dublin}
  \country{Ireland}}
\authornote{Work done while at Politecnico di Bari before joining IBM.}
\email{giandomenico.cornacchia1@ibm.com}

\author{Claudio Pomo}
\affiliation{%
  \institution{Politecnico di Bari}
  \city{Bari}
  \country{Italy}}
\email{claudio.pomo@poliba.it}

\author{Felice Antonio Merra}
\affiliation{%
  \institution{Amazon Science}
  \city{Berlin}
  \country{Germany}}
\authornote{Work done while at Politecnico di Bari before joining Amazon.}
\email{felmerra@amazon.de}

\author{Tommaso {Di Noia}}
\affiliation{%
  \institution{Politecnico di Bari}
  \city{Bari}
  \country{Italy}}
\email{tommaso.dinoia@poliba.it}

\author{Eugenio {Di Sciascio}}
\affiliation{%
  \institution{Politecnico di Bari}
  \city{Bari}
  \country{Italy}}
\email{eugenio.disciascio@poliba.it}

\renewcommand{\shortauthors}{Malitesta et al.}

\begin{abstract}
Recommender systems (RSs) provide customers with a personalized navigation experience within the vast catalogs of products and services offered on popular online platforms. Despite the substantial success of traditional RSs, recommendation remains a highly challenging task, especially in specific scenarios and domains. For example, human affinity for items described through multimedia content (e.g., images, audio, and text), such as fashion products, movies, and music, is multi-faceted and primarily driven by their diverse characteristics. Therefore, by leveraging all available signals in such scenarios, multimodality enables us to tap into richer information sources and construct more refined user/item profiles for recommendations. Despite the growing number of multimodal techniques proposed for multimedia recommendation, the existing literature lacks a shared and universal schema for modeling and solving the recommendation problem through the lens of multimodality. Given the recent advances in multimodal deep learning for other tasks and scenarios where precise theoretical and applicative procedures exist, we also consider it imperative to formalize a general multimodal schema for multimedia recommendation. In this work, we first provide a comprehensive literature review of multimodal approaches for multimedia recommendation from the last eight years. Second, we outline the theoretical foundations of a multimodal pipeline for multimedia recommendation by identifying and formally organizing recurring solutions/patterns; at the same time, we demonstrate its rationale by conceptually applying it to selected state-of-the-art approaches in multimedia recommendation. Third, we conduct a benchmarking analysis of recent algorithms for multimedia recommendation within Elliot, a rigorous framework for evaluating recommender systems, where we re-implement such multimedia recommendation approaches. Finally, we highlight the significant unresolved challenges in multimodal deep learning for multimedia recommendation and suggest possible avenues for addressing them. The primary aim of this work is to provide guidelines for designing and implementing the next generation of multimodal approaches in multimedia recommendation.
\end{abstract}



\received{20 February 2007}
\received[revised]{12 March 2009}
\received[accepted]{5 June 2009}

\maketitle

\section{Introduction}
\label{sec:introduction}

Over the last few decades, companies have increasingly been developing online platforms to reach their customers and offer them a more comprehensive selection of personalized products and services in various domains, from food and fashion to e-commerce and tourism. Recommender systems (RSs) are among the prominent technologies that work behind the scenes of such platforms to unveil the implicit preference patterns within the intricate set of users and items and curate the presentation of a list of products that customers may enjoy. Such technologies are an essential element of all major Internet businesses, driving up to 35\% of Amazon’s sales~\cite{mackenzie2013retailers} and more than 80\% of Netflix’s catalog~\cite{chhabra17}.

There exist recommendation scenarios, such as multimedia recommendation, where items naturally come with additional side information that may complement the knowledge conveyed by the historical user/item interaction matrix. Multimedia recommendation~\cite{deldjoo2022multimedia, DBLP:journals/csur/DeldjooSCP20} is the task of recommending products or services either described through multimedia content (e.g., a fashion item with a product image and description) or multimedia content themselves (e.g., a movie with its visuals, soundtrack, and subtitles). In such a context, any recorded user/item interaction may hide multiple possible reasons why that interaction occurred. A user could be interested in buying a fashion item due to the description on the item page and could enjoy a movie because of its soundtrack. Understanding these patterns means modeling users’ and items’ profiles through the \textit{multi-faceted} aspects of their interactions.

Our experience of daily life is intrinsically \textit{multimodal}. We interact with objects surrounding us through our five senses. For instance, watching a movie can involve three senses (i.e., modalities): we watch it (\textit{visual} modality) while listening to the dialogues (\textit{audio} modality) and possibly reading its subtitles (\textit{textual} modality). Multimodal learning has been one of the hot topics in deep learning for some years now, addressing applicative domains such as medical imaging~\cite{DBLP:journals/tmm/BhatnagarWL13, DBLP:journals/sigpro/HermessiMZ21, DBLP:journals/tip/TangHLD22, DBLP:conf/wacv/GeorgescuIMSRVK23}, autonomous driving~\cite{DBLP:conf/cvpr/CaesarBLVLXKPBB20, DBLP:conf/wacv/KawasakiS21, DBLP:journals/tits/XiaoCGUL22, DBLP:conf/mobicom/ZhengLCW023}, speech/emotion recognition~\cite{DBLP:conf/acl/Paraskevopoulos20, DBLP:conf/cvpr/LvCHDL21, DBLP:conf/acl/PanCGZWL22, DBLP:journals/ijon/LiWLZ23}, multimedia retrieval~\cite{DBLP:conf/sigir/HuZPL19, DBLP:conf/sigir/Li0YSCZS21, DBLP:conf/mm/ChenWC00P22, DBLP:conf/acl/HuGTKY23}, and, only recently, multimodal large language modelling~\cite{DBLP:journals/corr/abs-2306-13549}. Given the success and popularity it has encountered, some works have tried to outline, categorize, and formalize the core concepts behind multimodality in deep learning~\cite{DBLP:conf/icml/NgiamKKNLN11, DBLP:books/acm/18/BaltrusaitisAM18, DBLP:journals/pami/BaltrusaitisAM19}. Remarkably, the literature recognizes five steps and challenges when designing a multimodal deep learning pipeline~\cite{DBLP:books/acm/18/BaltrusaitisAM18}: \textit{representation}, \textit{translation}, \textit{alignment}, \textit{fusion}, and \textit{co-learning}.

Similarly to the cited domains and applications, approaches in multimedia recommendation have been shown to effectively apply multimodal deep learning techniques to the recommendation task. The idea is to model users’ and items’ profiles through the different modalities and suitably capture the multi-faceted nature of their interconnections. Recent works in the literature have brought multimodality to multimedia recommendation~\cite{DBLP:journals/jmlr/SalahTL20, DBLP:conf/recsys/TruongSL21, DBLP:journals/corr/abs-2302-03883, DBLP:journals/corr/abs-2302-04473} tackling (just to mention a few) micro-video recommendation~\cite{DBLP:conf/mm/WeiWN0HC19,DBLP:journals/tmm/ChenLXZ21,DBLP:journals/tmm/CaiQFX22}, food recommendation~\cite{DBLP:journals/tmm/MinJJ20,DBLP:journals/eswa/LeiHZSZ21,DBLP:journals/tomccap/WangDJJSN21}, outfit fashion compatibility~\cite{DBLP:conf/kdd/ChenHXGGSLPZZ19, DBLP:conf/aaai/YangDW20, DBLP:journals/tmm/ZhanLASDK22}, and artist/song recommendation~\cite{DBLP:conf/sigir/ChengSH16,DBLP:conf/recsys/OramasNSS17,DBLP:conf/bigmm/VaswaniAA21}. However, and differently from the other outlined domains and scenarios, recommendation lacks a \textit{shared} theoretical and applicative formalization to align the multimedia recommendation problem with the same formal pipeline proposed in multimodal deep learning~\cite{DBLP:conf/icml/NgiamKKNLN11, DBLP:books/acm/18/BaltrusaitisAM18, DBLP:journals/pami/BaltrusaitisAM19}. 

For these reasons, in this work, we first review the most popular and recent state-of-the-art approaches in multimedia recommendation. Indeed, it emerges that three main design choices are involved when proposing novel multimedia recommender systems leveraging multimodality: (i) \textbf{\textit{Which}} modalities to suitably describe the user/item input data; (ii) \textbf{\textit{How}} to extract and process meaningful multimodal representations; (iii) \textbf{\textit{When}} to integrate and inject multimodal data into the training/inference steps. While observing that many multimedia recommendation approaches are rarely aligned on the techniques to adopt for (i), (ii), and (iii), we maintain this could limit the future development of novel solutions in the field. This is true since each work claims to advance with respect to the state-of-the-art but it becomes cumbersome to distinguish which conceptual and implementation \textit{strategies} are contributing the most~\cite{DBLP:conf/kdd/MalitestaCPDN23}.    

Thus, inspired by the multimodal pipeline formalized in multimodal deep learning~\cite{DBLP:conf/icml/NgiamKKNLN11, DBLP:books/acm/18/BaltrusaitisAM18, DBLP:journals/pami/BaltrusaitisAM19}, we try to align the same schema with the three design choices recognized above. Our objective is to define a conceptual and theoretical schema that uses multimodality to encompass and summarize the most diffused solutions/patterns in the multimedia recommendation literature. To the best of our knowledge, this represents the first attempt that, differently from similar works in the literature~\cite{DBLP:journals/corr/abs-2302-03883, DBLP:journals/corr/abs-2302-04473}, \textbf{\textit{formalizes}} multimedia recommendation through the core concepts theorized in multimodal deep learning~\cite{DBLP:conf/icml/NgiamKKNLN11, DBLP:books/acm/18/BaltrusaitisAM18, DBLP:journals/pami/BaltrusaitisAM19}.

To sum up, we aim to answer the following research questions (RQs):

\begin{enumerate}[label=\textbf{RQ\arabic*.},left=0pt]
    \item \textit{\ul{Which are the main solutions in the related literature?}} We review existing works in multimedia recommendation adopting multimodal learning techniques, highlighting common and different architectural choices; in this respect, we categorize the reviewed papers according to the type of multimodal input (i.e., \textit{\textbf{Which}}), the technique for features processing (i.e., \textit{\textbf{How}}), and the moment to integrate modalities (i.e., \textit{\textbf{When}}).
    
   \item \textit{\ul{Can we summarize the observed strategies into a shared formal schema, which is also conceptually applicable to existing recommendation scenarios?}} On such basis, and following the related literature on multimodal deep learning, we revisit the multimedia recommendation task under the lens of multimodal deep learning; by mapping the multimodal pipeline outlined in~\cite{DBLP:conf/icml/NgiamKKNLN11, DBLP:books/acm/18/BaltrusaitisAM18, DBLP:journals/pami/BaltrusaitisAM19} to the threefold categorization from RQ1, we provide the general formulations for a formal schema involving three steps: multimodal input data, multimodal feature processing, and multimodal feature fusion. To conceptually validate the rationale of the introduced theoretical formulations, we also apply them to four selected recommendation systems spanning various tasks in multimedia recommendation.
    
    \item \textit{\ul{Can we integrate the multimodal schema into existing recommendation frameworks?}} We use Elliot~\cite{DBLP:conf/sigir/AnelliBFMMPDN21}, a rigorous framework for the reproducibility of recommender systems, and integrate our multimodal schema into it to benchmark six state-of-the-art multimodal techniques for multimedia recommendation (i.e., VBPR~\cite{DBLP:conf/aaai/HeM16}, MMGCN~\cite{DBLP:conf/mm/WeiWN0HC19}, GRCN~\cite{DBLP:conf/mm/WeiWN0C20}, LATTICE~\cite{DBLP:conf/mm/Zhang00WWW21}, BM3~\cite{DBLP:conf/www/ZhouZLZMWYJ23}, and FREEDOM~\cite{DBLP:conf/mm/ZhouS23}) and test them against four popular recommendation solutions which do not exploit multimodal features (i.e., BPR~\cite{DBLP:conf/uai/RendleFGS09}, NGCF~\cite{DBLP:conf/sigir/Wang0WFC19}, LightGCN~\cite{DBLP:conf/sigir/0001DWLZ020}, and SGL~\cite{DBLP:conf/sigir/WuWF0CLX21}); the measured recommendation metrics, accounting also for beyond-accuracy evaluation, open to future research directions.
    
    \item \textit{\ul{Which are the next challenges in multimodal learning for recommendation?}} Driven by the previous findings, we outline technical and conceptual challenges aimed to provide guidelines for future research in the field.
\end{enumerate}
The rest of this paper is organized as follows.~\Cref{sec:related} provides a comprehensive overview of the most recent works exploiting multimodality for multimedia recommendation, and highlights the main differences between this work and similar works in the literature. Then, in~\Cref{sec:framework}, we present our formal schema which tries to embody and generalize the different solutions reviewed in the literature, and show how to apply it to a valuable sample of state-of-the-art approaches. Under the same light, in~\Cref{sec:benchmarking}, we present the implementation of our formal schema, which we adopt to benchmark selected models from the literature. Furthermore, in~\Cref{sec:challenges}, we take advantage of the lesson learned from the literature and the benchmarking study to outline existing technical challenges that may be addressed in future directions (i.e.,~\Cref{sec:future-directions}). Finally, in~\Cref{sec:conclusion}, we sum up the main contributions of this paper. To foster the reproducibility of the current work, we release a GitHub repository with all the reviewed papers, along with the benchmarking framework and results at the following link:~\url{https://github.com/sisinflab/Formal-MultiMod-Rec}.
\section{Literature Review (RQ1)}\label{sec:related}

In this section, we present a literature review on recent multimodal applications for the task of multimedia recommendation.~\Cref{tab:papers} reports 43 papers collected from the proceedings of top-tier conferences and journals over the last eight years. A careful review and analysis aimed at outlining recurrent schematic and observed patterns suggests categorizing the retrieved papers according to three key questions: 
\begin{itemize}
    \item \textbf{\textit{Which}} modalities to choose for the input data?
    \item \textbf{\textit{How}} to process multimodal features in terms of feature extraction and representation?
    \item \textbf{\textit{When}} to fuse the different modalities to integrate them into the final recommendation framework?
\end{itemize}

To collect all reviewed papers, we also include a public GitHub repository\footnote{\url{https://github.com/sisinflab/Formal-MultiMod-Rec}.} to access their direct DOIs. We intend to update this repository with the most recent works leveraging multimodality for multimedia recommendation.

\begin{table}[!t]
\centering
\caption{Overview of the core questions which arise when modelling a multimedia recommender system based upon multimodality, as observed in the most updated literature. HFE: Handcrafted Feature Extraction, TFE: Trainable Feature Extraction, MMR: Multimodal Representation.}
\resizebox{\textwidth}{!}{%
\begin{tabular}{cccccccccccc}
\toprule
\textbf{Papers} &
\multicolumn{ 1}{l}{\textbf{Year}} &
\multicolumn{3}{c}{\textbf{Modalities (\textit{Which?})}} & \multicolumn{ 5}{c}{\textbf{Feature Processing (\textit{How?})}} & \multicolumn{ 2}{c}{\textbf{Fusion (\textit{When?})}} \\
\cmidrule(lr){3-5} \cmidrule(lr){6-10} \cmidrule(lr){11-12}
 \multicolumn{ 1}{c}{}  & \multicolumn{ 1}{c}{}   & 
 \multirow{2}{*}{ \begin{tabular}[c]{@{}c@{}}\multicolumn{ 1}{c}{\textit{Visual}} \end{tabular}}& 
 \multirow{2}{*}{ \begin{tabular}[c]{@{}c@{}}\multicolumn{ 1}{c}{\textit{Textual}} \end{tabular}}& 
 \multirow{2}{*}{ \begin{tabular}[c]{@{}c@{}}\multicolumn{ 1}{c}{\textit{Audio}} \end{tabular}}& 
 \multirow{2}{*}{ \begin{tabular}[c]{@{}c@{}}\multicolumn{1}{c}{HFE} \end{tabular}}&
 \multicolumn{ 2}{c}{TFE} & \multicolumn{ 2}{c}{MMR} & 
  \multirow{2}{*}{ \begin{tabular}[c]{@{}c@{}}\multicolumn{ 1}{c}{\textit{Early}} \end{tabular}}& \multirow{2}{*}{ \begin{tabular}[c]{@{}c@{}}\multicolumn{ 1}{c}{\textit{Late}} \end{tabular}}\\ 
  \cmidrule(lr){7-8} \cmidrule(lr){9-10}
&   \multicolumn{ 1}{c}{} &   &   &  \multicolumn{ 1}{c}{}   &\multicolumn{ 1}{c}{}    &  \multicolumn{1}{c}{\textit{Pretrained}} & \multicolumn{1}{c}{\textit{End-to-End}} & \multicolumn{1}{c}{\textit{Joint}} & \multicolumn{1}{c}{\textit{Coordinate}} &   &   \\ 
\midrule

\citet{DBLP:conf/mm/FerracaniPBMB15} & \multirow{3}{*}{2015} & \cmark & \cmark &  & \CC{20} & \cmark &  & \cmark &  & \multicolumn{2}{l}{\CC{20}}  \\ 
\citet{DBLP:conf/bigdataconf/JiaWLXXZ15} &  & \cmark & \cmark  &  & \cmark & \multicolumn{2}{l}{\CC{20}} & \cmark &  & \multicolumn{2}{l}{\CC{20}}  \\ 
\citet{DBLP:journals/mta/LiPGCZ15} &  & \cmark &  & \cmark & \cmark & \multicolumn{2}{l}{\CC{20}}  & \cmark &  & \multicolumn{2}{l}{\CC{20}}  \\ \midrule
\citet{DBLP:journals/mta/NieLZS16} & \multirow{2}{*}{2016} & \cmark & \cmark &  & \cmark & \multicolumn{2}{l}{\CC{20}} &  & \cmark & \cmark &  \\ 
\citet{DBLP:conf/mm/ChenHK16} & & \cmark & \cmark &  & \cmark & \multicolumn{2}{l}{\CC{20}} & \cmark &  &  \multicolumn{2}{l}{\CC{20}}  \\ \midrule
\citet{DBLP:conf/mm/HanWJD17} & \multirow{3}{*}{2017} & \cmark & \cmark  &  & \CC{20} &  & \cmark &  & \cmark & \cmark &  \\ 
\citet{DBLP:conf/recsys/OramasNSS17} &  &  & \cmark & \cmark  & \CC{20} &  & \cmark &  & \cmark & \cmark &  \\
\citet{DBLP:conf/ijcai/ZhangWHHG17} &  & \cmark & \cmark  &  & \CC{20} &  & \cmark &  & \cmark & \cmark &  \\ \midrule
\citet{DBLP:conf/kdd/YingHCEHL18} & \multirow{2}{*}{2018} & \cmark & \cmark &  & \CC{20} & \cmark &  & \cmark &  & \multicolumn{2}{l}{\CC{20}}  \\
\citet{DBLP:conf/emnlp/WangNL18} &  & \cmark & \cmark  &  & \CC{20} & \cmark &  & \cmark &  & \multicolumn{2}{l}{\CC{20}}  \\ \midrule
\citet{DBLP:conf/mm/LiuCSWNK19} & \multirow{7}{*}{2019} & \cmark & \cmark &  & \CC{20} & \cmark &  & \cmark &  & \multicolumn{2}{l}{\CC{20}}  \\
\citet{DBLP:conf/sigir/ChenCXZ0QZ19} &  & \cmark & \cmark &  & \CC{20} & \cmark &  & \multicolumn{2}{l}{\CC{20}} & \multicolumn{2}{l}{\CC{20}}  \\ 
\citet{DBLP:conf/mm/WeiWN0HC19} &  & \cmark & \cmark & \cmark  & \CC{20} & \cmark &  &  & \cmark & \cmark &  \\ 
\citet{DBLP:journals/tois/ChengCZKK19} &  & \cmark & \cmark &  & \CC{20} & \cmark &  & \multicolumn{2}{l}{\CC{20}} & \multicolumn{2}{l}{\CC{20}}  \\
\citet{DBLP:conf/mm/DongSFJXN19} &  & \cmark & \cmark &  & \CC{20} & \cmark &  &  & \cmark & \cmark &  \\ 
\citet{DBLP:conf/kdd/ChenHXGGSLPZZ19} &  & \cmark & \cmark &  & \CC{20} & \cmark &  & \cmark &  & \multicolumn{2}{l}{\CC{20}}  \\ 
\citet{DBLP:conf/mm/YuSZZJ19} & \multirow{12}{*}{2020} & \cmark & \cmark &  & \CC{20} & \cmark & \cmark & \cmark & \cmark & \cmark & \\ \midrule
\citet{DBLP:journals/tkde/CuiWLZW20} &  & \cmark & \cmark &  & \CC{20} & \cmark &  &  & \cmark &  & \cmark \\
\citet{DBLP:conf/mm/WeiWN0C20} &  & \cmark & \cmark & \cmark & \CC{20} & \cmark &  &  & \cmark & \cmark &  \\
\citet{DBLP:conf/cikm/SunCZWZZWZ20} &  & \cmark & \cmark &  & \CC{20} & \cmark &  & \cmark &  & \multicolumn{2}{l}{\CC{20}} \\ 
\citet{DBLP:conf/ijcai/Chen020} &  & \cmark & \cmark &  & \CC{20} & \cmark &  & \cmark &  & \multicolumn{2}{l}{\CC{20}} \\ 
\citet{DBLP:journals/tmm/MinJJ20} &  & \cmark & \cmark & \cmark & \CC{20} & \cmark &  & \cmark &  & \multicolumn{2}{l}{\CC{20}} \\ 
\citet{DBLP:conf/ijcnn/Shen0LWC20} &  & \cmark & \cmark &  & \cmark & \cmark &  &  & \cmark & \cmark &  \\ 
\citet{DBLP:conf/aaai/YangDW20} &  & \cmark & \cmark &  & \CC{20} & \cmark &  &  & \cmark &  & \cmark \\
\citet{DBLP:journals/ipm/TaoWWHHC20} &  & \cmark & \cmark & \cmark & \CC{20} & \cmark &  &  & \cmark & \cmark &  \\
\citet{DBLP:journals/tcss/YangWLLGDW20} &  & \cmark & \cmark &  & \CC{20} & \cmark &  &  & \cmark & \cmark &  \\ \midrule
\citet{DBLP:journals/tmm/SangXQMLW21} & \multirow{6}{*}{2021} & \cmark & \cmark &  & \CC{20} & \cmark &  &  & \cmark & \cmark & \\
\citet{DBLP:conf/mm/LiuYLWTZSM21} &  & \cmark & \cmark & \cmark & \CC{20} & \cmark &  &  & \cmark & \cmark & \\
\citet{DBLP:conf/mm/Zhang00WWW21} &  & \cmark & \cmark & & \CC{20} & \cmark &  &  & \cmark & \cmark & \\
\citet{DBLP:conf/bigmm/VaswaniAA21} &  &  & \cmark & \cmark & \CC{20} & \cmark &  &  & \cmark & \cmark & \\
\citet{DBLP:journals/eswa/LeiHZSZ21} &  & \cmark & \cmark & \cmark & \CC{20} & \cmark & \cmark &  &  &  \multicolumn{2}{l}{\CC{20}} \\
\citet{DBLP:journals/tomccap/WangDJJSN21} &  & \cmark & \cmark &  & \CC{20} & \cmark &  & \cmark & &  \multicolumn{2}{l}{\CC{20}} \\ \midrule
\citet{DBLP:journals/tmm/ZhanLASDK22} & \multirow{8}{*}{2022} & \cmark & \cmark  &  & \CC{20} & \cmark &  & \cmark &  & \multicolumn{2}{l}{\CC{20}} \\
\citet{DBLP:conf/sigir/WuWQZHX22} &  & \cmark & \cmark  &  & \CC{20} & \cmark &  & & \cmark & \cmark & \\
\citet{DBLP:journals/tmm/YiC22} &  & \cmark & \cmark  &  & \CC{20} & \cmark & & & \cmark &  & \cmark \\
\citet{DBLP:conf/sigir/Yi0OM22} &  & \cmark & \cmark  & \cmark & \CC{20} & \cmark &  & & \cmark &  & \cmark \\
\citet{DBLP:conf/mir/LiuMSO022} &  & \cmark & \cmark  &  & \CC{20} & \cmark &  & & \cmark & \cmark & \\
\citet{DBLP:conf/mm/MuZT0T22} &  & \cmark & \cmark  &  & \CC{20} & \cmark &  & & \cmark & \cmark & \\
\citet{DBLP:conf/mm/ChenWWZS22} &  & \cmark & \cmark  & \cmark & \CC{20} & \cmark &  & & \cmark & \cmark & \\ 
\midrule
\citet{DBLP:conf/mm/ZhouS23} & \multirow{4}{*}{2023} & \cmark & \cmark & & \CC{20} & \cmark &  &  & \cmark &  & \cmark \\
\citet{DBLP:journals/tmm/WangWYWSN23} & & \cmark & \cmark  & \cmark & \CC{20} & \cmark &  & & \cmark & \cmark & \\ 
\citet{DBLP:conf/www/WeiHXZ23} &  & \cmark & \cmark  & \cmark & \CC{20} & \cmark &  & & \cmark &  & \cmark \\ 
\citet{DBLP:conf/www/ZhouZLZMWYJ23} &  & \cmark & \cmark  &  & \CC{20} & \cmark &  & & \cmark &  & \cmark \\ 
\bottomrule
\end{tabular}
}
\label{tab:papers}
\end{table}

\subsection{\textit{Which} modalities?} 
In multimedia recommendation scenarios, input data generally comes in at least two of the three most common modalities in literature, namely \textit{Visual}, \textit{Textual}, and \textit{Audio} modalities. As evident from the collected papers, the vast majority of works consider the visual and textual modalities, which mainly refer to product images and descriptions (e.g.,~\cite{DBLP:conf/mm/HanWJD17, DBLP:conf/mm/LiuCSWNK19, DBLP:conf/mm/DongSFJXN19, DBLP:journals/tkde/CuiWLZW20, DBLP:conf/mm/Zhang00WWW21, DBLP:journals/tmm/ZhanLASDK22}), respectively, while fewer examples leverage such modalities to describe video frames and captions (e.g.,~\cite{DBLP:conf/bigdataconf/JiaWLXXZ15, DBLP:journals/tmm/SangXQMLW21, DBLP:journals/tomccap/WangDJJSN21}) or users' social media interactions through uploaded photographs together with texts (e.g.,~\cite{DBLP:conf/mm/ChenHK16, DBLP:conf/ijcai/ZhangWHHG17, DBLP:journals/tcss/YangWLLGDW20}). Another emerging trend from the literature is that audio is by far the most underrepresented modality, and it is usually coupled with the textual one to describe music in the form of audio signals and songs' descriptions (e.g.,~\cite{DBLP:conf/recsys/OramasNSS17, DBLP:conf/bigmm/VaswaniAA21}). Conversely, the related literature shows that the audio modality is frequently exploited for video input data (e.g.,~\cite{DBLP:conf/mm/WeiWN0HC19, DBLP:conf/mm/WeiWN0C20, DBLP:journals/ipm/TaoWWHHC20, DBLP:conf/sigir/Yi0OM22, DBLP:journals/tmm/WangWYWSN23}) which is also the unique scenario involving all modalities. 

The observed disparity in data modalities is not only linked to the specific task the various approaches address (e.g., product, song, or micro-video recommendation) but it is also found in each modality's different availability. In this respect, for example, datasets collecting user-item interactions on e-commerce platforms (e.g., the Amazon reviews dataset or IQON300) are more easily accessible than the ones involving social media videos. For instance, one may consider that a version of the TikTok dataset (introduced in~\cite{DBLP:conf/mm/WeiWN0HC19}) has been made available with pre-trained multimodal features involving visual, audio, and textual modalities only recently~\cite{DBLP:conf/www/WeiHXZ23}. This modality \textit{misalignment} is among the most discussed challenges in the community, so we decide to dedicate a section to it later (refer to~\Cref{sec:missing_modalities}).
 
\subsection{\textit{How} to process modalities?}
Once modalities have been selected for data inputs, two primary operations usually get involved in processing the multimodal data to be fed into the recommender system. First, high-level features are extracted from each of the available modalities. Interestingly, early approaches adopt handcrafted feature extraction (HFE) strategies (e.g., color histograms) as described in~\cite{DBLP:conf/bigdataconf/JiaWLXXZ15, DBLP:journals/mta/LiPGCZ15, DBLP:journals/mta/NieLZS16, DBLP:conf/mm/ChenHK16}. However, with the outbreak and the increasing popularity encountered by deep learning and deep neural models for image and text classification, object detection, and speech recognition, trainable feature extractors (TFE) soon became the de facto standard in the learning of latent features from the input data. In this respect, the literature~\cite{DBLP:conf/cvpr/DeldjooNMM21} indicates that the common approach is to use the activation of one of the final hidden layers of deep neural networks. For instance, the authors of~\cite{DBLP:conf/emnlp/WangNL18, DBLP:conf/mm/YuSZZJ19, DBLP:conf/cikm/SunCZWZZWZ20, DBLP:journals/tmm/MinJJ20, DBLP:conf/mm/LiuYLWTZSM21, DBLP:conf/mir/LiuMSO022, DBLP:conf/www/WeiHXZ23} exploit features extracted from deep networks. Furthermore, we categorize TFE strategies based on the use of \textit{Pretrained} deep networks and \textit{End-to-End} learned models. The former refer to the possibility of transferring the learned knowledge of already-trained deep networks to different domains, tasks, and datasets (e.g., see~\cite{DBLP:conf/icann/TanSKZYL18}), whereas the latter usually leverage custom deep neural networks trained in the downstream recommendation task. As evident from the collected papers, the pre-trained solution (e.g.,~\cite{DBLP:conf/mm/FerracaniPBMB15, DBLP:conf/kdd/YingHCEHL18, DBLP:conf/sigir/ChenCXZ0QZ19, DBLP:conf/mm/DongSFJXN19, DBLP:conf/mm/WeiWN0C20, DBLP:conf/ijcnn/Shen0LWC20, DBLP:conf/ijcai/Chen020, DBLP:journals/tmm/SangXQMLW21}) widely surpasses the end-to-end one (e.g.,~\cite{DBLP:conf/mm/HanWJD17, DBLP:conf/ijcai/ZhangWHHG17, DBLP:conf/recsys/OramasNSS17}) in terms of popularity, as the adoption of ready-to-use embedded features obtained from state-of-the-art deep learning models represents a more efficient and convenient approach than performing computationally-expensive and data-eager trained feature extractions. Nevertheless, an argument might be made that using features extracted through models already trained on different datasets and tasks could limit their expressiveness regarding the actual multimedia recommendation task. For this reason, we deepen into the issue in~\Cref{sec:pretrained-features-challenge}, trying to propose viable solutions in~\Cref{sec:domain-specific-features}. 

The second operation involved in the feature processing phase regards the implementation of a multimodal representation (MMR) solution to establish relations among the extracted modalities. We recognize two main approaches, namely, either combining all modalities so that they belong to a unique representation (\textit{Joint}) or keeping them separated to leverage the different influence they may have on recommendation (\textit{Coordinate}). From the collected papers, it follows that both the former (e.g.,~\cite{DBLP:conf/mm/FerracaniPBMB15, DBLP:conf/kdd/YingHCEHL18, DBLP:conf/kdd/ChenHXGGSLPZZ19, DBLP:conf/cikm/SunCZWZZWZ20, DBLP:journals/tmm/ZhanLASDK22, DBLP:conf/mm/ChenWWZS22, DBLP:conf/mm/MuZT0T22}), and the latter (e.g.,~\cite{DBLP:journals/mta/NieLZS16, DBLP:conf/mm/WeiWN0HC19, DBLP:conf/aaai/YangDW20, DBLP:conf/mm/Zhang00WWW21, DBLP:journals/tmm/YiC22}) are almost equally preferred; however, the coordinate multimodal representation is slightly more popular as learning different representations for each involved modality may help unveil the specific contribution it brings to the final personalized recommendation. Indeed, this could support \textit{explainability}, which is among the hottest topics in the community~\cite{DBLP:journals/corr/abs-1708-06409, DBLP:journals/ftir/ZhangC20}, and especially in multimedia recommendation scenarios, where user-item interactions may, by nature, be driven by non-evident and sometimes contrasting users' preferences and tastes~\cite{DBLP:conf/sigir/ChenZ0NLC17, DBLP:conf/sigir/ChenCXZ0QZ19, DBLP:conf/mm/WeiWN0C20}. Finally, the authors from~\cite{DBLP:conf/sigir/ChenCXZ0QZ19, DBLP:journals/tois/ChengCZKK19} do not integrate any multimodal representation approach since they exploit multimodality only for the optimization of the loss but not to predict user-item preferences.

\subsection{\textit{When} to fuse modalities?}
The last stage in the multimodal pipeline deals with the fusion of the different processed modalities so that they can be eventually integrated into the recommendation outcome as a single representation of multiple \textit{coordinated} modalities. This process may take place \textit{before} or \textit{after} the prediction of the user-item preference score. On this basis, the former and the latter approaches are usually known as \textit{Early} (e.g.,~\cite{DBLP:conf/mm/WeiWN0HC19, DBLP:conf/bigmm/VaswaniAA21}) and \textit{Late} (e.g.,~\cite{DBLP:journals/tkde/CuiWLZW20, DBLP:journals/tmm/YiC22}) fusion, respectively. It is worth pointing out that some solutions recognize a third strategy (i.e., \textit{Hybrid} fusion) that combines the two versions mentioned above, but for the sake of simplicity, we decide to categorize the works performing this kind of multimodal fusion as a particular case of late fusion. Additionally, we recognize that several approaches from the literature do not provide a precise differentiation between joint multimodal representation and early fusion. To better clarify this technical aspect, we propose to consider \textit{fusion} as an optional operation that takes place after the feature processing phase only in the case of \textit{coordinate} multimodal representation (you may refer to~\Cref{sec:modalities-representation-challenge}). Indeed, as evident from the table, \textit{Joint} multimodal representation and \textit{Early}/\textit{Late} fusion never occur in the same approach. What is more, we observe that early fusion, employed, for instance, in~\cite{DBLP:journals/mta/NieLZS16, DBLP:conf/ijcai/ZhangWHHG17, DBLP:conf/mm/DongSFJXN19, DBLP:conf/ijcnn/Shen0LWC20, DBLP:journals/tmm/SangXQMLW21, DBLP:conf/mm/Zhang00WWW21, DBLP:conf/sigir/WuWQZHX22, DBLP:conf/mir/LiuMSO022, DBLP:conf/mm/ChenWWZS22, DBLP:conf/mm/MuZT0T22}, is more popular than late fusion, used in~\cite{DBLP:journals/tkde/CuiWLZW20, DBLP:conf/aaai/YangDW20, DBLP:journals/tmm/YiC22, DBLP:conf/sigir/Yi0OM22, DBLP:conf/mm/ZhouS23, DBLP:conf/www/WeiHXZ23, DBLP:conf/www/ZhouZLZMWYJ23}. Motivating this tendency is an unanswered research question that we leave as a possible open issue to impact the design of recommender systems leveraging multimodality. In this respect, you may refer to~\Cref{sec:fusion-challenge} for our discussion on the current challenges about modality fusion, and to~\Cref{sec:fusion-future} where we sketch possible future research directions.

\subsection{Similar works to this paper} For the sake of completeness, we review the current literature works that provide similar contributions to ours to outline the main differences. As already mentioned, pioneer works such as~\cite{DBLP:conf/icml/NgiamKKNLN11, DBLP:books/acm/18/BaltrusaitisAM18, DBLP:journals/pami/BaltrusaitisAM19} introduce and formalize (for the first time) the core concepts and ideas behind the field of multimodal deep learning. After that, the recent years have seen a growing interest in systematically reviewing and schematizing techniques for multimodal fusion~\cite{DBLP:journals/neco/GaoLCZ20}, spanning different application domains such as medicine~\cite{DBLP:journals/mta/TawfikEFDE21}, conversational artificial intelligence~\cite{DBLP:conf/acl-convai/SundarH22}, and visual content syntesis~\cite{DBLP:journals/ijon/ZhangLWPD22}, up to addressing complex and novel machine learning strategies including meta-learning~\cite{DBLP:journals/kbs/MaZWLK22}. Although the cited works share similar rationales to ours, their focus is more general (e.g., deep learning) or heterogeneous (e.g., medicine) with respect to the multimedia recommendation task. 

In the recommendation domain, the study presented in~\cite{DBLP:journals/tmm/MinJJ20} is among the closest and most influential works to our proposal in the intention of introducing a unified framework for food recommendation which leverages the concept of multimodality; however, the work is different from ours in that: (i) it only addresses the task of food recommendation, and (ii) it does not provide either mathematical formalizations or benchmarking analyses of the proposed multimodal pipeline. Furthermore, it is worth recalling two surveys regarding the topic of multimodal recommender systems~\cite{DBLP:journals/corr/abs-2302-03883, DBLP:journals/corr/abs-2302-04473} on arXiv at the moment of this submission. Among the two, the work presented in~\cite{DBLP:journals/corr/abs-2302-04473} shows the major similarities to our paper, especially when recognizing a multimodal pipeline for multimedia recommendation and providing an extensive benchmarking study on numerous multimodal approaches in the literature. Nevertheless, our work stands out for the following novel contributions: (i) we systematically follow the multimodal pipeline outlined in~\cite{DBLP:conf/icml/NgiamKKNLN11, DBLP:books/acm/18/BaltrusaitisAM18, DBLP:journals/pami/BaltrusaitisAM19} in the attempt to adapt it to the three main questions arising in the multimedia recommendation literature, namely, \textit{\textbf{Which?}}, \textbf{\textit{How?}}, and \textit{\textbf{When?}}; (ii) we provide mathematical formalizations for each step of the proposed multimodal pipeline to sketch a formal schema for the next generation of multimodal approaches addressing multimedia recommendation; (iii) we identify a wider set of challenges regarding each step in the multimodal pipeline, and try to provide solutions to tackle them all; (iv) given the recent urge to evaluate the performance of recommender systems under other objectives apart from accuracy~\cite{DBLP:conf/sigir/Vargas14, DBLP:conf/recsys/VargasC11, DBLP:reference/rsh/ShaniG11, DBLP:journals/umuai/JannachLKJ15, DBLP:conf/recsys/AbdollahpouriBM17, DBLP:conf/recsys/Baeza-Yates20, DBLP:journals/ipm/BorattoFM21}, and specifically when investigating the impact of multimodality on novelty and diversity~\cite{DBLP:conf/kdd/MalitestaCPDN23} or popularity bias~\cite{DBLP:conf/mmir/MalitestaCPN23}, our benchmarking analysis (exploiting the evaluation pipeline from Elliot~\cite{DBLP:conf/sigir/AnelliBFMMPDN21}) represents the first effort to rigorously assess such large-scale performance measures in multimedia recommendation, opening to further research questions.
\section{A formal multimodal schema for multimedia recommendation (RQ2)}\label{sec:framework}

As previously outlined, the literature shows recurrent schematic patterns in adopting multimodal techniques for the task of multimedia recommendation. However, when considering the latest solutions in the field (\Cref{sec:related}) it appears evident that, differently from what happens for other applicative domains in machine learning, such approaches do not seem to follow any shared and officially recognized formal schema aligned with the principles of multimodal deep learning~\cite{DBLP:conf/icml/NgiamKKNLN11, DBLP:books/acm/18/BaltrusaitisAM18, DBLP:journals/pami/BaltrusaitisAM19}. 

\begin{figure*}[!t]
\centering
    \includegraphics[width=0.8\textwidth]{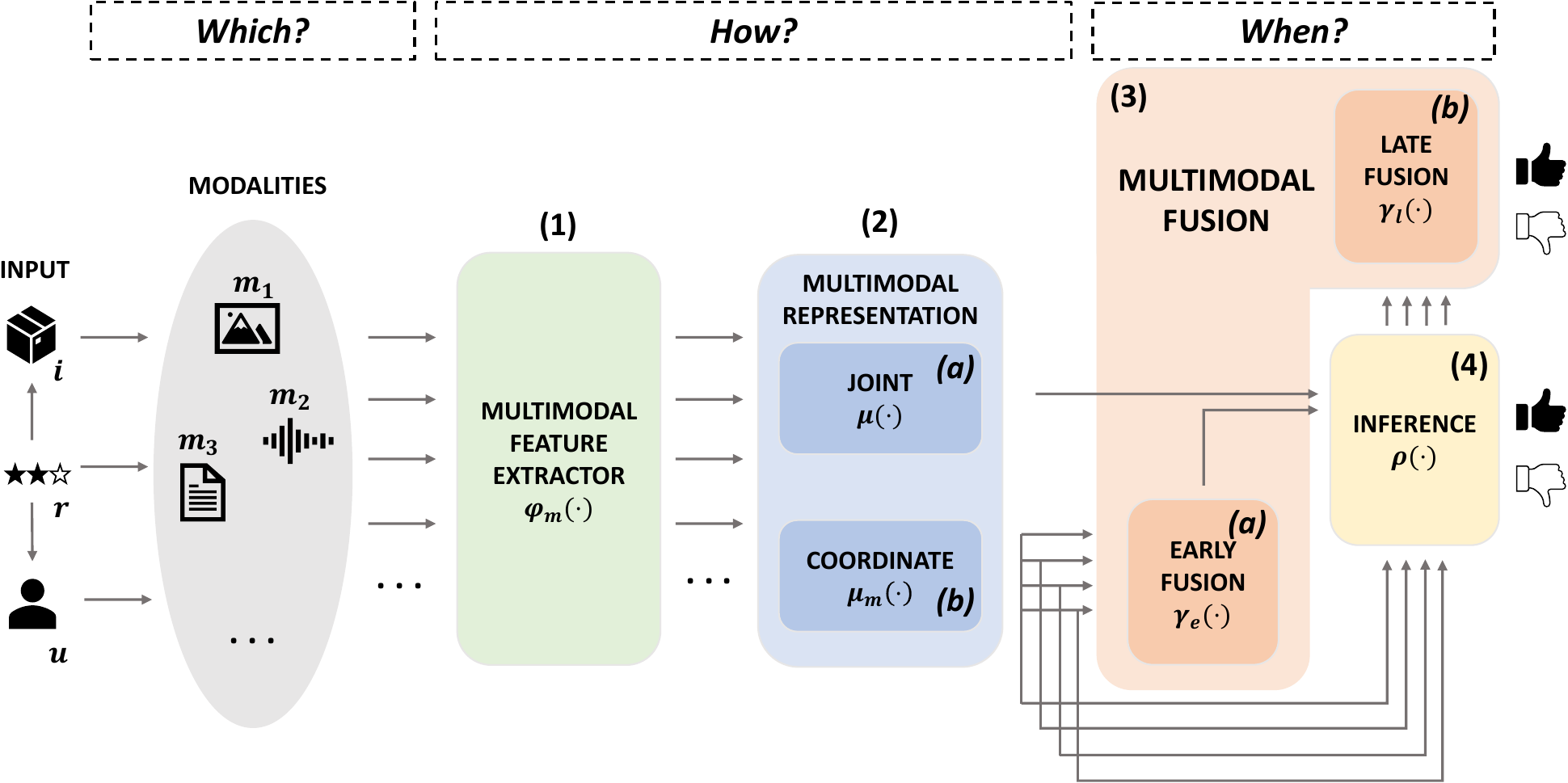}
    \caption{Our multimodal schema for multimedia recommendation. After (1) a modality-aware feature extraction, the extracted features may be either directly represented into a unique latent space (2\textit{a}) or projected into a different latent space for each modality (2\textit{b}). While in the former case, the multimodal representation is used to produce a prediction (4), in the latter case, all modalities must undergo a fusion phase (3). In the early fusion (3\textit{a}), we produce a final representation that is used for prediction (4). Otherwise, we first produce a different prediction for each modality (4), and then we fuse them (late fusion) into a single predicted value (3\textit{b}).}
    \label{fig:framework}
\end{figure*}

To sort things out, in this section, we propose to formally revisit multimedia recommendation under the lens of multimodal deep learning (\Cref{fig:framework}). First, we formalize the standard recommendation task. Then, we theoretically give answers to the core questions previously outlined, namely: \textbf{\textit{Which}} multimodal input data to adopt, \textbf{\textit{How}} to extract multimodal features and set relationships among them, and \textit{\textbf{When}} to fuse modalities. Finally, we specify the multimedia recommendation task through multimodality. 

Along with the formalization of the unified framework, we show how to apply the introduced theoretical notions to four selected state-of-the-art multimedia recommendation models (refer to the \textsc{\bfseries \underline{Applications}} paragraph). Specifically, we choose the proposed examples spanning a wide range of tasks, namely micro-video recommendation~\cite{DBLP:conf/mm/WeiWN0HC19}, food recommendation~\cite{DBLP:journals/tomccap/WangDJJSN21}, outfit fashion compatibility~\cite{DBLP:conf/mm/HanWJD17}, and artist/song recommendation~\cite{DBLP:conf/recsys/OramasNSS17}:

\begin{enumerate}[leftmargin=*]
    \item \textbf{Micro-video recommendation.} \citet{DBLP:conf/mm/WeiWN0HC19} build a bipartite user-item graph for micro-video personalized recommendation. The idea behind the approach is to exploit high-order users-items relations leveraging the multimodal nature of recommended items (i.e., micro-videos), to which users may experience different attitudes. The authors adopt a graph convolutional network~\cite{DBLP:conf/iclr/KipfW17} to refine user and item embeddings (conditioned on the graph topology). 
    \item \textbf{Food recommendation.} \citet{DBLP:journals/tomccap/WangDJJSN21} introduce a tripartite framework for food recommendation whose pipeline involves the retrieval of recipes according to user-generated videos, the profiling of users based upon their social media interactions, and the final health-aware food recommendation of recipes. 
    \item \textbf{Outfit fashion compatibility.} \citet{DBLP:conf/mm/HanWJD17} propose a framework to recommend the next fashion item that matches a set of already chosen ones to produce a visually appealing outfit. The authors address the task by considering the items composing a fashionable outfit as a temporal sequence, so they leverage a bidirectional LSTM~\cite{DBLP:series/sci/2012-385}. 
    \item \textbf{Artist and song recommendation.} The approach introduced by~\citet{DBLP:conf/recsys/OramasNSS17} deals with the task of music recommendation. Specifically, the authors propose to divide the problem into artist and song recommendations by learning their separate embeddings and leveraging the textual artist biography and audio spectrogram as inputs.
\end{enumerate}

Note that, in the following, we use the \textbf{bold} notation only when we explicitly define a vector for which we indicate its elements (i.e., scalars or other vectors).

\subsection{Classical recommendation task}
\label{sec:class_recsys}
We consider users, items, and user-item interactions as the inputs to the recommender system. We denote with $u \in \mathcal{U}$, $i \in \mathcal{I}$, and $r \in \mathcal{R}$ a user, an item, and a user-item interaction, respectively. To ease the notation, we say $x \in \mathcal{X}$ is a general input to the system, with $\mathcal{X} = \mathcal{U} \cup \mathcal{I} \cup \mathcal{R}$. Given a set of input data $\mathcal{X}$, and defined $\rho(\cdot)$ as the preference score prediction function, a recommender system aims to build a top@$k$ list of items maximizing the following posterior probability ($prob$): 
\begin{equation}
    \label{eq:theta_w}
        \hat{\Theta}_\rho = \argmax_{\bm{\Theta}_\rho} prob(\bm{\Theta}_\rho \;|\; \mathcal{X}),
\end{equation}
where $\bm{\Theta}_\rho = [\theta^{(0)}_\rho, \theta^{(1)}_\rho, \dots, \theta_{\rho}^{(|\mathcal{W}_{\rho}| - 1)}]$ is the vector collecting all weights for the inference function $\rho(\cdot)$, $\mathcal{W}_\rho = \{\theta^{(0)}_\rho, \theta^{(1)}_\rho, \dots, \theta_{\rho}^{(|\mathcal{W}_{\rho}| - 1)}\}$ is the set of such weights, and $|\mathcal{W}_\rho|$ its cardinality. 

For instance, in the case of latent factor models (e.g., matrix factorization~\cite{DBLP:journals/computer/KorenBV09}), which are the most popular ones in the literature, the set of trainable weights $\mathcal{W}_\rho$ involves the user and item embeddings: $\mathcal{W}_\rho = \{\theta_\rho^{(u_0)}, \theta_\rho^{(u_1)}, \dots, \theta_\rho^{(u_{|\mathcal{U}| - 1})}, \theta_\rho^{(i_0)},\\ \theta_\rho^{(i_1)}, \dots, \theta_\rho^{(i_{|\mathcal{I}| - 1})}\}$. To ease the notation, from now on, we will indicate with $\theta_\rho^{\mathcal{U}}$ and $\theta_\rho^{\mathcal{I}}$ as the embeddings of any user and item in the recommendation system, respectively. Thus, in the case of classical recommender system leveraging latent factor models, the inference function may be indicated as:
\begin{equation}
    \hat{y} = \rho(\theta_\rho^{\mathcal{U}}, \theta_\rho^{\mathcal{I}}, \mathcal{X}),
\end{equation} 
where $\hat{y}$ is the predicted value through the inference function $\rho(\cdot)$, such as a rating score for any user-item pair.

\subsection{Multimodal input data}
\label{sec:multimodal_input_data}
As shown in~\Cref{fig:framework}, the first step of our multimodal schema is to identify input modalities. A common list of modalities for each input data (i.e., user, item, user-item interaction) in multimedia scenarios may be defined as follows:
\begin{itemize}
    \item visual (\textbf{\texttt{v}}), e.g., images, video frames;
    \item textual (\textbf{\texttt{t}}), e.g., image captions, video subtitles, song lyrics, reviews;
    \item audio (\textbf{\texttt{a}}), e.g., songs, podcasts, movie soundtracks.  
\end{itemize}
Formally, we define $m \in \mathcal{M}$ as an admissible modality for the system (i.e., $\mathcal{M} = \{\textbf{\texttt{v}}, \textbf{\texttt{t}}, \textbf{\texttt{a}}\}$). We should mention that data may come with all such modalities or just a subset. For instance, videos from video streaming platforms (such as Netflix or Amazon Prime Video) have frames (\textbf{\texttt{v}}), subtitles and/or descriptions (\textbf{\texttt{t}}), and an audio track and/or soundtrack (\textbf{\texttt{a}}). Similarly, e-commerce platforms (such as Amazon or Zalando) sell products that may come with photographs (\textbf{\texttt{v}}) and reviews which stand for the textual feedback users express towards those products (\textbf{\texttt{t}}).

Let $x \in \mathcal{X}$ be an input to the recommender system, whose set of available modalities is indicated as $\mathcal{M}_x \subseteq \mathcal{M}$. We represent the \textit{content} data of input $x$ in modality $m$ as $c^{(m)}_x$, with $m \in \mathcal{M}_x$, and the vector of content data for input $x$ in all modalities as $\mathbf{c}_x$. Concerning the examples from above, a video item $x$ may be described through three modalities (i.e., $\mathcal{M}_x = \{\textbf{\texttt{v}}, \textbf{\texttt{a}}, \textbf{\texttt{t}}\}$) and, for example, its visual content data (a frame) is an RGB image indicated as $c^{(\textbf{\texttt{v}})}_x$. Similarly, a fashion item $x$ may be described through two modalities (i.e., $\mathcal{M}_x = \{\textbf{\texttt{v}}, \textbf{\texttt{t}}\}$) and, for example, its textual content data (the description) is a set of words indicated as $c^{(\textbf{\texttt{t}})}_x$.

\vspace{1em}
\noindent\textsc{\bfseries \underline{Applications}}
\begin{enumerate}[leftmargin=*]
    \item \textbf{Micro-video recommendation~\cite{DBLP:conf/mm/WeiWN0HC19}.}  Micro-videos (the items) are described via three modalities, namely: \textit{visual} (i.e., frames), \textit{textual} (i.e., user-generated captions and descriptions) and \textit{audio} (i.e., the audio track, that is not always available). It is worth pointing out that also users are described through three embeddings representing how each item modality might influence them differently. Nevertheless, they cannot be formally considered as multimodal input data (we do not report any information about the multimodal input data and feature extraction columns in the table).
    \item \textbf{Food recommendation~\cite{DBLP:journals/tomccap/WangDJJSN21}.} On the user side, it should be noticed that the input data does not properly follow the above definition we provide about multimodality, as users are profiled only according to the textual description of their generated tweets. However, we maintain the importance of this example since it represents one of the few approaches in the literature that proposes to model users through a multimodality-like solution. On the other side, items' description is multimodal because it integrates frames of user-generated videos to retrieve recipes from (i.e., \textit{visual} modality) and descriptions of recipe ingredients (i.e., \textit{textual} modality).
    \item \textbf{Outfit fashion compatibility~\cite{DBLP:conf/mm/HanWJD17}.} Recommendation is multimodal because the authors adopt both product images (i.e., \textit{visual} modality) and text descriptions of the fashion items extracted from the product details (i.e., \textit{textual} modality).
    \item \textbf{Artist and song recommendation~\cite{DBLP:conf/recsys/OramasNSS17}.} Multimodality is to be found in the item's description, which is based upon artist biography (i.e., \textit{textual} modality) and audio spectrogram derived from songs (i.e., \textit{audio} modality). 
\end{enumerate}

\subsection{Multimodal feature processing}
As in~\Cref{fig:framework}, multimodal inputs are processed to be transferred into a low-dimensional representation. This step runs through a multimodal feature extractor and a component that constructs a multimodal feature representation.

\subsubsection{Feature extraction} 
\label{sec:feature_extraction}
Content input data is generally not exploitable as it is in a recommender model (e.g., the matrix of pixels from an image is not meant to be directly integrated into a recommender). Hence, our schema introduces a \textit{Feature Extractor} (FE) component to extract features, which should follow two principles, being (i) \textit{high-level} (i.e., meaningful for the recommender system) and (ii) \textit{functional} to the final recommendation task. Indeed, choosing the most suitable feature extractor for each modality may affect the performance.

Let $c^{(m)}_x$ be the content data for input $x$ in modality $m \in \mathcal{M}_x$. Then, let $\varphi_{m}(\cdot)$ be the feature extractor function for the modality $m$. We define the feature extraction process in the modality $m$ as:
\begin{equation}
    \label{eq:feature_extraction}
    \overline{c}^{(m)}_x = \varphi_{m}(c_x^{(m)}) \quad \forall m \in \mathcal{M}_x,
\end{equation}
where $\overline{c}^{(m)}_x$ is the extracted feature for input $x$ in modality $m$. We use the notation $\overline{\mathbf{c}}_x = [\overline{c}^{(0)}_x, \overline{c}^{(1)}_x, \dots, \overline{c}_x^{(|\mathcal{M}_x| - 1)}]$ to refer to the vector of extracted features for input $x$ in all modalities. Generally speaking, $\varphi_m(\cdot)$ may refer either to a handcrafted extractor, HFE (e.g., SIFT and color histogram for visual, and MFCCs for audio), or to a trainable extractor, TFE (e.g., deep learning-based models such as CNNs for visual, audio, and textual). In the latter case, $\varphi_m(\cdot)$ can be either pre-trained or trained end-to-end along with the recommender system. Regarding the extraction of multimodal features for recommendation, the authors in~\cite{DBLP:conf/mm/MalitestaGPN23, DBLP:journals/corr/abs-2403-04503} propose Ducho, which aims to standardize and unify the whole process.

\vspace{1em}
\noindent\textsc{\bfseries \underline{Applications}}
\begin{enumerate}[leftmargin=*]
    \item \textbf{Micro-video recommendation~\cite{DBLP:conf/mm/WeiWN0HC19}.} Visual features are extracted through a pretrained ResNet50~\cite{DBLP:conf/cvpr/HeZRS16} only from key video frames. Textual features are derived from Sentence2Vector~\cite{DBLP:conf/iclr/AroraLM17}. Audio features are extracted using a pre-trained VGGish~\cite{DBLP:conf/icassp/HersheyCEGJMPPS17}.
    \item \textbf{Food recommendation~\cite{DBLP:journals/tomccap/WangDJJSN21}.} Users' tweets are the input to what the authors define as a word-class interaction-based recurrent convolutional network (WIRCNN), which involves a recurrent neural network (RNN) and a convolutional neural network (CNN) to classify user tags. As for items, sampled video frames are encoded through a pre-trained VGGNet-19~\cite{DBLP:journals/corr/SimonyanZ14a}, while textual recipe ingredients are processed via TextCNN~\cite{DBLP:conf/emnlp/Kim14}. 
    \item \textbf{Outfit fashion compatibility~\cite{DBLP:conf/mm/HanWJD17}.} The visual features of fashion items are extracted from a GoogleNet InceptionV3~\cite{DBLP:conf/cvpr/SzegedyVISW16} pretrained network (the TFE), whose dimensionality is 2048. As for the textual description, each word is a one-hot-encoded vector.
    \item \textbf{Artist and song recommendation~\cite{DBLP:conf/recsys/OramasNSS17}.} Artist biographies are processed through the state-of-the-art CNN, which is re-trained using word2vec word embeddings pre-trained on the Google News dataset~\cite{DBLP:journals/corr/abs-1301-3781}. As for the song latent factors, a custom CNN with 256, 512, and 1024 convolutional filters is trained on the time axis, having as output the 4096-dense layer.
\end{enumerate}

\subsubsection{Multimodal representation} 
\label{sec:multimodal_representation}

\begin{figure*}[!t]
\centering
    \includegraphics[width=0.8\textwidth]{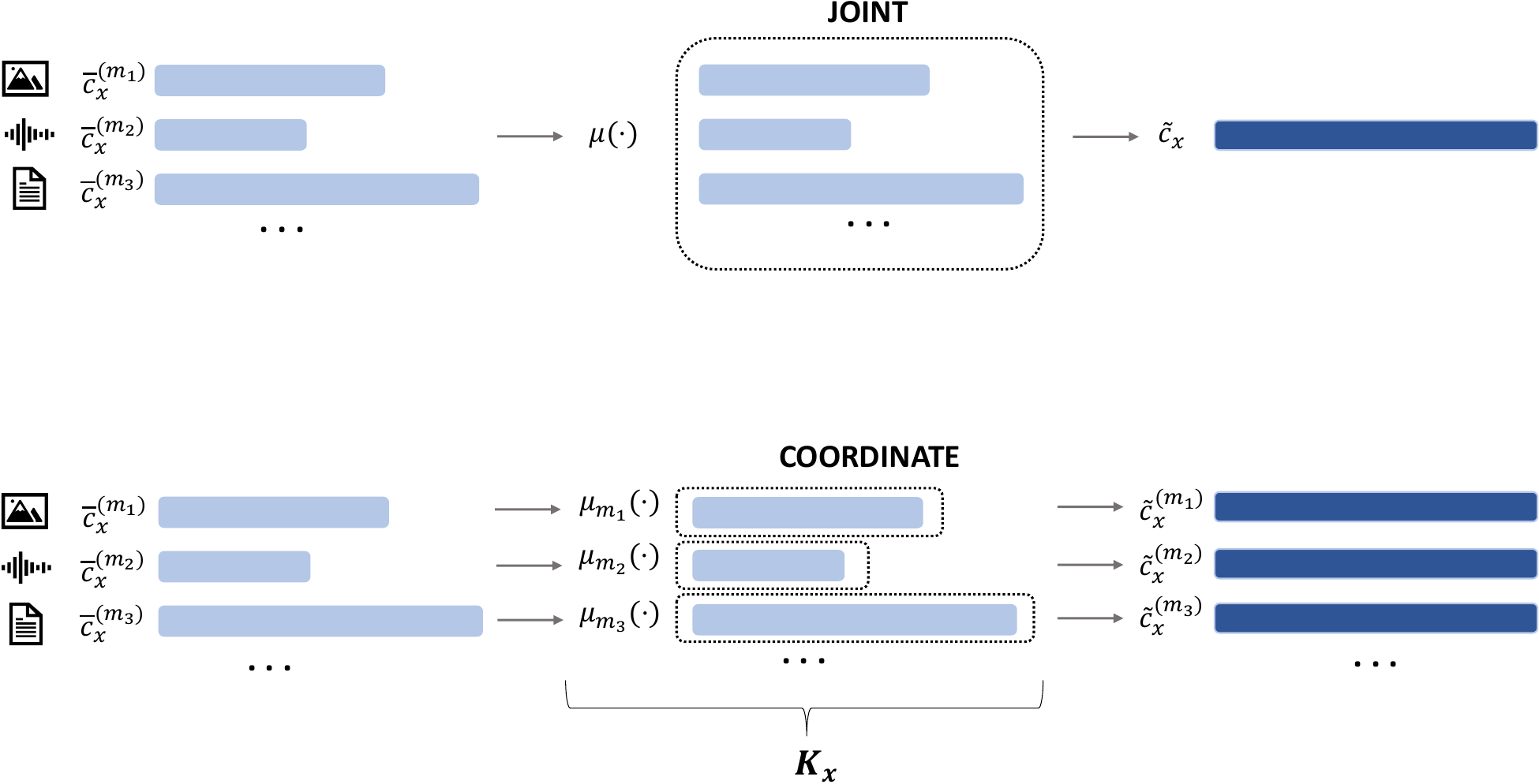}
    \caption{A visual representation of \textit{Joint} and \textit{Coordinate} multimodal representation (above and below, respectively).}
    \label{fig:multimodal_representation}
\end{figure*}

Once high-level features have been extracted from each modality of the input data, the next step is to design a \textit{Representation} strategy to handle the relationships among modalities and eventually inject such data into the recommender system. As shown in~\Cref{sec:related}, the literature follows two main approaches: \textit{Joint} and \textit{Coordinate} (\Cref{fig:multimodal_representation}). The former relies on projecting multimodal features into a shared latent space to produce a unique final representation (e.g., concatenation is usually the simplest approach). Conversely, the latter involves adopting a different latent space for each modality, with the possibility of setting specific constraints among modalities that are expressed, for instance, through similarity metrics. In the following, we mathematically formalize the two strategies.

\noindent \textbf{Joint representation.} Let $\overline{\mathbf{c}}_x$ be the vector of extracted features for input $x$ in all modalities. In the case of \textit{Joint} representation, we assume $\mu(\cdot)$ is the function to produce the multimodal representation of the extracted features. Thus:
\begin{equation}
    \label{eq:joint_representation}
    \Tilde{c}_x = \mu(\overline{\mathbf{c}}_x),
\end{equation}
where $\Tilde{c}_x$ is the multimodal representation for input $x$. 

\noindent \textbf{Coordinate representation.} Let $\overline{c}_x^{(m)}$ be the extracted feature for input $x$ in modality $m \in \mathcal{M}_x$. In the case of \textit{Coordinate} representation, we assume $\mu_m(\cdot)$ is the multimodal representation function for modality $m$, and let $\mathcal{K}_x$ be a set of constraints on multimodal representations of input $x$. Thus, we say:
\begin{equation}
   \Tilde{c}^{(m)}_x = \mu_m(\overline{c}^{(m)}_x) \text{ \textbf{subject to} } \mathcal{K}_x, \text{ \textbf{with} } |\mathcal{K}_x|  \geq 0,
   \label{eq:correlate_representation}
\end{equation}
where $\Tilde{c}_x^{(m)}$ is the coordinate multimodal representation for input $x$ in modality $m$. Note that with~\Cref{eq:correlate_representation} we are referring to $\mathbf{\Tilde{c}}_x = [\Tilde{c}_x^{(0)}, \Tilde{c}_x^{(1)}, \dots, \Tilde{c}_x^{(|\mathcal{M}_x| - 1)}]$ as the vector of coordinate multimodal representations for input $x$ in all modalities. 

\vspace{1em}
\noindent\textsc{\bfseries \underline{Applications}}
\begin{enumerate}[leftmargin=*]
    \item \textbf{Micro-video recommendation~\cite{DBLP:conf/mm/WeiWN0HC19}.} The framework leverages three versions of the same bipartite user-item graph (i.e., one for each micro-video modality). The graph convolutional layer first aggregates the neighborhood features of the ego node and then combines the result of such aggregation with the collaborative embedding and the multimodal representation from the previous iteration. Given the formalism introduced above, we might say this approach goes under the definition of \textit{Coordinate} representation. The model adopts a linear projection for each modality to map the input into a modality-specific latent space, both in the aggregation and combination steps. No explicit constraints are introduced.
    \item \textbf{Food recommendation~\cite{DBLP:journals/tomccap/WangDJJSN21}.} Given that the user profile is not multimodal \textit{per se}, we do not recognize any multimodal representation stage. On the item side, the multimodal representation is \textit{Coordinate}, but no particular operation is performed on the extracted multimodal features. 
    \item  \textbf{Outfit fashion compatibility~\cite{DBLP:conf/mm/HanWJD17}.} Visual and textual extracted features are projected into a unique latent space, whose dimensionality is 512. According to the earlier formalism, this approach follows a \textit{Joint} multimodal representation. On the one hand, 2048-dimensional visual features are compressed into a 512-dimensional embedding through a fully connected neural network layer, which is trained end-to-end with the recommendation model. On the other hand, the textual features for each description are first projected into the 512-dimensional latent space through linear mapping (i.e., adopting a projection matrix, which is also trained end-to-end). Then, the authors adopt bag-of-words to obtain a unique representation for the description of each fashion item.
    \item \textbf{Artist and song recommendation~\cite{DBLP:conf/recsys/OramasNSS17}.} Before further processing the extracted multimodal features, both textual and audio features are normalized. Afterward, they optionally go through two separate MLPs (i.e., \textit{Coordinate} representation). 
\end{enumerate}

\subsection{Multimodal feature fusion}
\label{sec:multimodal_fusion}
As an optional third step, when \textit{Coordinate} representation is used, our multimodal schema allows an additional \textit{Fusion} step to combine all produced multimodal representations.
In the following, we describe the inference step in the two cases of \textit{Early} and \textit{Late} fusion.

\noindent \textbf{Early fusion.} Let $\Tilde{\mathbf{c}}_x$ be the vector of coordinate multimodal representations for input $x$ in all modalities. Then, let $\gamma_{e}(\cdot)$ be the function for \textit{Early} fusion. We generate the multimodal representation for input $x$ as:
\begin{equation}
    \label{eq:early_fusion}
    \Tilde{c}_x = \gamma_e(\Tilde{\mathbf{c}}_x).
\end{equation}
Note that after applying~\Cref{eq:early_fusion}, everything we describe in the following also applies to \textit{Joint} representation. We obtain the predicted output $\hat{y}$ for input $x$ as:
\begin{equation}
    \label{eq:early_fusion_predict}
    \hat{y} = \rho(\Tilde{c}_x).
\end{equation}

\noindent \textbf{Late fusion.} Let $\Tilde{c}^{(m)}_x$ be the coordinate multimodal representation for input $x$ in modality $m \in \mathcal{M}_x$. We first predict the different output values for each modality as:
\begin{equation}
    \label{eq:late_fusion_predict_1}
    \hat{y}^{(m)} = \rho(\Tilde{c}^{(m)}_x) \quad \forall m \in \mathcal{M}_x.
\end{equation}
Let $\mathbf{\hat{y}}$ be the vector of multimodal predicted outputs in all modalities. If we denote $\gamma_{l}(\cdot)$ as the function for \textit{Late} fusion, we finally aggregate (fuse) all modality-aware predictions:
\begin{equation}
    \label{eq:late_fusion_predict_2}
    \hat{y} = \gamma_{l}(\mathbf{\hat{y}}).
\end{equation}
Whatever the type of \textit{Fusion}, the literature shows that various works perform this operation differently, from more straightforward solutions such as concatenation and element-wise addition, multiplication, or average to more refined techniques (i.e., neural-based ones, like attention mechanisms). Note that, in this work, we consider \textit{Late} fusion also when multimodal representations are exploited for some specific components of the loss function; indeed, in such settings, multimodal fusion does not occur even until the very last stage of the recommendation pipeline (i.e., the calculation and optimization of the loss function). 

\vspace{1em}
\noindent\textsc{\bfseries \underline{Applications}}
\begin{enumerate}[leftmargin=*]
    \item \textbf{Micro-video recommendation~\cite{DBLP:conf/mm/WeiWN0HC19}.} The adoption of a multimodal coordinate representation requires a modality fusion phase. This is performed through element-wise addition among modalities for users and items. As this occurs before feeding them into the inference function, we categorize it as \textit{Early} fusion.
    \item \textbf{Food recommendation~\cite{DBLP:journals/tomccap/WangDJJSN21}.} No modality fusion is run over user profiles. Regarding items, authors adopt an \textit{Early} modality fusion by concatenating the visual and textual features.
    \item \textbf{Outfit fashion compatibility~\cite{DBLP:conf/mm/HanWJD17}.} As the method adopts \textit{Joint} multimodal representation (see above) no fusion takes place.
    \item \textbf{Artist and song recommendation~\cite{DBLP:conf/recsys/OramasNSS17}.} The authors explore two possibilities: if no MLP processing occurred in the multimodal representation stage, then normalized features are concatenated and fed into a one-layer MLP; otherwise, multimodal representations are connected to the one-layer MLP. They adopt an \textit{Early} multimodal fusion in both cases. 
\end{enumerate}

\subsection{Multimodal recommendation task}
Let $\mathcal{W}_\varphi$, $\mathcal{W}_\mu$, and $\mathcal{W}_\gamma$ be the sets of the additional model trainable weights from (i) feature extraction, (ii) multimodal representation, and (iii) multimodal fusion, respectively. Note that they could be empty, as the correspondent functions may be non-trainable. Then, given the set of multimodal input data $\mathcal{X}$, we extend~\Cref{eq:theta_w}:
\begin{equation}
    \label{eq:theta}
        \hat{\Theta} = \argmax_{\bm{\Theta}} prob(\bm{\Theta} | \mathcal{X}),
\end{equation}
where $\bm{\Theta} = [\bm{\Theta}_\rho, \bm{\Theta}_\varphi, \bm{\Theta}_\mu, \bm{\Theta}_\gamma]$, with
\begin{equation}
    \label{eq:weights}
    \bm{\Theta}_\varphi = [\theta_\varphi^{(0)}, \theta_\varphi^{(1)}, \dots, \theta_\varphi^{(|\mathcal{W_\varphi}| - 1)}], \quad \bm{\Theta}_\mu = [\theta_\mu^{(0)}, \theta_\mu^{(1)}, \dots, \theta_\mu^{(|\mathcal{W_\mu}| - 1)}], \quad \bm{\Theta}_\gamma = [\theta_\gamma^{(0)}, \theta_\gamma^{(1)}, \dots, \theta_\gamma^{(|\mathcal{W_\gamma}| - 1)}],
\end{equation}
as the vectors of the model's feature extractor weights, multimodal representation weights, and multimodal fusion weights, respectively. For instance, let us consider the case of simple (but popular) latent factor models leveraging multimodal features (e.g., visual Bayesian personalized ranking~\cite{DBLP:conf/aaai/HeM16} in the multimodal version proposed in~\cite{DBLP:conf/mm/Zhang00WWW21}). Specifically, we have a set of trainable weights for: (i) the inference function comprising the users' and items' embeddings $\mathcal{W}_\rho = \{\theta_\rho^{\mathcal{U}}, \theta_\rho^{\mathcal{I}}\}$ (as previously defined, see again~\Cref{sec:class_recsys}); (ii) the multimodal \textit{Coordinate} representation function comprising the projection matrices, each translating the items' multimodal input features into the same latent space as users' and items' embeddings $\mathcal{W}_\mu = \{\theta_\mu^{\mathcal{M}}\}$, with $\theta_\mu^{\mathcal{M}}$ representing such projection matrices for all modalities. Thus, in the case of multimodal recommender systems leveraging latent factor models, the inference function may be indicated:
\begin{equation}
    \hat{y} = \rho(\theta_{\rho}^{\mathcal{U}}, \theta_{\rho}^{\mathcal{I}}, \theta_{\mu}^{\mathcal{M}}, \mathcal{X}).
\end{equation}
Generally, we solve~\Cref{eq:theta} by optimizing the loss $L$:
\begin{equation}
    \label{eq:loss}
    L = L_{rec}(\bm{\Theta}, \hat{y}, y) + \alpha L_{reg}(\bm{\Theta}),
\end{equation}
where $y$ is the ground-truth value corresponding to the predicted output $\hat{y}$, and $\alpha$ is a model hyper-parameter to weight the \textit{regularization} component of the loss function (i.e., $L_{reg}$).

\vspace{1em}
\noindent\textsc{\bfseries \underline{Applications}}
\begin{enumerate}[leftmargin=*]
    \item \textbf{Micro-video recommendation~\cite{DBLP:conf/mm/WeiWN0HC19}.} As in several collaborative filtering approaches, the inference is performed through the inner product between the multimodal representations of users and items. Regarding the loss function, the authors use the broadly-adopted BPR~\cite{DBLP:conf/uai/RendleFGS09} optimization framework, maximizing the distance between predicted ratings for positive items (i.e., the ones interacted by users) and negative items (i.e., the ones not already interacted by users).
    \item \textbf{Food recommendation~\cite{DBLP:journals/tomccap/WangDJJSN21}.} User embeddings are learned for tag prediction, with a one-layer MLP used to predict scores and cross-entropy as a loss function. The final user embeddings are eventually exploited for the main task of food recommendation. Contrarily, item embeddings are directly adopted for the score prediction, run with a one-layer MLP trained on a binary cross-entropy loss.
    \item \textbf{Outfit fashion compatibility~\cite{DBLP:conf/mm/HanWJD17}.} Only the visual modality is adopted as input for the recommendation inference. However, the textual modality is exploited jointly with the visual input to minimize the contrastive component of the loss function, for whom the cosine similarity measures the distance between visual and textual modalities in the shared latent space.
    \item \textbf{Artist and song recommendation~\cite{DBLP:conf/recsys/OramasNSS17}.} Inference is run through the inner product between user and item final embeddings, while cosine distance is the chosen loss function as the learned latent embeddings are l2-normalized. 
\end{enumerate}

To conclude, \Cref{algorithm} provides a general overview of the overall multimodal schema we presented, while~\Cref{tab:validation} summarizes the main contributions of the four selected recommender systems to conceptually validate our theoretical framework.

\begin{algorithm}[!t]
\SetAlgoLined
\textbf{Input:} Set of available modalities $\mathcal{M}$; set of multimodal input data $\mathcal{X}$ and admissible modalities $\mathcal{M}_x$, $\forall x \in \mathcal{X}$.\\
\textbf{Output:} Trained model's weights $\hat{\Theta}$.\\
Initialize all model's trainable weights  $\bm{\Theta}$.\\
\Repeat{convergence}{
    extract features according to~\Cref{eq:feature_extraction}\\
    \uIf{Joint representation}{
        get joint representation according to~\Cref{eq:joint_representation}\\
        get model's prediction according to~\Cref{eq:early_fusion_predict}
    }
    \uElseIf{Coordinate representation}{
        get coordinate multimodal representations according to~\Cref{eq:correlate_representation}\\
        \uIf{Early fusion}{
            get multimodal representation according to~\Cref{eq:early_fusion}\\
            get model's prediction according to~\Cref{eq:early_fusion_predict}
        }
        \uElseIf{Late fusion}{
            get predictions for each modality according to~\Cref{eq:late_fusion_predict_1}\\
            get model's prediction according to~\Cref{eq:late_fusion_predict_2}
        }
    }
    \For{$\hat{\theta} \in \hat{\Theta}$}{
        update $\hat{\theta}$ according to~\Cref{eq:theta}, by optimizing the loss function $L$ in~\Cref{eq:loss}
    }
}
Return $\hat{\Theta}$.
\caption{Multimodal schema for multimedia recommendation}
\label{algorithm}
\end{algorithm}

\begin{table}[!t]
\caption{Four literature examples of multimodal frameworks for multimedia recommendation. For each work, we report the performed task, the considered modalities for each input to the system (e.g., user and item), the feature extraction and multimodal representation strategies, the multimodal fusion, and the adopted inference/loss functions.}
\label{tab:validation}
\centering
\resizebox{\textwidth}{!}{%
\begin{tabular}{lccccclll}
\toprule
\multicolumn{2}{c}{\textbf{Paper}} & \textbf{Input} & \textbf{Modalities} & \makecell[c]{\textbf{Multimodal}\\\textbf{Input Data}} & \makecell[c]{\textbf{Feature}\\\textbf{Extraction}} & \makecell[c]{\textbf{Multimodal}\\\textbf{Representation}} & \multicolumn{1}{c}{\textbf{Multimodal Fusion}} & \multicolumn{1}{c}{\textbf{Inference and Loss}} \\ \cmidrule{1-9}
\multirow{6}{*}{\rotatebox[origin=*]{90}{\parbox[c]{3cm}{\textbf{Micro-video}\\\textbf{recommendation}}}} & \multirow{6}{*}{\citet{DBLP:conf/mm/WeiWN0HC19}} & \multirow{3}{*}{User} & Visual & \CC{20} & \CC{20} & \multirow{6}{*}{\parbox[t]{3.5cm}{Coordinate, aggregation of node's neighborhood and combination with the ego node. Projection in both aggregation and combination.\\\phantom{ }}} & \multirow{6}{*}{\parbox[t]{4cm}{Early, modalities are combined through addition.\\\phantom{ }\\\phantom{ }\\\phantom{ }\\\phantom{ }}} & \multirow{6}{*}{\parbox[t]{4cm}{Inference via inner-product between final user and item embeddings. BPR is the optimization function.\\\phantom{ }\\\phantom{ }}}  \\ \cmidrule{4-4}
& & & Textual & \CC{20} & \CC{20} & & \\ \cmidrule{4-4}
& & & Audio & \CC{20} & \CC{20} & & \\ \cmidrule{3-6}
& & \multirow{3}{*}{Item} & Visual & Video frames & ResNet50 & & & \\ \cmidrule{4-6}
& & & Textual & \parbox[t]{2.5cm}{Captions and descriptions} & Sentence2Vector & & & \\ \cmidrule{4-6}
& & & Audio & Audio track & VGGish & & & \\ \cmidrule{1-9}
\multirow{3}{*}{\rotatebox[origin=*]{90}{\parbox[c]{3cm}{\textbf{Food}\\\textbf{recommendation}}}} & \multirow{3}{*}{\citet{DBLP:journals/tomccap/WangDJJSN21}} & User & Textual & Users' tweets & Bi-RNN \& CNN & \CC{20} & \CC{20} & \parbox[t]{4cm}{Score prediction with MLP for user tags. Loss is cross-entropy. User embedding is later adopted for recommendation.\\\phantom{ }}\\ \cmidrule{3-6} \cmidrule{9-9}
& & \multirow{2}{*}{Item} & Visual & Video Frames & VGGNet-19 & \multirow{2}{*}{\parbox[t]{3.5cm}{Coordinate, no particular operation performed.}} & \multirow{2}{*}{\parbox[t]{4cm}{Early, modalities are combined via concatenation.}} & \multirow{2}{*}{\parbox[t]{4cm}{User-item score prediction with MLP and cross-entropy.}}\\ \cmidrule{4-6} &  &  & \makecell[c]{Textual\\\phantom{ }} & \makecell[c]{Recipe ingredients\\\phantom{  }}& \makecell[c]{TextCNN\\\phantom{ }} & & & \\ \cmidrule{1-9}
\multirow{2}{*}{\rotatebox[origin=*]{90}{\parbox[c]{2.3cm}{\textbf{Outfit fashion}\\ \textbf{compatibility}}}} & \multirow{2}{*}{\citet{DBLP:conf/mm/HanWJD17}} &  \multirow{2}{*}{Item} & Visual & Product images & InceptionV3 & \multirow{2}{*}{\parbox[t]{3.5cm}{Joint, features are projected into a shared embedding space.}} & \CC{20} & \parbox[t]{4cm}{The visual modality is used for the inference.\\\phantom{ }} \\ \cmidrule{4-6} \cmidrule{9-9}
& & & Textual & Descriptions & One-hot-encode & & \CC{20} & \parbox[t]{4cm}{The textual modality is used for the contrastive component of the loss function.\\\phantom{ }} \\ \cmidrule{1-9}
\multirow{2}{*}{\rotatebox{90}{\parbox[t]{2.7cm}{\textbf{Artist and song}\\ \textbf{recommendation}}}} & \multirow{2}{*}{\citet{DBLP:conf/recsys/OramasNSS17}} & \multirow{2}{*}{Item} & Textual & Artist biography & Custom CNN & \multirow{2}{*}{\parbox[t]{3.5cm}{Coordinate, textual and audio features normalized and optionally fed into two separate MLPs.}} & \parbox[t]{4cm}{Early, either normalized features are concatenated and fed into a one-layer MLP, or multimodal representations are connected to the one-layer MLP.} & \parbox[t]{4cm}{Inference via inner-product between user and final item embeddings. Cosine distance is the loss function.} \\ \cmidrule{4-6}
& & & Audio & \makecell[c]{Audio\\spectrogram} & Custom CNN & & & \\ \bottomrule
\end{tabular}
}
\end{table}

\section{Implementation and benchmarking (RQ3)}
\label{sec:benchmarking}
In this section, we show how we integrate the proposed multimodal schema for multimedia recommendation into Elliot~\cite{DBLP:conf/sigir/AnelliBFMMPDN21}, a framework for the rigorous reproducibility and evaluation of recommender systems. Specifically, we use this implementation to benchmark six state-of-the-art multimedia recommendation approaches (i.e., VBPR~\cite{DBLP:conf/aaai/HeM16}, MMGCN~\cite{DBLP:conf/mm/WeiWN0HC19}, GRCN~\cite{DBLP:conf/mm/WeiWN0C20}, LATTICE~\cite{DBLP:conf/mm/Zhang00WWW21}, BM3~\cite{DBLP:conf/www/ZhouZLZMWYJ23}, and FREEDOM~\cite{DBLP:conf/mm/ZhouS23}). To complement our benchmarking analysis, we also select four classical recommender systems which do not leverage multimodal features, namely: BPR~\cite{DBLP:conf/uai/RendleFGS09}, NGCF~\cite{DBLP:conf/sigir/Wang0WFC19}, LightGCN~\cite{DBLP:conf/sigir/0001DWLZ020}, and SGL~\cite{DBLP:conf/sigir/WuWF0CLX21}; this is useful to assess the performance improvement obtained through the adoption of multimodal features. Additionally, Elliot provides evaluation metrics accounting for recommendation accuracy and beyond-accuracy measures, we maintain the importance of assessing the performance of such models especially on the latter category of metrics. Indeed, the related literature about multimedia recommendation has mainly focused on the performance evaluation through standard accuracy metrics, often disregarding the potentially huge relevance that other evaluation dimensions could have by measuring the novelty and diversity of the produced recommendations~\cite{DBLP:conf/kdd/MalitestaCPDN23}, or the portion of popular items recommended to users~\cite{DBLP:conf/mmir/MalitestaCPN23}.

In the following, we report on the datasets we used, the technical aspects of the (multimedia) recommender systems involved, the evaluation metrics we adopted (spanning both accuracy and beyond-accuracy recommendation metrics), the reproducibility details of our framework, and the results of the benchmarking analysis. 

\subsection{Datasets}
For the benchmarking, we use five popular~\cite{DBLP:conf/sigir/ChenCXZ0QZ19, DBLP:conf/mm/Zhang00WWW21, DBLP:conf/cikm/KimLSK22, DBLP:conf/www/ZhouZLZMWYJ23} datasets which collect the purchase history from five product categories of the Amazon catalog~\cite{DBLP:conf/www/HeM16, DBLP:conf/sigir/McAuleyTSH15}, namely, Office Products (i.e., Office), Toys \& Games (i.e., Toys), All Beauty (i.e., Beauty), Sports \& Outdoors (i.e., Sports), and Clothing Shoes \& Jewelry (i.e., Clothing). Besides containing the records of user-product interactions with timestamps and other metadata, such datasets come with the visual features extracted from the product images, stored as 4,096-dimensional embeddings which are publicly available at the same URL of the datasets\footnote{\url{https://cseweb.ucsd.edu/~jmcauley/datasets/amazon/links.html}.}. Conversely, in terms of textual modality, we adopt the same procedure indicated in~\cite{DBLP:conf/mm/Zhang00WWW21}, and concatenate the item's title, descriptions, categories, and brand, to extract the 1,024-dimensional textual embeddings through sentence transformers~\cite{DBLP:conf/emnlp/ReimersG19}. Overall dataset information can be found in~\Cref{tab:datasetInfo}.

\subsection{Recommender systems baselines}
We decide to benchmark the results of six state-of-the-art multimedia recommender systems, namely, VBPR~\cite{DBLP:conf/aaai/HeM16}, MMGCN~\cite{DBLP:conf/mm/WeiWN0HC19}, GRCN~\cite{DBLP:conf/mm/WeiWN0C20}, LATTICE~\cite{DBLP:conf/mm/Zhang00WWW21}, BM3~\cite{DBLP:conf/www/ZhouZLZMWYJ23}, and FREEDOM~\cite{DBLP:conf/mm/ZhouS23}. Such approaches represent a group of techniques that are widely recognized in the related literature as strong baselines in multimedia recommendation exploiting multimodality, as well as recently proposed solutions at top-tier conferences (\Cref{tab:baselines}). Moreover, we also present the four selected recommendation solutions which do not leverage multimodal features: BPR~\cite{DBLP:conf/uai/RendleFGS09}, NGCF~\cite{DBLP:conf/sigir/Wang0WFC19}, LightGCN~\cite{DBLP:conf/sigir/0001DWLZ020}, and SGL~\cite{DBLP:conf/sigir/WuWF0CLX21}. In the following, we describe the main technical details for each of them. 

\vspace{1em}
\noindent \ul{Classical recommendation}
\subsubsection{BPR} Bayesian-personalized ranking~\cite{DBLP:conf/uai/RendleFGS09}, also known as BPR, is among the pioneer optimization algorithms for recommendation. Specifically, it works by maximizing the distance of predicted ratings for positive and negative items for the same user. Coupled with the matrix factorization~\cite{DBLP:journals/computer/KorenBV09} (MF) strategy, which decouples the user-item interaction matrix into latent factors (i.e., embeddings) representing users and items, it has been exploited as building block for a large plethora of recommender systems from the last decades.

\subsubsection{NGCF} Neural graph collaborative filtering~\cite{DBLP:conf/sigir/Wang0WFC19}, indicated as NGCF, is among the first techniques in recommendation leveraging the representational power of graph neural networks (GNNs). Its message-passing formulation works by aggregating the information coming from the neighborhood nodes into the ego nodes at different distance hops, and it also leverages the inter-dependencies among the ego and the neighborhood nodes.

\subsubsection{LightGCN} Light graph convolutional network~\cite{DBLP:conf/sigir/0001DWLZ020}, always referred to as LightGCN, propose a light-weight formulation of the graph convolutional layer proposed by~\citet{DBLP:conf/iclr/KipfW17}, suggesting (and empirically demonstrating) that such a variation could lead to superior performance in personalized recommendation. To be specific, the authors design the model such that by dropping feature transformations and non-linearities from the message-passing.

\subsubsection{SGL} Self-supervised graph learning~\cite{DBLP:conf/sigir/WuWF0CLX21}, presented as SGL, proposes to apply self-supervised~\cite{DBLP:conf/cikm/HuangXW0Y22} and contrastive~\cite{DBLP:conf/nips/KhoslaTWSTIMLK20} learning to GNN-based recommendation. The model is based upon a LightGCN backbone, but is designed to learn different views of nodes by performing node/edge dropout and random walk operations on the underlying user-item graph structure. At this stage, a self-supervised contrastive loss component is added to promote the consistency and divergence among same and different views of the same and different nodes, respectively.

\vspace{1em}
\noindent \ul{Multimedia recommendation}
\subsubsection{VBPR}
Visual-bayesian personalized ranking~\cite{DBLP:conf/aaai/HeM16}, abbreviated as VBPR, represents one of the pioneering efforts to integrate visual-aware content such as high-level visual features derived from product images, into the BPR-MF recommendation algorithm. In addition to the collaborative embeddings for users and items, the authors introduce a pair of visual embeddings for users and items. The latter captures high-level features extracted from product images using a pre-trained convolutional neural network. The prediction score is determined by summing the inner products of the collaborative and visual embeddings. While VBPR was initially conceived as a recommendation system focused on a single modality, we have followed the approach outlined in~\cite{DBLP:conf/mm/Zhang00WWW21} to introduce additional user and item embeddings for each additional modality, mirroring the methodology applied to the visual modality.

\begin{table}[!t]
    \caption{Statistics of the tested datasets.}\label{tab:datasetInfo}
    \centering
    \begin{tabular}{lcccc}
    \toprule
        \textbf{Datasets} & $\bm{|\mathcal{U}|}$ & $\bm{|\mathcal{I}|}$ & $\bm{|\mathcal{R}|}$ & \textbf{Sparsity (\%)}\\ \cmidrule{1-5}
        Office & 4,905 & 2,420 & 53,258 & 99.55\% \\
        Toys & 19,412 & 11,924 & 167,597 & 99.93\% \\ 
        Beauty & 22,363 & 12,101 & 198,502 & 99.93\% \\
        Sports & 35,598 & 18,357 & 296,337 & 99.95\% \\
        Clothing & 39,387 & 23,033 & 278,677 & 99.97\% \\
        \bottomrule
    \end{tabular}
\end{table}

\subsubsection{MMGCN}
Multimodal graph convolution network~\cite{DBLP:conf/mm/WeiWN0HC19}, referred to as MMGCN, introduces the concept of multimodality in graph-based recommendation systems. In this approach, the authors suggest training distinct graph convolutional networks for each modality under consideration. This results in three sets of user and item representations, accommodating the various perspectives that users may have toward each modality. Ultimately, all the embeddings related to different modalities, both for users and items, are merged through element-wise addition, and the preference prediction score is computed using an inner product. 

\subsubsection{GRCN}
Graph-refined convolutional network~\cite{DBLP:conf/mm/WeiWN0C20}, denoted as GRCN leverages the information from various modalities to refine the values of the adjacency matrix. This is particularly important due to the implicit nature of user-item interactions in the bipartite graph. The primary objective is to identify and potentially remove edges that do not accurately reflect each user's actual preferences. The multimodal representations of users and items that are learned are then fused through concatenation to produce a comprehensive representation for predicting preference scores. 

\subsubsection{LATTICE}
Latent structure mining method for multimodal recommendation~\cite{DBLP:conf/mm/Zhang00WWW21}, dubbed LATTICE, creates a similarity graph between items for each modality and enhances this structure using graph structure learning techniques. These improved adjacency matrices are then merged through a weighted sum, assigning varying importance weights to each modality. The resulting adjacency matrix is employed to refine item embeddings using a graph convolutional network. Ultimately, the obtained item embeddings can serve as the building blocks for various user and item latent factor-based models, such as BPR-MF.

\subsubsection{BM3} Bootstrapped multimodal model~\cite{DBLP:conf/www/ZhouZLZMWYJ23}, indicated as BM3, proposes a self-supervised multimodal technique for recommendation. Different from previous approaches using computationally expensive augmentations, BM3 leverages dropout as a simple operation for generating contrastive views of the same embeddings. In detail, the loss function consists of three components, where a reconstruction loss minimizes the similarity between the contrastive views of user and item embeddings, while an inter- and intra-modality alignment loss works to minimize the distance between the contrastive views generated for the same or different modalities.

\subsubsection{FREEDOM}
The authors from~\cite{DBLP:conf/mm/ZhouS23} demonstrate that freezing the item-item multimodal similarity graph (derived from LATTICE) and denoising the user-item graph can lead to improved recommendation performance (the proposed model is named FREEDOM). As for the denoising operation of the user-item graph, the authors propose a degree-sensitive edge pruning to remove noisy edges from the user-item adjacency matrix. Moreover, and differently from LATTICE, the model optimizes a double BPR-like loss function, where the first component of the loss integrates a multimodal-enhanced representation of the item embedding, while the second component explicitly leverages the item projected multimodal features.

\begin{table}[!t]
    \caption{An overview on the selected multimedia recommender systems, along with: (i) their year and publication venue; (ii) whether they represent users and/or items through multimodal embeddings; (iii) whether they use GNNs and (in case) on which type of graph (U-U and I-I stand for user-user and item-item, respectively); (iv) a non-exhaustive set of papers where they are used as baselines.}\label{tab:baselines}
    \centering
    \begin{tabular}{lcccccccc}
    \toprule
        \multirow{3}{*}{\textbf{Models}} & \multirow{3}{*}{\textbf{Year}} & \multirow{3}{*}{\textbf{Venue}} & 
        \multicolumn{2}{c}{\makecell[c]{\textbf{Multimodal}\\\textbf{embeddings}}} & \multirow{3}{*}{\textbf{GNN}} & \multicolumn{2}{c}{\makecell[c]{\textbf{Multimodal}\\\textbf{graphs}}} & 
        \multirow{3}{*}{\textbf{Baseline in}}\\ \cmidrule{4-5} \cmidrule{7-8}
        & & & \textbf{Users} & \textbf{Items} & & \textbf{U-I} & \textbf{I-I} & \\ \cmidrule{1-9}
        VBPR~\cite{DBLP:conf/aaai/HeM16} & 2016 & AAAI & \cmark & \cmark & \xmark & \multicolumn{2}{c}{\CC{20}} & \cite{DBLP:conf/sigir/ChenCXZ0QZ19,DBLP:journals/tois/ChengCZKK19,DBLP:conf/mm/LiuCSWNK19,DBLP:journals/ipm/TaoWWHHC20,DBLP:conf/mm/Zhang00WWW21,DBLP:journals/tois/LiuXGWLH23}\\
        MMGCN~\cite{DBLP:conf/mm/WeiWN0HC19} & 2019 & MM & \cmark & \cmark & \cmark & \cmark & \xmark & \cite{DBLP:conf/cikm/SunCZWZZWZ20,DBLP:journals/tmm/WeiWHNRC22,DBLP:journals/tmm/WangWYWSN23,DBLP:journals/tmm/ZhanLASDK22,DBLP:conf/www/ZhouZLZMWYJ23,DBLP:conf/www/WeiHXZ23} \\
        
        GRCN~\cite{DBLP:conf/mm/WeiWN0C20} & 2020 & MM & \xmark & \cmark & \cmark & \cmark & \xmark &\cite{10.1145/3573010,DBLP:conf/mm/Zhang00WWW21,DBLP:journals/tois/LiuXGWLH23,DBLP:conf/www/ZhouZLZMWYJ23,DBLP:conf/www/WeiHXZ23,10075502} \\
        LATTICE~\cite{DBLP:conf/mm/Zhang00WWW21} & 2021 & MM & \xmark & \cmark & \cmark & \xmark & \cmark &\cite{DBLP:conf/www/ZhouZLZMWYJ23,DBLP:conf/www/WeiHXZ23,DBLP:journals/corr/abs-2111-00678,DBLP:conf/cikm/KimLSK22,DBLP:conf/mm/MuZT0T22} \\
        BM3~\cite{DBLP:conf/www/ZhouZLZMWYJ23} & 2023 & WWW & \xmark & \cmark & \cmark & \cmark & \xmark &\cite{DBLP:journals/corr/abs-2302-04473, DBLP:conf/mm/LiuC0NK23, DBLP:conf/mm/Yu0LB23} \\
        FREEDOM~\cite{DBLP:conf/mm/ZhouS23} & 2023 & MM & \xmark & \cmark & \cmark & \xmark & \cmark &\cite{DBLP:journals/corr/abs-2302-04473} \\
        \bottomrule
    \end{tabular}
\end{table}

\subsection{Evaluation metrics}
To conduct the benchmarking analysis, we measure the recommendation performance through accuracy and beyond-accuracy metrics. While the former are commonly adopted measures in the related literature on multimedia recommendation, the latter have been brought to the attention of the community only in recent works, especially to assess the novelty and diversity of the recommended items~\cite{DBLP:conf/kdd/MalitestaCPDN23}, or how multimedia recommender systems are biased towards suggesting popular items~\cite{DBLP:conf/mmir/MalitestaCPN23}. For the recommendation accuracy, we consider the Recall@\textit{k} and the nDCG@\textit{k}; for the novelty~\cite{DBLP:conf/recsys/VargasC11} and diversity~\cite{DBLP:conf/www/SunKNS19}, we measure the EFD@\textit{k} and the Gini@\textit{k}, respectively; for the popularity bias~\cite{DBLP:conf/recsys/AbdollahpouriBM17}, we calculate the APTL@\textit{k}; finally, as a general index of how recommendations cover the entire catalog of products, we adopt the iCov@\textit{k}. In the following, we present each of these metrics along with their mathematical formulations.

\subsubsection{Recall} The recall (Recall) evaluates to what extent the recommender systems can retrieve relevant items from the recommendation list: 
\begin{equation}\label{eq:Recall}
\mathrm{Recall} @ k=\frac{1}{|\mathcal{U}|} \sum_{u \in \mathcal{U}} \frac{\left|\operatorname{Rel}_u @ k\right|}{\left|\operatorname{Rel}_u\right|},
\end{equation}
with $\operatorname{Rel}_u$ as the set of relevant items for user $u$, and $\operatorname{Rel}_u @ k$ as the set of relevant recommended items in the top@$k$.

\subsubsection{Normalized discount cumulative gain} The normalized discount cumulative gain (nDCG) measures both the relevance and the ranking position of items from the recommendation lists, considering the various relevance degrees:
\begin{equation}\label{eq:nDCG}
\mathrm{nDCG}@k=\frac{1}{|\mathcal{U}|} \sum_u \frac{{\mathrm{DCG}_u@k}}{\mathrm{IDCG}_u @ k}, 
\end{equation}
where $\mathrm{DCG}@k = \sum_{i=1}^{k} \frac{{2^{rel_{u,i}} - 1}}{{\log_2(i+1)}}$ is the cumulative gain of relevance scores in the recommendation list, and $rel_{u,i} \in \operatorname{Rel}_u$, and $\mathrm{IDCG}$ is the cumulative gain of relevance scores for an ideal recommender system.

\subsubsection{Expected free discovery} The expected free discovery (EFD), as introduced in~\cite{DBLP:conf/recsys/VargasC11}, is a metric accounting for novelty in recommendation that utilizes the \textit{inverse collection frequency}. Specifically, it provides a quantification on how a recommender system can suggest relevant \textit{long-tail} items (i.e., niche products):

\begin{equation}
    \mathrm{EFD}@k=C \sum_{i_k \in R} \operatorname{disc}(k) P \left(\operatorname{Rel}_u @ k \mid i_k, u\right) \cdot \left(-\log_2 p(i \mid \text { seen, } \theta)\right).
\end{equation}

\subsubsection{Gini index} The Gini index (Gini) indicates the disparity in items' popularity when considering the recommendation lists for each user. In detail, it provides a measurement of how certain items are consistently favored by a large portion of users with respect to other items. For the sake of this analysis, we use a normalized version of Gini according to which high values stand for a wide set of items being suggested to users, and low values indicate that a few popular items are generally recommended, leading to a less diverse recommendation experience~\cite{DBLP:conf/www/SunKNS19}. Its (normalized) formulation is:
\begin{equation}
\mathrm{Gini}@ k = 1 - \left(\frac{\sum_{i=1}^{|\mathcal{I}|}(2 i-|\mathcal{I}|-1) P|_{@k}(i)}{|\mathcal{I}| \sum_{i=1}^{|\mathcal{I}|} P|_{@k}(i)}\right),
\end{equation}
where $P|_{@k}(i)$ represents the popularity of item $i$, in the top@$k$ recommendation lists, sorted in non-decreasing order (i.e., $P|_{@k}(i) \leq P|_{@k}(i+1)$).

\subsubsection{Average percentage of long-tail items} \label{sec:aplt}
The average percentage of long-tail items (APLT) assesses the presence of popularity bias in recommendation~\cite{DBLP:conf/recsys/AbdollahpouriBM17}. When referring to popularity bias, we indicate the tendency of recommender systems to boost the recommendation of popular/mainstream items (i.e., \textit{short-head}) at the detriment of unpopular/niche products (i.e., \textit{long-tail}), thus limiting the exposure of certain item categories to users. The APLT calculates the percentage of \textit{long-tail} items belonging to the recommendation lists (averaged over all users):
\begin{equation}
\mathrm{APLT}@k=\frac{1}{\left|\mathcal{U}\right|} \sum_{u \in \mathcal{U}} \frac{|\{i\; |\; i \in(\hat{\mathcal{I}}_u@k \;\cap\; \sim\Phi)\}|}{k},
\end{equation}
where $\hat{\mathcal{I}}_u@k$ is the list of top@$k$ recommended items for user $u$, and $\Phi$ is the set of items belonging to the \textit{short-tail} distribution while $\sim\Phi$ stands for the remaining \textit{long-tail} items.

\subsubsection{Item coverage}
The item coverage (iCov) is a percentage estimating how the recommended items span the entire catalog of products in the recommendation system:
\begin{equation}\label{eq:Coverage}
\mathrm{iCov}@k = \frac{{|\bigcup_{u} \hat{\mathcal{I}}_u@k|}}{{|\mathcal{I}_{train}|}}\%,
\end{equation}
where $\mathcal{I}_{train}$ is the set of items in the training set.

\subsection{Reproducibility}
First, we pre-process the datasets following the 5-core filtering on users and items to remove cold-start users and items as done in~\cite{DBLP:conf/mm/Zhang00WWW21}. Second, we split them according to the 80:20 hold-out strategy for the training and test sets, where the former and the latter contain the 80\% and the 20\% of interactions recorded for each user, respectively. Then, we decide to train the recommendation models so that the number of epochs (i.e., 200) and the batch size (i.e., 1024) are the same for all of them to ensure fair comparison. As for the other models' hyper-parameters, we empirically find that, even if we consider (partially) different datasets with respect to the ones from the original works, exploring only the learning rate and regularization coefficient while fixing the remaining (model-specific) hyper-parameters to the indicated best values is enough to generally obtain performance trends comparable with the original works. For this reason, we explore only the learning rate (i.e., [0.0001, 0.0005, 0.001, 0.005, 0.01]) and the regularization coefficient (i.e., [$10^{-5}$, $10^{-2}$]) for a total of 10 explorations. Note that the search spaces for the learning rate and the regularization coefficient overlap across all baselines because this (i) is what usually happens in the original works, and (ii) ensures (once again) fair comparison. Finally, to select the best configuration for each model and dataset, we remove the 50\% of the test set for the validation set (following again~\cite{DBLP:conf/mm/Zhang00WWW21}), and select the hyper-parameter setting providing the highest Recall@20 value on the validation data measured for a specific epoch (maximum 200 epochs). To foster the reproducibility of the proposed benchmarks, we provide the codes, datasets, and configuration files to replicate our results at:~\url{https://github.com/sisinflab/Formal-MultiMod-Rec}, where we integrated the selected multimedia recommender systems into Elliot~\cite{DBLP:conf/sigir/AnelliBFMMPDN21}. Explored hyper-parameter values are also reported in~\Cref{tab:hyperparameters}.

\begin{table*}[!t]
\caption{Hyper-parameter values as explored in our benchmarking. Note that values reported within ``$[\dots]$'' are explored in different settings, while those with ``$(\dots)$'' refer to different values for the same configuration (e.g., multimodal factors for each modality).}\label{tab:hyperparameters}
\centering
\begin{adjustbox}{width=\textwidth,center}
\begin{tabular}{lll}
\toprule
\textbf{Families} & \textbf{Models} & \textbf{Hyper-parameters} \\  \cmidrule{1-3}
\multirow{10}{*}{Classical} & BPR & \parbox[t]{13cm}{\texttt{learning\_rate}: [0.0001, 0.0005, 0.001, 0.005, 0.01], \texttt{factors}: 64, \texttt{regularization}: [1e-2, 1e-5]} \\ \\
& NGCF & \parbox[t]{13cm}{\texttt{learning\_rate}: [0.0001, 0.0005, 0.001, 0.005, 0.01], \texttt{factors}: 64, \texttt{regularization}: [1e-2, 1e-5], \texttt{n\_layers}: 3, \texttt{weight\_size}: 64, \texttt{node\_dropout}: 0.1, \texttt{message\_dropout}: 0.1, \texttt{adj\_normalization}: True} \\ \\
& LightGCN & \parbox[t]{13cm}{\texttt{learning\_rate}: [0.0001, 0.0005, 0.001, 0.005, 0.01], \texttt{factors}: 64, \texttt{regularization}: [1e-2, 1e-5], \texttt{n\_layers}: 3, \texttt{adj\_normalization}: True} \\ \\
& SGL & \parbox[t]{13cm}{\texttt{learning\_rate}: [0.0001, 0.0005, 0.001, 0.005, 0.01], \texttt{factors}: 64, \texttt{regularization}: [1e-2, 1e-5], \texttt{n\_layers}: 3, \texttt{node\_dropout}: 0.1, \texttt{ssl\_temp}: 0.2, \texttt{ssl\_reg}: 0.1, \texttt{ssl\_ratio}: 0.1, \texttt{sampling}: edge\_dropout} \\  \cmidrule{1-3}
\multirow{18}{*}{Multimedia} & VBPR & \parbox[t]{13cm}{\texttt{learning\_rate}: [0.0001, 0.0005, 0.001, 0.005, 0.01], \texttt{factors}: 64, \texttt{regularization}: [1e-2, 1e-5], \texttt{comb\_mod}: concat} \\ \\
& MMGCN & \parbox[t]{13cm}{\texttt{learning\_rate}: [0.00001, 0.00003, 0.0001, 0.001, 0.01], \texttt{factors}: 64, \texttt{regularization}: [1e-2, 1e-5], \texttt{n\_layers}: 3, \texttt{factors\_mm}: (256, None), \texttt{aggregation}: mean, \texttt{concatenation}: False, \texttt{has\_id}: True}\\ \\
& GRCN & \parbox[t]{13cm}{\texttt{learning\_rate}: [0.0001, 0.001, 0.01, 0.1, 1], \texttt{factors}: 64, \texttt{regularization}: [1e-2, 1e-5], \texttt{n\_layers}: 2, \texttt{n\_routings}: 3, \texttt{factors\_mm}: 128, \texttt{aggregation}: add, \texttt{weight\_mode}: confid, \texttt{pruning}: True, \texttt{has\_act}: False, \texttt{fusion\_mode}: concat} \\ \\
& LATTICE & \parbox[t]{13cm}{\texttt{learning\_rate}: [0.0001, 0.0005, 0.001, 0.005, 0.01], \texttt{factors}: 64, \texttt{regularization}: [1e-2, 1e-5], \texttt{n\_layers}: 1, \texttt{n\_ui\_layers}: 2 \texttt{top\_k}: 20, \texttt{l\_m}: 0.7, \texttt{factors\_mm}: 64} \\ \\
& BM3 & \parbox[t]{13cm}{\texttt{learning\_rate}: [0.0001, 0.0005, 0.001, 0.005, 0.01], \texttt{factors}: 64, \texttt{regularization}: [1e-1, 1e-2], \texttt{n\_layers}: 2, \texttt{cl\_weight}: 2.0, \texttt{dropout}: 0.3, \texttt{lr\_sched}: (1.0,50), \texttt{factors\_mm}: 64} \\ \\
& FREEDOM & \parbox[t]{13cm}{\texttt{learning\_rate}: [0.0001, 0.0005, 0.001, 0.005, 0.01], \texttt{factors}: 64, \texttt{regularization}: [1e-2, 1e-5], \texttt{n\_layers}: 1, \texttt{n\_ui\_layers}: 2 \texttt{top\_k}: 10, \texttt{factors\_mm}: 64, \texttt{mw}: (0.1,0.9), \texttt{dropout}: 0.8, \texttt{lr\_sched}: (1.0,50)} \\
\bottomrule
\end{tabular}
\end{adjustbox}
\end{table*}

\begin{table*}[!t]
\caption{Benchmarking results on selected datasets and multimedia recommenders for accuracy and beyond-accuracy recommendation metrics, on top@10 and top@20 lists. For each metric-dataset pair, \textbf{boldface} and \underline{underlined} indicate best and second-to-best values.}\label{tab:benchmarking}
\centering
\begin{adjustbox}{width=\textwidth,center}
\begin{tabular}{lllcccccrcccccr}
\toprule
\multirow{3}{*}{\textbf{Datasets}} & & \multirow{3}{*}{\textbf{Models}} & \multicolumn{6}{c}{top@10} & \multicolumn{6}{c}{top@20} \\ \cmidrule(lr){4-9} \cmidrule(lr){10-15} \multicolumn{3}{c}{} & \multicolumn{2}{c}{\textbf{Accuracy}} & \multicolumn{4}{c}{\textbf{Beyond-accuracy}} & \multicolumn{2}{c}{\textbf{Accuracy}} & \multicolumn{4}{c}{\textbf{Beyond-accuracy}} \\ \cmidrule(lr){4-5} \cmidrule(lr){6-9} \cmidrule(lr){10-11} \cmidrule(lr){12-15} \multicolumn{3}{c}{} & \textbf{Recall} & \textbf{nDCG} & \textbf{EFD} & \textbf{Gini} & \textbf{APLT} & \textbf{iCov} & \textbf{Recall} & \textbf{nDCG} & \textbf{EFD} & \textbf{Gini} & \textbf{APLT} & \textbf{iCov} \\ \cmidrule{1-15}
\multirow{11}{*}{\textbf{Office}} & \multirow{4}{*}{\rotatebox[origin=*]{90}{\parbox[c]{1.2cm}{Classical}}} & BPR & 0.0604 & 0.0390 & 0.1655 & 0.3675 & 0.2292 & 93.95\% & 0.0958 & 0.0501 & 0.1387 & 0.4047 & 0.2378 & \underline{97.64\%}\\
& & NGCF & 0.0582 & 0.0365 & 0.1604 & \underline{0.4428} & 0.2726 & \underline{97.47\%} & 0.0958 & 0.0478 & 0.1365 & \underline{0.4757} & 0.2817 & 99.59\%\\
& & LightGCN & \textbf{0.0802} & \textbf{0.0520} & \textbf{0.2099} & 0.1684 & 0.1594 & 76.14\% & \textbf{0.1237} & \textbf{0.0655} & \textbf{0.1740} & 0.2144 & 0.1817 & 86.70\%\\
& & SGL & 0.0695 & \underline{0.0469} & \underline{0.1991} & 0.3479 & \underline{0.2833} & 84.92\% & 0.1012 & 0.0566 & \underline{0.1592} & 0.3992 & \underline{0.2928} & 91.55\% \\
\cmidrule{3-15}
& \multirow{6}{*}{\rotatebox[origin=*]{90}{\parbox[c]{1.5cm}{Multimedia}}} & VBPR & 0.0652 & 0.0419 & 0.1753 & 0.3634 & 0.2321 & 93.83\% & 0.1025 & 0.0533 & 0.1479 & 0.3960 & 0.2375 & 97.51\% \\
& & MMGCN & 0.0455 & 0.0300 & 0.1140 & 0.0128 & 0.0016 & 3.07\% & 0.0798 & 0.0405 & 0.1027 & 0.0231 & 0.0078 & 4.64\% \\
& & GRCN & 0.0393 & 0.0253 & 0.1215 & \textbf{0.4587} & \textbf{0.3438} & \textbf{99.01\%} & 0.0667 & 0.0339 & 0.1051 & \textbf{0.4892} & \textbf{0.3469} & \textbf{99.79\%} \\
& & LATTICE & 0.0664 & 0.0449 &  0.1827 & 0.2128 & 0.1752 & 87.86\% & 0.1029 & 0.0566 & 0.1513 & 0.2652 & 0.2039 & 95.90\% \\
& & BM3 & \underline{0.0701} & 0.0460 & 0.1837 & 0.1407 & 0.1427 & 77.13\% & \underline{0.1081} & \underline{0.0583} & 0.1550 & 0.1900 & 0.1715 & 91.55\% \\
& & FREEDOM & 0.0560 & 0.0365 & 0.1493 & 0.1922 & 0.1875 & 79.12\% & 0.0884 & 0.0469 & 0.1282 & 0.2439 & 0.2080 & 90.64\% \\ \cmidrule{1-15}
\multirow{11}{*}{\textbf{Toys}} & \multirow{4}{*}{\rotatebox[origin=*]{90}{\parbox[c]{1.2cm}{Classical}}} & BPR & 0.0643 & 0.0404 & 0.1712 & 0.2673 & 0.1166 & 84.32\% & 0.0906 & 0.0482 & 0.1345 & 0.3049 & 0.1268 & 92.12\% \\
& & NGCF & 0.0622 & 0.0403 & 0.1715 & \underline{0.3535} & \underline{0.1609} & \underline{92.36\%} & 0.0902 & 0.0485 & 0.1365 & \underline{0.3960} & \underline{0.1760} & \underline{97.24\%} \\
& & LightGCN & 0.0714 & 0.0455 & 0.1810 & 0.0740 & 0.0288 & 52.69\% & 0.1036 & 0.0550 & 0.1446 & 0.1037 & 0.0374 & 68.41\% \\
& & SGL & 0.0789 & \underline{0.0519} & \underline{0.2203} & 0.2352 & 0.0977 & 78.27\% & 0.1061 & 0.0601 & \underline{0.1681} & 0.2838 & 0.1106 & 88.77\% \\
\cmidrule{3-15}
& \multirow{6}{*}{\rotatebox[origin=*]{90}{\parbox[c]{1.5cm}{Multimedia}}} & VBPR & 0.0710 & 0.0458 & 0.1948 & 0.2645 & 0.1064 & 84.90\% & 0.1006 & 0.0545 & 0.1527 & 0.3011 & 0.1180 & 92.82\% \\
& & MMGCN & 0.0256 & 0.0150 & 0.0648 & 0.0989 & 0.0961 & 37.87\% & 0.0426 & 0.0200 & 0.0570 & 0.1450 & 0.1058  & 52.51\% \\
& & GRCN & 0.0554 & 0.0354 & 0.1604 & \textbf{0.3954} & \textbf{0.2368} & \textbf{92.66\%} & 0.0831 & 0.0436 & 0.1298 & \textbf{0.4329} & \textbf{0.2482} & \textbf{97.73\%} \\
& & LATTICE & \underline{0.0805} & 0.0512 &  0.2090 & 0.1656 & 0.0546 & 73.80\% & \underline{0.1165} & \underline{0.0617} & 0.1665 & 0.2026 & 0.0684 & 86.58\% \\
& & BM3 & 0.0613 & 0.0393 & 0.1582 & 0.0776 & 0.0486 & 56.23\% & 0.0901 & 0.0478 & 0.1270 & 0.1154 & 0.0658 & 73.50\% \\
& & FREEDOM & \textbf{0.0870} & \textbf{0.0548} & \textbf{0.2284} & 0.1474 & 0.0756 & 62.09\% & \textbf{0.1249} & \textbf{0.0660} & \textbf{0.1820} & 0.2007 & 0.0951 & 78.42\% \\ \cmidrule{1-15}
\multirow{11}{*}{\textbf{Beauty}} & \multirow{4}{*}{\rotatebox[origin=*]{90}{\parbox[c]{1.2cm}{Classical}}} & BPR & 0.0676 & 0.0414 & 0.1869 & 0.2348 & \underline{0.1099} & 83.78\% & 0.0993 & 0.0511 & 0.1519 & 0.2689 & \underline{0.1195} & 91.51\% \\
& & NGCF & 0.0661 & 0.0408 & 0.1787 & \underline{0.2375} & 0.0929 & \underline{86.47\%} & 0.0987 & 0.0505 & 0.1473 & \underline{0.2754} & 0.1071 & \underline{94.08\%} \\
& & LightGCN & 0.0785 & 0.0493 & 0.2044 & 0.0555 & 0.0290 & 49.92\% & 0.1141 & 0.0599 & 0.1647 & 0.0794 & 0.0343 & 65.74\% \\
& & SGL & 0.0810 & 0.0524 & \textbf{0.2291} & 0.1554 & 0.0807 & 66.78\% & 0.1144 & 0.0626 & 0.1824 & 0.1935 & 0.0870 & 78.36\% \\
\cmidrule{3-15}
& \multirow{6}{*}{\rotatebox[origin=*]{90}{\parbox[c]{1.5cm}{Multimedia}}} & VBPR & 0.0760 & 0.0483 & 0.2119 & 0.2076 & 0.0833 & 83.06\% & 0.1102 & 0.0586 & 0.1700 & 0.2376 & 0.0915 & 91.41\% \\
& & MMGCN & 0.0496 & 0.0294 & 0.1300 & 0.0252 & 0.0282 & 13.75\% & 0.0772 & 0.0379 & 0.1105 & 0.0423 & 0.0345 & 21.37\% \\
& & GRCN & 0.0575 & 0.0370 & 0.1817 & \textbf{0.3823} & \textbf{0.2497} & \textbf{94.59\%} & 0.0892 & 0.0466 & 0.1498 & \textbf{0.4178} & \textbf{0.2608} & \textbf{98.56\%}  \\
& & LATTICE & \textbf{0.0867} & \textbf{0.0544} &  0.2272 & 0.1153 & 0.0386 & 65.82\% & \underline{0.1259} & \underline{0.0661} & \underline{0.1830} & 0.1558 & 0.0511 & 81.60\% \\
& & BM3 & 0.0713 & 0.0443 & 0.1831 & 0.0245 & 0.0179 & 32.31\% & 0.1051 & 0.0545 & 0.1490 & 0.0414 & 0.0228 & 48.75\% \\
& & FREEDOM & \underline{0.0864} & \underline{0.0539} & \underline{0.2279} & 0.0921 & 0.0486 & 55.89\% & \textbf{0.1286} & \textbf{0.0666} & \textbf{0.1868} & 0.1359 & 0.0653 & 72.96\% \\ \cmidrule{1-15}
\multirow{11}{*}{\textbf{Sports}} & \multirow{4}{*}{\rotatebox[origin=*]{90}{\parbox[c]{1.2cm}{Classical}}} & BPR & 0.0411 & 0.0251 & 0.1047 & 0.1713 & \underline{0.0709} & 76.60\% & 0.0627 & 0.0314 & 0.0861 & 0.1963 & \underline{0.0774} & 86.62\% \\
& & NGCF & 0.0419 & 0.0260 & 0.1073 & \underline{0.1821} & 0.0616 & \underline{82.50\%} & 0.0633 & 0.0323 & 0.0876 & \underline{0.2093} & 0.0708 & \underline{92.28\%} \\
& & LightGCN & 0.0558 & 0.0346 & 0.1324 & 0.0215 & 0.0086 & 33.59\% & 0.0839 & 0.0428 & 0.1081 & 0.0342 & 0.0113 & 49.49\%\\
& & SGL & 0.0545 & 0.0347 & 0.1413 & 0.0680 & 0.0305 & 45.99\% & 0.0788 & 0.0419 & 0.1122 & 0.0888 & 0.0346 & 58.65\%\\
\cmidrule{3-15}
& \multirow{6}{*}{\rotatebox[origin=*]{90}{\parbox[c]{1.5cm}{Multimedia}}} & VBPR & 0.0450 & 0.0281 & 0.1167 & 0.1501 & 0.0497 & 75.77\% & 0.0677 & 0.0349 & 0.0949 & 0.1722 & 0.0552 & 86.54\% \\
& & MMGCN & 0.0342 & 0.0207 & 0.0791 & 0.0095 & 0.0046 & 5.10\% & 0.0551 & 0.0269 & 0.0678 & 0.0168 & 0.0065 & 8.39\% \\
& & GRCN & 0.0330 & 0.0202 & 0.0885 & \textbf{0.3087} & \textbf{0.2190} & \textbf{91.28\%} & 0.0523 & 0.0259 & 0.0746 & \textbf{0.3386} & \textbf{0.2273} & \textbf{97.09\%} \\
& & LATTICE & \textbf{0.0610} & \underline{0.0372} & \underline{0.1465} & 0.0573 & 0.0129 & 48.44\% & \underline{0.0898} & \underline{0.0456} & \underline{0.1185} & 0.0802 & 0.0185 & 64.90\% \\
& & BM3 & 0.0548 & 0.0349 & 0.1372 & 0.0776 & 0.0283 & 59.13\% & 0.0825 & 0.0430 & 0.1118 & 0.1120 & 0.0385 & 76.75\% \\
& & FREEDOM & \underline{0.0603} & \textbf{0.0375} & \textbf{0.1494} & 0.0621 & 0.0319 & 48.37\% & \textbf{0.0911} & \textbf{0.0465} & \textbf{0.1219} & 0.0926 & 0.0442 & 65.81\% \\ \cmidrule{1-15}
\multirow{11}{*}{\textbf{Clothing}} & \multirow{4}{*}{\rotatebox[origin=*]{90}{\parbox[c]{1.2cm}{Classical}}} & BPR & 0.0303 & 0.0156 & 0.0427 & 0.2260 & 0.0729 & 80.00\% & 0.0459 & 0.0195 & 0.0347 & 0.2600 & 0.0824 & 89.42\% \\
& & NGCF & 0.0284 & 0.0146 & 0.0395 & \underline{0.2438} & 0.0633 & \underline{87.55\%} & 0.0443 & 0.0186 & 0.0327 & 0.2788 & 0.0743 & \underline{95.65\%} \\
& & LightGCN & 0.0393 & 0.0210 & 0.0534 & 0.0379 & 0.0086 & 41.61\% & 0.0602 & 0.0262 & 0.0438 & 0.0569 & 0.0111 & 58.59\%\\
& & SGL & 0.0408 & 0.0220 & 0.0583 & 0.1296 & 0.0352 & 66.77\% & 0.0580 & 0.0263 & 0.0457 & 0.1699 & 0.0435 & 81.32\%\\
\cmidrule{3-15}
& \multirow{6}{*}{\rotatebox[origin=*]{90}{\parbox[c]{1.5cm}{Multimedia}}} & VBPR & 0.0339 & 0.0181
& 0.0502 & 0.2437 & \underline{0.0809} & 83.40\% & 0.0529 &
0.0229 & 0.0413 & \underline{0.2791} & \underline{0.0915} & 92.33\% \\
& & MMGCN & 0.0227 & 0.0119 & 0.0292 & 0.0136 & 0.0044 & 7.58\% & 0.0348 & 0.0150 & 0.0240 & 0.0236 & 0.0066 & 12.44\% \\
& & GRCN & 0.0319 & 0.0164 & 0.0481 & \textbf{0.3990} & \textbf{0.2358} & \textbf{93.37\%} & 0.0496 & 0.0209 & 0.0397 & \textbf{0.4368} & \textbf{0.2459} & \textbf{97.77\%} \\
& & LATTICE & \underline{0.0502} & \underline{0.0275} & \underline{0.0738} & 0.1022 & 0.0134 & 58.49\% & \underline{0.0744} & \underline{0.0336} & \underline{0.0589} & 0.1384 & 0.0207 & 76.20\% \\
& & BM3 & 0.0418 & 0.0226 & 0.0596 & 0.1348 & 0.0319 & 72.88\% & 0.0633 & 0.0281 & 0.0486 & 0.1825 & 0.0449 & 88.65\% \\
& & FREEDOM & \textbf{0.0547} & \textbf{0.0294} & \textbf{0.0805} & 0.1509 & 0.0600 & 65.54\% & \textbf{0.0822} & \textbf{0.0363} & \textbf{0.0652} & 0.2078 & 0.0843 & 81.91\% \\
\bottomrule
\end{tabular}
\end{adjustbox}
\end{table*}

\subsection{Benchmarking results}
We organize the proposed benchmarking analysis as follows. First, we present the results regarding the overall performance of both classical and multimedia recommender systems in their best configurations. Then, we provide a finer-grained investigation on the impact of multimodal features under different settings.

\subsubsection{Overall performance}
\Cref{tab:benchmarking} reports on the results of the extensive benchmarking analysis we conduct on the selected datasets and state-of-the-art classical and multimedia recommendation systems (in the Table categorized under the `Classical' and `Multimedia' fields, respectively). The calculated metrics involve both accuracy (i.e., Recall and nDCG) and beyond-accuracy (i.e., EFD, Gini, APLT, and iCov) measures when considering top@10 and top@20 recommendation lists. Based on how we defined all the recommendation metrics, higher values indicate better performance.

In terms of \textbf{accuracy} performance, we observe that one of LATTICE, BM3, and FREEDOM is steadily among the two best recommendation models, outperforming the best classical recommendation approaches by a great margin; indeed, this is something that also emerges from the related literature. This holds across all datasets and top@\textit{k} under analysis, with the only exception of Office. Nevertheless, we notice an interesting trend when comparing VBPR to the other multimedia recommendation techniques. Particularly, we see how this model is almost always among the top-3 recommendation techniques despite being one of the shallowest approaches compared to other more recent and complex models. As already stated in recent works~\cite{DBLP:conf/kdd/MalitestaCPDN23, DBLP:conf/mmir/MalitestaCPN23}, this finding demonstrates how even a not-so-deep, but still careful hyper-parameter exploration (such as the one we performed) may help uncover unexpected results with respect to what described in the literature.

However, the most surprising behaviors involves the analysis of the \textbf{beyond-accuracy} performance. First, we notice that, differently from what observed in the accuracy analysis, the classical recommendation approaches almost always settle as second-to-best solutions under beyond-accuracy metrics, surpassing most of the multimedia counterparts. Additionally, when considering the multimedia recommendation scenario only, models dominating the accuracy performance such as LATTICE, BM3, and FREEDOM cannot provide similar results on the other metrics accounting for novelty, diversity, and popularity bias. Indeed, the only observable trend in this setting is that GRCN steadily settle as best-performing algorithms. Particularly, it is worth pointing out how this approach can strike a sufficient trade-off between accuracy and beyond-accuracy measures. Once again, such observations corroborate what has recently been pointed out in similar works~\cite{DBLP:conf/kdd/MalitestaCPDN23, DBLP:conf/mmir/MalitestaCPN23}, by extending the analysis to additional datasets and recommendation systems.

\subsubsection{Impact of different multimodal feature settings} To assess the impact of injecting multimodal features into the recommendation pipeline, we select BPR~\cite{DBLP:conf/uai/RendleFGS09} as recommender system, and calculate the performance variation in three different settings: (i) \textbf{visual} modality only (i.e., VBPR~\cite{DBLP:conf/aaai/HeM16} in its original formulation), (ii) \textbf{textual} modality only, and (iii) \textbf{visual + textual} modalities (the one reported in the overall performance from above). Overall (\Cref{tab:ablation}), results show an interesting trend, which complements the observation from the previous analysis. While on the accuracy metric (i.e., Recall@20) the integration of (multi)modal features generally leads to improved recommendation performance, the same is not true on the other two beyond-accuracy measures (i.e., APLT@20 and iCov@20). Indeed, in the latter setting, it is evident how even the base configuration (i.e., without multimodal features) can reach best or second-to-best performance. The only exception to the outlined observations is on the Clothing dataset, where (as normally acknowledged in the literature) the textual and visual+textual (i.e., multimodal) settings are the best ones. We deem this different trend to be ascribed to the specific dataset characteristics of the Clothing dataset (as already suggested by other previous works~\cite{DBLP:journals/corr/abs-2308-10778, DBLP:conf/sigir/DeldjooNSM20}) and not to the multimodal characteristics of such a dataset.

Summing up, the proposed benchmarking studies indicate that training and evaluating multimedia recommender systems remain an open challenge in the literature, showing (in same cases) unexpected outcomes that pave the way to more rigorous and careful analyses to be conducted in future work (see \Cref{sec:future-reproducibility}).

\begin{table*}[!t]
\caption{Recommendation results calculated as Recall@20, APLT@20, and iCov@20 on all tested datasets for BPR in four configurations: (i) without multimodal features (i.e., the original BPR~\cite{DBLP:conf/uai/RendleFGS09}), (ii) BPR + visual (i.e., VBPR in its original version~\cite{DBLP:conf/aaai/HeM16}), (iii) BPR + textual, and (iv) BPR + (visual + textual). \textbf{Boldface} and \underline{underlined} stand for best and second-to-best values.}\label{tab:ablation}
\centering
\begin{tabular}{llccc}
\toprule
\textbf{Datasets} & \textbf{Models} & \textbf{Recall@20} & \textbf{APLT@20} & \textbf{iCov@20} \\ \cmidrule{1-5}
\multirow{4}{*}{\textbf{Office}} & \cellcolor{gray_ablation} BPR & \cellcolor{gray_ablation} 0.0958 & \cellcolor{gray_ablation} \textbf{0.2378} & \cellcolor{gray_ablation} \textbf{97.64\%} \\
& + visual  & \underline{0.1002} & 0.2361 & 97.43\% \\
& + textual & 0.0997 & 0.2256 & 97.47\% \\
& + (visual + textual) & \textbf{0.1025} & \underline{0.2375} & \underline{97.51\%}\\ \cmidrule{1-5}
\multirow{4}{*}{\textbf{Toys}} & \cellcolor{gray_ablation} BPR & \cellcolor{gray_ablation} 0.0906 & \cellcolor{gray_ablation} \textbf{0.1268} & \cellcolor{gray_ablation} 92.12\% \\
& + visual & 0.1002 & 0.1164 & 92.33\% \\
& + textual & \textbf{0.1033} & \underline{0.1227} & \underline{92.50\%} \\
& + (visual + textual) & \underline{0.1006} & 0.1180 & \textbf{92.82\%} \\ \cmidrule{1-5}
\multirow{4}{*}{\textbf{Beauty}} & \cellcolor{gray_ablation} BPR & \cellcolor{gray_ablation} 0.0993 & \cellcolor{gray_ablation} \textbf{0.1195} & \cellcolor{gray_ablation} 91.51\% \\
& + visual & \underline{0.1074} & 0.1116 & \underline{92.14\%} \\
& + textual & 0.1066 & \underline{0.1174} & \textbf{92.48\%} \\
& + (visual + textual) & \textbf{0.1102} & 0.0915 & 91.41\%\\ \cmidrule{1-5}
\multirow{4}{*}{\textbf{Sports}} & \cellcolor{gray_ablation} BPR & \cellcolor{gray_ablation} 0.0627 & \cellcolor{gray_ablation} \underline{0.0774} & \cellcolor{gray_ablation} \underline{86.62\%}\\
& + visual & \textbf{0.0691} & 0.0702 & 86.14\%\\
& + textual & \underline{0.0678} & \textbf{0.0875} & \textbf{89.11\%}\\
& + (visual + textual) & 0.0677 & 0.0552 & 86.54\%\\ \cmidrule{1-5}
\multirow{4}{*}{\textbf{Clothing}} & \cellcolor{gray_ablation} BPR & \cellcolor{gray_ablation} 0.0459 & \cellcolor{gray_ablation} 0.0824 & \cellcolor{gray_ablation} 89.42\% \\
& + visual & 0.0526 & 0.0833 & 90.99\%\\
& + textual & \underline{0.0528} & \textbf{0.0947} & \textbf{92.71\%}\\
& + (visual + textual) & \textbf{0.0529} & \underline{0.0915} & \underline{92.33\%}\\
\bottomrule
\end{tabular}
\end{table*}
\section{Technical Challenges (RQ4)}\label{sec:challenges}
This section aims to overview the main technical challenges we recognize in multimodal approaches for multimedia recommendation. Starting from the proposed schema we presented in the previous sections, we outline the evident (or even less evident) issues emerging from the literature. 

\subsection{Missing modalities in the input data} \label{sec:missing_modalities}
Describing data under the lens of multimodality may be a two-sided coin. From one perspective, multimodality helps enrich the informative content carried by the input, thus exploring data's multi-faceted nature to learn better-tailored user-item preference patterns~\cite{DBLP:journals/corr/abs-2403-04503, DBLP:conf/mm/MalitestaGPN23}. On the other side, the need to provide descriptive content for every input modality may come at the expense of some missing modalities (e.g., a video dataset could integrate videos having no textual content, for example, subtitles or descriptions may be only sometimes available). Tackling the modality misalignment in the data is a recent and widely discussed challenge in other domains~\cite{DBLP:conf/kdd/ZhangCMZWWZ22, DBLP:conf/sigir/ZengL022, DBLP:conf/cvpr/0002R0T022, DBLP:conf/cvpr/LeeTCL23, DBLP:conf/log/RossiK0C0B22}, and requires ad-hoc techniques to provide equal representation of all involved modalities to fully exploit their informative richness~\cite{DBLP:conf/aaai/MaRZTWP21, DBLP:conf/icassp/SunMLHN021}. Nevertheless, apart from very recent preliminary attempts~\cite{DBLP:journals/corr/abs-2403-19841}, to the best of our knowledge, the issue remains open in recommendation.

\subsection{Pre-trained feature extractors} \label{sec:pretrained-features-challenge}
Deep learning models processing images, texts, or audio have been shown to enrich the informative content of items' profiles in several recommendation algorithms. In most solutions, such architectures are used as pre-trained blocks to extract high-level features from the input data, thus exploiting the capability of deep neural networks to transfer knowledge among different datasets and/or tasks. Despite the ease of adopting ready-to-use feature extraction networks, we seek to underline a conceptual limitation that, to the best of our knowledge, is only partially investigated in the literature. Indeed, pre-trained representations extracted through state-of-the-art deep learning models are not necessarily supposed to capture those semantic features, which will likely captivate users for their final decision-making process. As an example, the embedding feature extracted from a product image (e.g., a bag) through a pre-trained deep convolutional network (e.g., ResNet50) is carrying high-level informative content driven by the task of \textit{image classification}, but this does not mean the same knowledge will be helpful to predict whether the product could be \textit{recommended} to a user.

\subsection{Modalities representation} 
\label{sec:modalities-representation-challenge}
The multimodal representation of the extracted input data is among the main stages in the multimodal schema we described since it establishes the relationships for the selected input modalities. Nonetheless, the literature is not generally aligned on its definition since most of the works usually refer to \textit{Joint} representation and \textit{Early} fusion interchangeably. We recognize this as a conceptual issue because the two stages (i.e., representation and fusion) should be considered separately. We maintain that the former stands for the initial step to set interconnections among early-extracted multimodal features, while the latter, despite dealing again with modalities relationships, involves features that have been further processed towards the task optimization (i.e., recommendation in our case), thus embodying different rationales and techniques. Furthermore, the related literature suggests two possible solutions to multimodal representation, either \textit{Joint} or \textit{Coordinate}, where the latter additionally requires the subsequent fusion step. However, each of the paradigms' advantages and whether they might depend on the task remain under investigation. 

\subsection{Multimodal-aware fusion and optimization}\label{sec:fusion-challenge} While multimodal representation builds on input modalities in the early stages of the schema, multimodal fusion accounts for multimodal features that have already been processed, with a specific focus on the last steps (i.e., inference and model optimization). Similarly to what was observed above, multimodal fusion may come in the form of \textit{Early} or \textit{Late} fusion. The significant difference between the two approaches lies in preserving or not modalities separation during the inference (i.e., \textit{Late} and \textit{Early}, respectively). The literature demonstrates the vast predominance of \textit{Early} solutions, whereas several works quite often refer to \textit{Late} fusion by mistaking it for \textit{Early} fusion. Indeed, providing a precise definition for the two is of paramount importance because the two approaches may serve different purposes. The rationale of \textit{Late} fusion is to keep the modalities separation explicit during the inference phase so that the contribution of each modality is observable up to the last operation. Moreover, the literature is not aligned on the operation to fuse modalities. Non-trainable fusion functions (e.g., element-wise addition) are usually the preferred direction given that it is more lightweight and easy to perform to trainable approaches (e.g., neural networks) which (on their side) may allow to better tailor user-item preference prediction. 

\section{Future directions (RQ4 --- extension)}
\label{sec:future-directions}

The scope of this section is to outline possible future directions for the application of multimodality for multimedia recommendation. While some of the presented solutions may apply to the above-raised challenges, we also discuss different research paths we suggest to follow in future work. 

\subsection{Domain-specific multimodal features} \label{sec:domain-specific-features} Given the limitations imposed by the adoption of pre-trained multimodal features (see again~\Cref{sec:pretrained-features-challenge}), we wish to underline the benefits of \textit{domain-specific} features in the multimodal schema we have outlined. Extracting such high-level features from input data would entail injecting meaningful and task-aware informative content into the recommendation system, thereby better-profiling items and users on the platform to generate more tailored recommendations. Domain-specific features should necessitate domain-specific extraction models, which may have been previously trained and optimized on similar tasks to the one we are pursuing. Regarding \textit{fashion} recommendation, for instance, we recall the work by~\citet{DBLP:conf/cvpr/GeZWTL19}, a pre-trained architecture for the comprehensive visual analysis of clothing photos. Another example is the \textit{food} recommendation system proposed by~\citet{DBLP:journals/tois/YangHYPDBCE17}, which analyzes food-related photos.

Furthermore, in the field of \textit{audio} and \textit{text} understanding and classification,~\citet{DBLP:conf/icassp/ChoiFSC17} construct a deep model based on convolutional recurrent neural networks for music tagging by taking into account songs' local features and temporal characteristics, whereas~\citet{DBLP:conf/emnlp/BarbieriCAN20} tackle the task of sentiment analysis in user-generated tweets.

\subsection{Multimodality on user-item interactions}
Multimodality is the most intuitive approach to describe the multifaceted nature of items in multimedia recommendation~\cite{deldjoo2022multimedia,DBLP:journals/csur/DeldjooSCP20}, but this does not hold for the users' profile.

First, from a technical point of view, profiling each user through multimodal features (e.g., her voice, her visual appearance) would require sophisticated technologies that users' digital devices could not necessarily support (e.g., smartphones). Second, from a practical point of view, it is likely that users would not be disposed to share such personal data on online platforms, primarily for privacy concerns. Despite the raised critical aspects, a few examples from the literature~\cite{DBLP:conf/mm/WeiWN0HC19, DBLP:journals/ipm/TaoWWHHC20} propose to model the user profile in such a way that her preferences toward each multimodal aspect of items are made explicit and learned during model's training. However, these systems rely solely on the multimodal profiles of the items, disregarding alternative information sources. \textit{Product reviews}, which express opinions and comments about items that have been clicked, watched, or purchased, could be a valuable tool for revealing users' nuanced preferences toward each item in the recommendation system. 

\begin{figure}[!t]
\centering
    \includegraphics[width=0.45\textwidth]{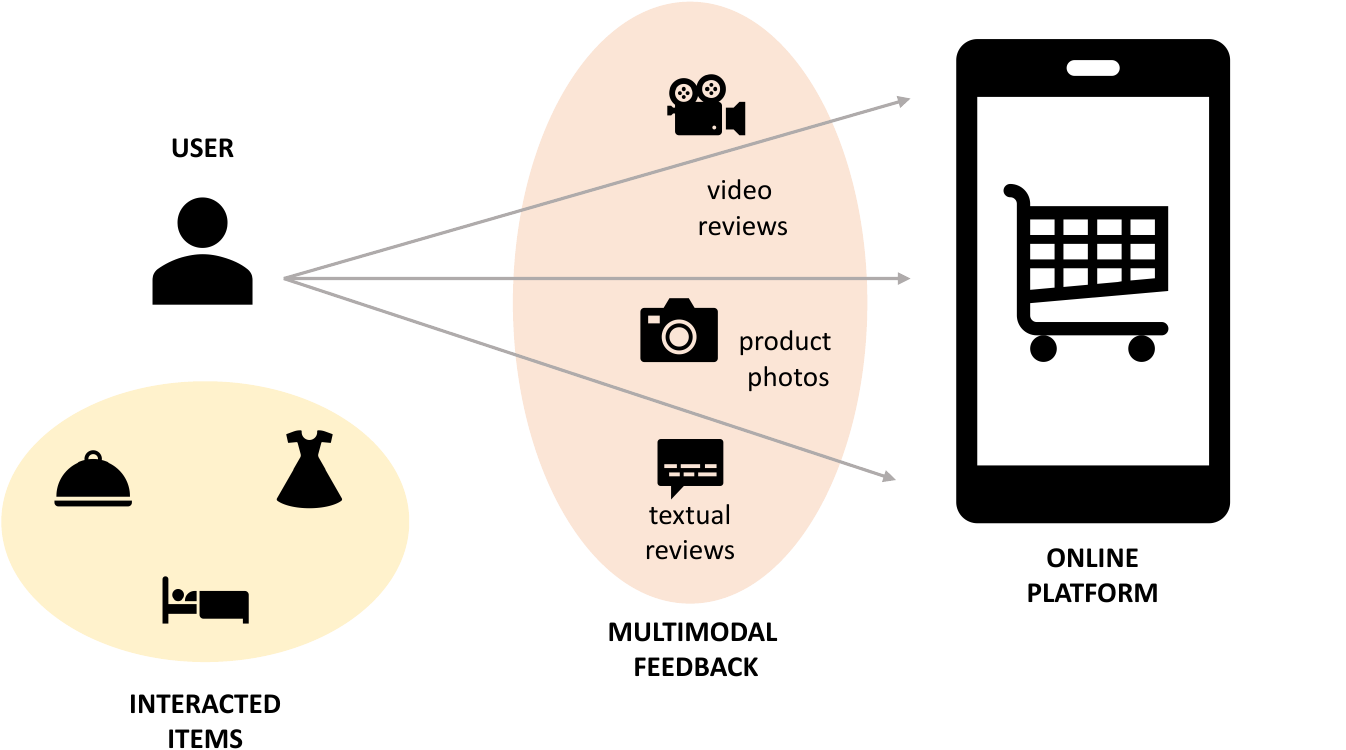}
    \caption{ \small An example of how users generate and upload \textit{multimodal feedback} about interacted items (e.g., textual reviews, product photos, or even video reviews) on online platforms. Such \textit{user-item} sources of information may be suitably exploited to better profile user' preferences~\cite{DBLP:conf/cikm/AnelliDNSFMP22}.}
    \label{fig:edge_multimodal}
\end{figure}

Existing review-based approaches~\cite{DBLP:conf/wsdm/ZhengNY17, DBLP:conf/www/0012OM21} work by integrating reviews as the \textit{textual modality} to represent \textit{items}. However, we believe that a more logical and effective way to integrate reviews would be to view them as a medium to represent \textit{user} preference \textit{over items}, thereby providing additional and complementary preference scores in addition to numeric ratings or implicit feedback that are typically used to compute recommendations. Such reasoning may be easily generalized to include user-generated data regarding interacting things (such as images or videos of delivered products), which we can characterize as \textit{multimodal feedback} (see~\Cref{fig:edge_multimodal}). When compared to numerical feedback, which tends to be \textit{atomic} (single-faceted), multimodal feedback could be considered as \textit{composite}, revealing nuance and the user's multi-faceted opinion of the products~\cite{DBLP:conf/cikm/AnelliDNSFMP22}.

\subsection{Fine-grained multimodal features} \label{sec:fusion-future} Multimodality is a way to effectively profile the multi-faceted aspects of items and users' preference (e.g., I bought this smartphone because its technical \textit{description} is quite exhaustive and its \textit{display} amazes me; I like this song since I love the \textit{music} and the \textit{lyrics}). Nevertheless, analyzing and learning users' tastes at modalities' granularity might not be enough to uncover all aspects underlying every user-item interaction. In contexts where modalities bring a great source of heterogeneous information, a finer-grained feature processing could help better unveil hidden facets. For instance, when it comes to the recommendation of fashion items (e.g., dresses, shoes, jewelry), user attention may be captivated by specific item visual characteristics, such as colors, shapes, and particular patterns and motifs~\cite{DBLP:conf/ecir/DeldjooNMM22}. Similarly, a song involves several features~\cite{DBLP:conf/mm/LiangLCYSC20} (i.e., pitch, rhythm, and dynamics), which could differently influence users' attitudes towards it. Uncovering and understanding details at this finer granularity should be one of the main directions toward the novel recommendation approaches of multimedia products and services.

\subsection{An extensive and fair evaluation of multimedia recommender systems}
\label{sec:future-reproducibility}
To date, very limited effort has been put into the extensive evaluation of multimedia recommender systems. The principal reason is that, apart from some recent frameworks~\cite{DBLP:journals/jmlr/SalahTL20, DBLP:journals/corr/abs-2302-04473} which integrate multimedia recommender systems into their pipelines, each novel multimedia recommender system introduces its own implementation of the proposed approach with different dataset pre-processing solutions, sampling strategies, and evaluation protocols. Indeed, this may undermine the fair comparison of multimedia recommender systems, which cannot benefit from shared and unified training and evaluation frameworks to run rigorous and reproducible experiments as in other recommendation domains and scenarios~\cite{DBLP:conf/cikm/ZhaoHPYZLZBTSCX22, DBLP:conf/sigir/AnelliBFMMPDN21}. To this end, we plan to start from the initial benchmarking analysis we proposed in this work to further assess the reproducibility of the tested baselines. On such basis, the next steps would be to evaluate the recommendation performance under more comprehensive experimental settings involving, for instance, (i) a larger plethora of pre-trained deep learning models for the extraction of multimodal features; (ii) other multimodal datasets involving all modalities (as our framework offers the possibility to inject visual, textual, and audio features); (iii) a more careful evaluation of such models under beyond-accuracy recommendation metrics~\cite{DBLP:conf/kdd/MalitestaCPDN23, DBLP:conf/mmir/MalitestaCPN23}. 
\section{Conclusion}\label{sec:conclusion}

In this paper, we highlighted the importance of formalizing the multimedia recommendation task under the lens of multimodal deep learning. By recognizing how the main recommendation approaches in the related literature fall into some recurrent strategy patterns, we outlined a unified multimodal schema that, following the established multimodal deep learning pipeline, formalizes the core stages of multimedia recommendation as (i) multimodal input data, (ii) multimodal feature processing, (iii) multimodal feature fusion, and (iv) the multimodal recommendation task. By applying each of the outlined phases to four selected multimedia recommendation scenarios, we conceptually validated its rationale. Then, we integrated the same schema into Elliot to benchmark the results of six state-of-the-art multimedia recommender systems. The obtained results, which assess the recommendation performance in terms of accuracy and beyond-accuracy measures, along with the proposed formal schema, allowed highlighting technical challenges as well as possible avenues to address such challenges in future directions. 
\section*{ACKNOWLEDGMENTS}
This work was partially supported by the following projects: CT\_FINCONS\_III, OVS Fashion Retail Reloaded, LUTECH DIGITALE 4.0, VAI2C, IDENTITA, REACH-XY. We acknowledge the CINECA award under the ISCRA initiative, for the availability of high-performance computing resources and support.

\bibliographystyle{ACM-Reference-Format}
\bibliography{references}


\begin{thebibliography}{145}


\ifx \showCODEN    \undefined \def \showCODEN     #1{\unskip}     \fi
\ifx \showDOI      \undefined \def \showDOI       #1{#1}\fi
\ifx \showISBNx    \undefined \def \showISBNx     #1{\unskip}     \fi
\ifx \showISBNxiii \undefined \def \showISBNxiii  #1{\unskip}     \fi
\ifx \showISSN     \undefined \def \showISSN      #1{\unskip}     \fi
\ifx \showLCCN     \undefined \def \showLCCN      #1{\unskip}     \fi
\ifx \shownote     \undefined \def \shownote      #1{#1}          \fi
\ifx \showarticletitle \undefined \def \showarticletitle #1{#1}   \fi
\ifx \showURL      \undefined \def \showURL       {\relax}        \fi
\providecommand\bibfield[2]{#2}
\providecommand\bibinfo[2]{#2}
\providecommand\natexlab[1]{#1}
\providecommand\showeprint[2][]{arXiv:#2}

\bibitem[Abdollahpouri et~al\mbox{.}(2017)]%
        {DBLP:conf/recsys/AbdollahpouriBM17}
\bibfield{author}{\bibinfo{person}{Himan Abdollahpouri}, \bibinfo{person}{Robin
  Burke}, {and} \bibinfo{person}{Bamshad Mobasher}.}
  \bibinfo{year}{2017}\natexlab{}.
\newblock \showarticletitle{Controlling Popularity Bias in Learning-to-Rank
  Recommendation}. In \bibinfo{booktitle}{\emph{RecSys}}.
  \bibinfo{publisher}{{ACM}}, \bibinfo{pages}{42--46}.
\newblock


\bibitem[Anelli et~al\mbox{.}(2021)]%
        {DBLP:conf/sigir/AnelliBFMMPDN21}
\bibfield{author}{\bibinfo{person}{Vito~Walter Anelli},
  \bibinfo{person}{Alejandro Bellog{\'{\i}}n}, \bibinfo{person}{Antonio
  Ferrara}, \bibinfo{person}{Daniele Malitesta},
  \bibinfo{person}{Felice~Antonio Merra}, \bibinfo{person}{Claudio Pomo},
  \bibinfo{person}{Francesco~Maria Donini}, {and} \bibinfo{person}{Tommaso~Di
  Noia}.} \bibinfo{year}{2021}\natexlab{}.
\newblock \showarticletitle{Elliot: {A} Comprehensive and Rigorous Framework
  for Reproducible Recommender Systems Evaluation}. In
  \bibinfo{booktitle}{\emph{{SIGIR}}}. \bibinfo{publisher}{{ACM}},
  \bibinfo{pages}{2405--2414}.
\newblock


\bibitem[Anelli et~al\mbox{.}(2022)]%
        {DBLP:conf/cikm/AnelliDNSFMP22}
\bibfield{author}{\bibinfo{person}{Vito~Walter Anelli}, \bibinfo{person}{Yashar
  Deldjoo}, \bibinfo{person}{Tommaso~Di Noia}, \bibinfo{person}{Eugenio~Di
  Sciascio}, \bibinfo{person}{Antonio Ferrara}, \bibinfo{person}{Daniele
  Malitesta}, {and} \bibinfo{person}{Claudio Pomo}.}
  \bibinfo{year}{2022}\natexlab{}.
\newblock \showarticletitle{Reshaping Graph Recommendation with Edge Graph
  Collaborative Filtering and Customer Reviews}. In
  \bibinfo{booktitle}{\emph{DL4SR@CIKM}} \emph{(\bibinfo{series}{{CEUR}
  Workshop Proceedings}, Vol.~\bibinfo{volume}{3317})}.
  \bibinfo{publisher}{CEUR-WS.org}.
\newblock


\bibitem[Arora et~al\mbox{.}(2017)]%
        {DBLP:conf/iclr/AroraLM17}
\bibfield{author}{\bibinfo{person}{Sanjeev Arora}, \bibinfo{person}{Yingyu
  Liang}, {and} \bibinfo{person}{Tengyu Ma}.} \bibinfo{year}{2017}\natexlab{}.
\newblock \showarticletitle{A Simple but Tough-to-Beat Baseline for Sentence
  Embeddings}. In \bibinfo{booktitle}{\emph{{ICLR} (Poster)}}.
  \bibinfo{publisher}{OpenReview.net}.
\newblock


\bibitem[Attimonelli et~al\mbox{.}(2024)]%
        {DBLP:journals/corr/abs-2403-04503}
\bibfield{author}{\bibinfo{person}{Matteo Attimonelli}, \bibinfo{person}{Danilo
  Danese}, \bibinfo{person}{Daniele Malitesta}, \bibinfo{person}{Claudio Pomo},
  \bibinfo{person}{Giuseppe Gassi}, {and} \bibinfo{person}{Tommaso~Di Noia}.}
  \bibinfo{year}{2024}\natexlab{}.
\newblock \showarticletitle{Ducho 2.0: Towards a More Up-to-Date Unified
  Framework for the Extraction of Multimodal Features in Recommendation}.
\newblock \bibinfo{journal}{\emph{CoRR}}  \bibinfo{volume}{abs/2403.04503}
  (\bibinfo{year}{2024}).
\newblock


\bibitem[Baeza{-}Yates(2020)]%
        {DBLP:conf/recsys/Baeza-Yates20}
\bibfield{author}{\bibinfo{person}{Ricardo Baeza{-}Yates}.}
  \bibinfo{year}{2020}\natexlab{}.
\newblock \showarticletitle{Bias in Search and Recommender Systems}. In
  \bibinfo{booktitle}{\emph{RecSys}}. \bibinfo{publisher}{{ACM}},
  \bibinfo{pages}{2}.
\newblock


\bibitem[Baltrusaitis et~al\mbox{.}(2018)]%
        {DBLP:books/acm/18/BaltrusaitisAM18}
\bibfield{author}{\bibinfo{person}{Tadas Baltrusaitis},
  \bibinfo{person}{Chaitanya Ahuja}, {and} \bibinfo{person}{Louis{-}Philippe
  Morency}.} \bibinfo{year}{2018}\natexlab{}.
\newblock \showarticletitle{Challenges and applications in multimodal machine
  learning}.
\newblock In \bibinfo{booktitle}{\emph{The Handbook of Multimodal-Multisensor
  Interfaces, Volume 2 {(2)}}}. \bibinfo{publisher}{Association for Computing
  Machinery}, \bibinfo{pages}{17--48}.
\newblock


\bibitem[Baltrusaitis et~al\mbox{.}(2019)]%
        {DBLP:journals/pami/BaltrusaitisAM19}
\bibfield{author}{\bibinfo{person}{Tadas Baltrusaitis},
  \bibinfo{person}{Chaitanya Ahuja}, {and} \bibinfo{person}{Louis~Philippe
  Morency}.} \bibinfo{year}{2019}\natexlab{}.
\newblock \showarticletitle{Multimodal Machine Learning: {A} Survey and
  Taxonomy}.
\newblock \bibinfo{journal}{\emph{{IEEE} Trans. Pattern Anal. Mach. Intell.}}
  \bibinfo{volume}{41}, \bibinfo{number}{2} (\bibinfo{year}{2019}),
  \bibinfo{pages}{423--443}.
\newblock


\bibitem[Barbieri et~al\mbox{.}(2020)]%
        {DBLP:conf/emnlp/BarbieriCAN20}
\bibfield{author}{\bibinfo{person}{Francesco Barbieri},
  \bibinfo{person}{Jos{\'{e}} Camacho{-}Collados},
  \bibinfo{person}{Luis~Espinosa Anke}, {and} \bibinfo{person}{Leonardo
  Neves}.} \bibinfo{year}{2020}\natexlab{}.
\newblock \showarticletitle{TweetEval: Unified Benchmark and Comparative
  Evaluation for Tweet Classification}. In \bibinfo{booktitle}{\emph{{EMNLP}
  (Findings)}} \emph{(\bibinfo{series}{Findings of {ACL}},
  Vol.~\bibinfo{volume}{{EMNLP} 2020})}. \bibinfo{publisher}{Association for
  Computational Linguistics}, \bibinfo{pages}{1644--1650}.
\newblock


\bibitem[Bhatnagar et~al\mbox{.}(2013)]%
        {DBLP:journals/tmm/BhatnagarWL13}
\bibfield{author}{\bibinfo{person}{Gaurav Bhatnagar},
  \bibinfo{person}{Q.~M.~Jonathan Wu}, {and} \bibinfo{person}{Zheng Liu}.}
  \bibinfo{year}{2013}\natexlab{}.
\newblock \showarticletitle{Directive Contrast Based Multimodal Medical Image
  Fusion in {NSCT} Domain}.
\newblock \bibinfo{journal}{\emph{{IEEE} Trans. Multim.}} \bibinfo{volume}{15},
  \bibinfo{number}{5} (\bibinfo{year}{2013}), \bibinfo{pages}{1014--1024}.
\newblock


\bibitem[Boratto et~al\mbox{.}(2021)]%
        {DBLP:journals/ipm/BorattoFM21}
\bibfield{author}{\bibinfo{person}{Ludovico Boratto}, \bibinfo{person}{Gianni
  Fenu}, {and} \bibinfo{person}{Mirko Marras}.}
  \bibinfo{year}{2021}\natexlab{}.
\newblock \showarticletitle{Connecting user and item perspectives in popularity
  debiasing for collaborative recommendation}.
\newblock \bibinfo{journal}{\emph{Inf. Process. Manag.}} \bibinfo{volume}{58},
  \bibinfo{number}{1} (\bibinfo{year}{2021}), \bibinfo{pages}{102387}.
\newblock


\bibitem[Caesar et~al\mbox{.}(2020)]%
        {DBLP:conf/cvpr/CaesarBLVLXKPBB20}
\bibfield{author}{\bibinfo{person}{Holger Caesar}, \bibinfo{person}{Varun
  Bankiti}, \bibinfo{person}{Alex~H. Lang}, \bibinfo{person}{Sourabh Vora},
  \bibinfo{person}{Venice~Erin Liong}, \bibinfo{person}{Qiang Xu},
  \bibinfo{person}{Anush Krishnan}, \bibinfo{person}{Yu Pan},
  \bibinfo{person}{Giancarlo Baldan}, {and} \bibinfo{person}{Oscar Beijbom}.}
  \bibinfo{year}{2020}\natexlab{}.
\newblock \showarticletitle{nuScenes: {A} Multimodal Dataset for Autonomous
  Driving}. In \bibinfo{booktitle}{\emph{{CVPR}}}. \bibinfo{publisher}{Computer
  Vision Foundation / {IEEE}}, \bibinfo{pages}{11618--11628}.
\newblock


\bibitem[Cai et~al\mbox{.}(2022)]%
        {DBLP:journals/tmm/CaiQFX22}
\bibfield{author}{\bibinfo{person}{Desheng Cai}, \bibinfo{person}{Shengsheng
  Qian}, \bibinfo{person}{Quan Fang}, {and} \bibinfo{person}{Changsheng Xu}.}
  \bibinfo{year}{2022}\natexlab{}.
\newblock \showarticletitle{Heterogeneous Hierarchical Feature Aggregation
  Network for Personalized Micro-Video Recommendation}.
\newblock \bibinfo{journal}{\emph{{IEEE} Trans. Multim.}}  \bibinfo{volume}{24}
  (\bibinfo{year}{2022}), \bibinfo{pages}{805--818}.
\newblock


\bibitem[Chen et~al\mbox{.}(2022a)]%
        {DBLP:conf/mm/ChenWC00P22}
\bibfield{author}{\bibinfo{person}{Dapeng Chen}, \bibinfo{person}{Min Wang},
  \bibinfo{person}{Haobin Chen}, \bibinfo{person}{Lin Wu},
  \bibinfo{person}{Jing Qin}, {and} \bibinfo{person}{Wei Peng}.}
  \bibinfo{year}{2022}\natexlab{a}.
\newblock \showarticletitle{Cross-Modal Retrieval with Heterogeneous Graph
  Embedding}. In \bibinfo{booktitle}{\emph{{ACM} Multimedia}}.
  \bibinfo{publisher}{{ACM}}, \bibinfo{pages}{3291--3300}.
\newblock


\bibitem[Chen et~al\mbox{.}(2022b)]%
        {DBLP:conf/mm/ChenWWZS22}
\bibfield{author}{\bibinfo{person}{Feiyu Chen}, \bibinfo{person}{Junjie Wang},
  \bibinfo{person}{Yinwei Wei}, \bibinfo{person}{Hai{-}Tao Zheng}, {and}
  \bibinfo{person}{Jie Shao}.} \bibinfo{year}{2022}\natexlab{b}.
\newblock \showarticletitle{Breaking Isolation: Multimodal Graph Fusion for
  Multimedia Recommendation by Edge-wise Modulation}. In
  \bibinfo{booktitle}{\emph{{ACM} Multimedia}}. \bibinfo{publisher}{{ACM}},
  \bibinfo{pages}{385--394}.
\newblock


\bibitem[Chen and Li(2020)]%
        {DBLP:conf/ijcai/Chen020}
\bibfield{author}{\bibinfo{person}{Huiyuan Chen} {and} \bibinfo{person}{Jing
  Li}.} \bibinfo{year}{2020}\natexlab{}.
\newblock \showarticletitle{Neural Tensor Model for Learning Multi-Aspect
  Factors in Recommender Systems}. In \bibinfo{booktitle}{\emph{{IJCAI}}}.
  \bibinfo{publisher}{ijcai.org}, \bibinfo{pages}{2449--2455}.
\newblock


\bibitem[Chen et~al\mbox{.}(2017)]%
        {DBLP:conf/sigir/ChenZ0NLC17}
\bibfield{author}{\bibinfo{person}{Jingyuan Chen}, \bibinfo{person}{Hanwang
  Zhang}, \bibinfo{person}{Xiangnan He}, \bibinfo{person}{Liqiang Nie},
  \bibinfo{person}{Wei Liu}, {and} \bibinfo{person}{Tat{-}Seng Chua}.}
  \bibinfo{year}{2017}\natexlab{}.
\newblock \showarticletitle{Attentive Collaborative Filtering: Multimedia
  Recommendation with Item- and Component-Level Attention}. In
  \bibinfo{booktitle}{\emph{{SIGIR}}}. \bibinfo{publisher}{{ACM}},
  \bibinfo{pages}{335--344}.
\newblock


\bibitem[Chen et~al\mbox{.}(2016)]%
        {DBLP:conf/mm/ChenHK16}
\bibfield{author}{\bibinfo{person}{Tao Chen}, \bibinfo{person}{Xiangnan He},
  {and} \bibinfo{person}{Min{-}Yen Kan}.} \bibinfo{year}{2016}\natexlab{}.
\newblock \showarticletitle{Context-aware Image Tweet Modelling and
  Recommendation}. In \bibinfo{booktitle}{\emph{{ACM} Multimedia}}.
  \bibinfo{publisher}{{ACM}}, \bibinfo{pages}{1018--1027}.
\newblock


\bibitem[Chen et~al\mbox{.}(2019b)]%
        {DBLP:conf/kdd/ChenHXGGSLPZZ19}
\bibfield{author}{\bibinfo{person}{Wen Chen}, \bibinfo{person}{Pipei Huang},
  \bibinfo{person}{Jiaming Xu}, \bibinfo{person}{Xin Guo},
  \bibinfo{person}{Cheng Guo}, \bibinfo{person}{Fei Sun}, \bibinfo{person}{Chao
  Li}, \bibinfo{person}{Andreas Pfadler}, \bibinfo{person}{Huan Zhao}, {and}
  \bibinfo{person}{Binqiang Zhao}.} \bibinfo{year}{2019}\natexlab{b}.
\newblock \showarticletitle{{POG:} Personalized Outfit Generation for Fashion
  Recommendation at Alibaba iFashion}. In \bibinfo{booktitle}{\emph{{KDD}}}.
  \bibinfo{publisher}{{ACM}}.
\newblock


\bibitem[Chen et~al\mbox{.}(2019a)]%
        {DBLP:conf/sigir/ChenCXZ0QZ19}
\bibfield{author}{\bibinfo{person}{Xu Chen}, \bibinfo{person}{Hanxiong Chen},
  \bibinfo{person}{Hongteng Xu}, \bibinfo{person}{Yongfeng Zhang},
  \bibinfo{person}{Yixin Cao}, \bibinfo{person}{Zheng Qin}, {and}
  \bibinfo{person}{Hongyuan Zha}.} \bibinfo{year}{2019}\natexlab{a}.
\newblock \showarticletitle{Personalized Fashion Recommendation with Visual
  Explanations based on Multimodal Attention Network: Towards Visually
  Explainable Recommendation}. In \bibinfo{booktitle}{\emph{{SIGIR}}}.
  \bibinfo{publisher}{{ACM}}, \bibinfo{pages}{765--774}.
\newblock


\bibitem[Chen et~al\mbox{.}(2021)]%
        {DBLP:journals/tmm/ChenLXZ21}
\bibfield{author}{\bibinfo{person}{Xusong Chen}, \bibinfo{person}{Dong Liu},
  \bibinfo{person}{Zhiwei Xiong}, {and} \bibinfo{person}{Zheng{-}Jun Zha}.}
  \bibinfo{year}{2021}\natexlab{}.
\newblock \showarticletitle{Learning and Fusing Multiple User Interest
  Representations for Micro-Video and Movie Recommendations}.
\newblock \bibinfo{journal}{\emph{{IEEE} Trans. Multim.}}  \bibinfo{volume}{23}
  (\bibinfo{year}{2021}), \bibinfo{pages}{484--496}.
\newblock


\bibitem[Cheng et~al\mbox{.}(2019)]%
        {DBLP:journals/tois/ChengCZKK19}
\bibfield{author}{\bibinfo{person}{Zhiyong Cheng}, \bibinfo{person}{Xiaojun
  Chang}, \bibinfo{person}{Lei Zhu}, \bibinfo{person}{Rose~Catherine
  Kanjirathinkal}, {and} \bibinfo{person}{Mohan~S. Kankanhalli}.}
  \bibinfo{year}{2019}\natexlab{}.
\newblock \showarticletitle{{MMALFM:} Explainable Recommendation by Leveraging
  Reviews and Images}.
\newblock \bibinfo{journal}{\emph{{ACM} Trans. Inf. Syst.}}
  \bibinfo{volume}{37}, \bibinfo{number}{2} (\bibinfo{year}{2019}),
  \bibinfo{pages}{16:1--16:28}.
\newblock


\bibitem[Cheng et~al\mbox{.}(2016)]%
        {DBLP:conf/sigir/ChengSH16}
\bibfield{author}{\bibinfo{person}{Zhiyong Cheng}, \bibinfo{person}{Jialie
  Shen}, {and} \bibinfo{person}{Steven C.~H. Hoi}.}
  \bibinfo{year}{2016}\natexlab{}.
\newblock \showarticletitle{On Effective Personalized Music Retrieval by
  Exploring Online User Behaviors}. In \bibinfo{booktitle}{\emph{Proceedings of
  the 39th International {ACM} {SIGIR} conference on Research and Development
  in Information Retrieval, {SIGIR} 2016, Pisa, Italy, July 17-21, 2016}},
  \bibfield{editor}{\bibinfo{person}{Raffaele Perego},
  \bibinfo{person}{Fabrizio Sebastiani}, \bibinfo{person}{Javed~A. Aslam},
  \bibinfo{person}{Ian Ruthven}, {and} \bibinfo{person}{Justin Zobel}} (Eds.).
  \bibinfo{publisher}{{ACM}}, \bibinfo{pages}{125--134}.
\newblock
\urldef\tempurl%
\url{https://doi.org/10.1145/2911451.2911491}
\showDOI{\tempurl}


\bibitem[Chhabra({[n.\,d.]})]%
        {chhabra17}
\bibfield{author}{\bibinfo{person}{Sameer Chhabra}.}
  \bibinfo{year}{[n.\,d.]}\natexlab{}.
\newblock \bibinfo{title}{Netflix says 80 percent of watched content is based
  on algorithmic recommendations}.
\newblock
  \bibinfo{howpublished}{\url{https://mobilesyrup.com/2017/08/22/80-percent-netflix-shows-discovered-recommendation/}}.
\newblock
\newblock
\shownote{Accessed: 2021-03-13}.


\bibitem[Choi et~al\mbox{.}(2017)]%
        {DBLP:conf/icassp/ChoiFSC17}
\bibfield{author}{\bibinfo{person}{Keunwoo Choi}, \bibinfo{person}{Gy{\"{o}}rgy
  Fazekas}, \bibinfo{person}{Mark~B. Sandler}, {and} \bibinfo{person}{Kyunghyun
  Cho}.} \bibinfo{year}{2017}\natexlab{}.
\newblock \showarticletitle{Convolutional recurrent neural networks for music
  classification}. In \bibinfo{booktitle}{\emph{{ICASSP}}}.
  \bibinfo{publisher}{{IEEE}}, \bibinfo{pages}{2392--2396}.
\newblock


\bibitem[Cui et~al\mbox{.}(2020)]%
        {DBLP:journals/tkde/CuiWLZW20}
\bibfield{author}{\bibinfo{person}{Qiang Cui}, \bibinfo{person}{Shu Wu},
  \bibinfo{person}{Qiang Liu}, \bibinfo{person}{Wen Zhong}, {and}
  \bibinfo{person}{Liang Wang}.} \bibinfo{year}{2020}\natexlab{}.
\newblock \showarticletitle{{MV-RNN:} {A} Multi-View Recurrent Neural Network
  for Sequential Recommendation}.
\newblock \bibinfo{journal}{\emph{{IEEE} Trans. Knowl. Data Eng.}}
  \bibinfo{volume}{32}, \bibinfo{number}{2} (\bibinfo{year}{2020}),
  \bibinfo{pages}{317--331}.
\newblock


\bibitem[Deldjoo et~al\mbox{.}(2021)]%
        {DBLP:conf/cvpr/DeldjooNMM21}
\bibfield{author}{\bibinfo{person}{Yashar Deldjoo}, \bibinfo{person}{Tommaso~Di
  Noia}, \bibinfo{person}{Daniele Malitesta}, {and}
  \bibinfo{person}{Felice~Antonio Merra}.} \bibinfo{year}{2021}\natexlab{}.
\newblock \showarticletitle{A Study on the Relative Importance of Convolutional
  Neural Networks in Visually-Aware Recommender Systems}. In
  \bibinfo{booktitle}{\emph{{CVPR} Workshops}}. \bibinfo{publisher}{Computer
  Vision Foundation / {IEEE}}, \bibinfo{pages}{3961--3967}.
\newblock


\bibitem[Deldjoo et~al\mbox{.}(2022a)]%
        {DBLP:conf/ecir/DeldjooNMM22}
\bibfield{author}{\bibinfo{person}{Yashar Deldjoo}, \bibinfo{person}{Tommaso~Di
  Noia}, \bibinfo{person}{Daniele Malitesta}, {and}
  \bibinfo{person}{Felice~Antonio Merra}.} \bibinfo{year}{2022}\natexlab{a}.
\newblock \showarticletitle{Leveraging Content-Style Item Representation for
  Visual Recommendation}. In \bibinfo{booktitle}{\emph{{ECIR} {(2)}}}
  \emph{(\bibinfo{series}{Lecture Notes in Computer Science},
  Vol.~\bibinfo{volume}{13186})}. \bibinfo{publisher}{Springer},
  \bibinfo{pages}{84--92}.
\newblock


\bibitem[Deldjoo et~al\mbox{.}(2020a)]%
        {DBLP:conf/sigir/DeldjooNSM20}
\bibfield{author}{\bibinfo{person}{Yashar Deldjoo}, \bibinfo{person}{Tommaso~Di
  Noia}, \bibinfo{person}{Eugenio~Di Sciascio}, {and}
  \bibinfo{person}{Felice~Antonio Merra}.} \bibinfo{year}{2020}\natexlab{a}.
\newblock \showarticletitle{How Dataset Characteristics Affect the Robustness
  of Collaborative Recommendation Models}. In
  \bibinfo{booktitle}{\emph{{SIGIR}}}. \bibinfo{publisher}{{ACM}},
  \bibinfo{pages}{951--960}.
\newblock


\bibitem[Deldjoo et~al\mbox{.}(2020b)]%
        {DBLP:journals/csur/DeldjooSCP20}
\bibfield{author}{\bibinfo{person}{Yashar Deldjoo}, \bibinfo{person}{Markus
  Schedl}, \bibinfo{person}{Paolo Cremonesi}, {and} \bibinfo{person}{Gabriella
  Pasi}.} \bibinfo{year}{2020}\natexlab{b}.
\newblock \showarticletitle{Recommender Systems Leveraging Multimedia Content}.
\newblock \bibinfo{journal}{\emph{{ACM} Comput. Surv.}} \bibinfo{volume}{53},
  \bibinfo{number}{5} (\bibinfo{year}{2020}), \bibinfo{pages}{106:1--106:38}.
\newblock


\bibitem[Deldjoo et~al\mbox{.}(2022b)]%
        {deldjoo2022multimedia}
\bibfield{author}{\bibinfo{person}{Yashar Deldjoo}, \bibinfo{person}{Markus
  Schedl}, \bibinfo{person}{Balasz Hidasi}, \bibinfo{person}{Xiangnan He},
  {and} \bibinfo{person}{Yinwei Wei}.} \bibinfo{year}{2022}\natexlab{b}.
\newblock \showarticletitle{Multimedia Recommender Systems: Algorithms and
  Challenges}.
\newblock In \bibinfo{booktitle}{\emph{Recommender Systems Handbook}}.
  \bibinfo{publisher}{Springer US}.
\newblock


\bibitem[Dong et~al\mbox{.}(2019)]%
        {DBLP:conf/mm/DongSFJXN19}
\bibfield{author}{\bibinfo{person}{Xue Dong}, \bibinfo{person}{Xuemeng Song},
  \bibinfo{person}{Fuli Feng}, \bibinfo{person}{Peiguang Jing},
  \bibinfo{person}{Xin{-}Shun Xu}, {and} \bibinfo{person}{Liqiang Nie}.}
  \bibinfo{year}{2019}\natexlab{}.
\newblock \showarticletitle{Personalized Capsule Wardrobe Creation with Garment
  and User Modeling}. In \bibinfo{booktitle}{\emph{{ACM} Multimedia}}.
  \bibinfo{publisher}{{ACM}}, \bibinfo{pages}{302--310}.
\newblock


\bibitem[Ferracani et~al\mbox{.}(2015)]%
        {DBLP:conf/mm/FerracaniPBMB15}
\bibfield{author}{\bibinfo{person}{Andrea Ferracani}, \bibinfo{person}{Daniele
  Pezzatini}, \bibinfo{person}{Marco Bertini}, \bibinfo{person}{Saverio
  Meucci}, {and} \bibinfo{person}{Alberto~Del Bimbo}.}
  \bibinfo{year}{2015}\natexlab{}.
\newblock \showarticletitle{A System for Video Recommendation using Visual
  Saliency, Crowdsourced and Automatic Annotations}. In
  \bibinfo{booktitle}{\emph{{ACM} Multimedia}}. \bibinfo{publisher}{{ACM}}.
\newblock


\bibitem[Gao et~al\mbox{.}(2020)]%
        {DBLP:journals/neco/GaoLCZ20}
\bibfield{author}{\bibinfo{person}{Jing Gao}, \bibinfo{person}{Peng Li},
  \bibinfo{person}{Zhikui Chen}, {and} \bibinfo{person}{Jianing Zhang}.}
  \bibinfo{year}{2020}\natexlab{}.
\newblock \showarticletitle{A Survey on Deep Learning for Multimodal Data
  Fusion}.
\newblock \bibinfo{journal}{\emph{Neural Comput.}} \bibinfo{volume}{32},
  \bibinfo{number}{5} (\bibinfo{year}{2020}), \bibinfo{pages}{829--864}.
\newblock


\bibitem[Ge et~al\mbox{.}(2019)]%
        {DBLP:conf/cvpr/GeZWTL19}
\bibfield{author}{\bibinfo{person}{Yuying Ge}, \bibinfo{person}{Ruimao Zhang},
  \bibinfo{person}{Xiaogang Wang}, \bibinfo{person}{Xiaoou Tang}, {and}
  \bibinfo{person}{Ping Luo}.} \bibinfo{year}{2019}\natexlab{}.
\newblock \showarticletitle{DeepFashion2: {A} Versatile Benchmark for
  Detection, Pose Estimation, Segmentation and Re-Identification of Clothing
  Images}. In \bibinfo{booktitle}{\emph{{CVPR}}}. \bibinfo{publisher}{Computer
  Vision Foundation / {IEEE}}, \bibinfo{pages}{5337--5345}.
\newblock


\bibitem[Georgescu et~al\mbox{.}(2023)]%
        {DBLP:conf/wacv/GeorgescuIMSRVK23}
\bibfield{author}{\bibinfo{person}{Mariana{-}Iuliana Georgescu},
  \bibinfo{person}{Radu~Tudor Ionescu}, \bibinfo{person}{Andreea{-}Iuliana
  Miron}, \bibinfo{person}{Olivian Savencu}, \bibinfo{person}{Nicolae{-}Catalin
  Ristea}, \bibinfo{person}{Nicolae Verga}, {and}
  \bibinfo{person}{Fahad~Shahbaz Khan}.} \bibinfo{year}{2023}\natexlab{}.
\newblock \showarticletitle{Multimodal Multi-Head Convolutional Attention with
  Various Kernel Sizes for Medical Image Super-Resolution}. In
  \bibinfo{booktitle}{\emph{{WACV}}}. \bibinfo{publisher}{{IEEE}},
  \bibinfo{pages}{2194--2204}.
\newblock


\bibitem[Graves(2012)]%
        {DBLP:series/sci/2012-385}
\bibfield{author}{\bibinfo{person}{Alex Graves}.}
  \bibinfo{year}{2012}\natexlab{}.
\newblock \bibinfo{booktitle}{\emph{Supervised Sequence Labelling with
  Recurrent Neural Networks}}. \bibinfo{series}{Studies in Computational
  Intelligence}, Vol.~\bibinfo{volume}{385}.
\newblock \bibinfo{publisher}{Springer}.
\newblock


\bibitem[Han et~al\mbox{.}(2017)]%
        {DBLP:conf/mm/HanWJD17}
\bibfield{author}{\bibinfo{person}{Xintong Han}, \bibinfo{person}{Zuxuan Wu},
  \bibinfo{person}{Yu{-}Gang Jiang}, {and} \bibinfo{person}{Larry~S. Davis}.}
  \bibinfo{year}{2017}\natexlab{}.
\newblock \showarticletitle{Learning Fashion Compatibility with Bidirectional
  LSTMs}. In \bibinfo{booktitle}{\emph{{ACM} Multimedia}}.
  \bibinfo{publisher}{{ACM}}.
\newblock


\bibitem[He et~al\mbox{.}(2016)]%
        {DBLP:conf/cvpr/HeZRS16}
\bibfield{author}{\bibinfo{person}{Kaiming He}, \bibinfo{person}{Xiangyu
  Zhang}, \bibinfo{person}{Shaoqing Ren}, {and} \bibinfo{person}{Jian Sun}.}
  \bibinfo{year}{2016}\natexlab{}.
\newblock \showarticletitle{Deep Residual Learning for Image Recognition}. In
  \bibinfo{booktitle}{\emph{{CVPR}}}. \bibinfo{publisher}{{IEEE} Computer
  Society}, \bibinfo{pages}{770--778}.
\newblock


\bibitem[He and McAuley(2016a)]%
        {DBLP:conf/www/HeM16}
\bibfield{author}{\bibinfo{person}{Ruining He} {and} \bibinfo{person}{Julian~J.
  McAuley}.} \bibinfo{year}{2016}\natexlab{a}.
\newblock \showarticletitle{Ups and Downs: Modeling the Visual Evolution of
  Fashion Trends with One-Class Collaborative Filtering}. In
  \bibinfo{booktitle}{\emph{{WWW}}}. \bibinfo{publisher}{{ACM}},
  \bibinfo{pages}{507--517}.
\newblock


\bibitem[He and McAuley(2016b)]%
        {DBLP:conf/aaai/HeM16}
\bibfield{author}{\bibinfo{person}{Ruining He} {and} \bibinfo{person}{Julian~J.
  McAuley}.} \bibinfo{year}{2016}\natexlab{b}.
\newblock \showarticletitle{{VBPR:} Visual Bayesian Personalized Ranking from
  Implicit Feedback}. In \bibinfo{booktitle}{\emph{{AAAI}}}.
  \bibinfo{publisher}{{AAAI} Press}, \bibinfo{pages}{144--150}.
\newblock


\bibitem[He et~al\mbox{.}(2020)]%
        {DBLP:conf/sigir/0001DWLZ020}
\bibfield{author}{\bibinfo{person}{Xiangnan He}, \bibinfo{person}{Kuan Deng},
  \bibinfo{person}{Xiang Wang}, \bibinfo{person}{Yan Li},
  \bibinfo{person}{Yong{-}Dong Zhang}, {and} \bibinfo{person}{Meng Wang}.}
  \bibinfo{year}{2020}\natexlab{}.
\newblock \showarticletitle{LightGCN: Simplifying and Powering Graph
  Convolution Network for Recommendation}. In
  \bibinfo{booktitle}{\emph{{SIGIR}}}. \bibinfo{publisher}{{ACM}},
  \bibinfo{pages}{639--648}.
\newblock


\bibitem[Hermessi et~al\mbox{.}(2021)]%
        {DBLP:journals/sigpro/HermessiMZ21}
\bibfield{author}{\bibinfo{person}{Haithem Hermessi}, \bibinfo{person}{Olfa
  Mourali}, {and} \bibinfo{person}{Ezzeddine Zagrouba}.}
  \bibinfo{year}{2021}\natexlab{}.
\newblock \showarticletitle{Multimodal medical image fusion review: Theoretical
  background and recent advances}.
\newblock \bibinfo{journal}{\emph{Signal Process.}}  \bibinfo{volume}{183}
  (\bibinfo{year}{2021}), \bibinfo{pages}{108036}.
\newblock


\bibitem[Hershey et~al\mbox{.}(2017)]%
        {DBLP:conf/icassp/HersheyCEGJMPPS17}
\bibfield{author}{\bibinfo{person}{Shawn Hershey}, \bibinfo{person}{Sourish
  Chaudhuri}, \bibinfo{person}{Daniel P.~W. Ellis}, \bibinfo{person}{Jort~F.
  Gemmeke}, \bibinfo{person}{Aren Jansen}, \bibinfo{person}{R.~Channing Moore},
  \bibinfo{person}{Manoj Plakal}, \bibinfo{person}{Devin Platt},
  \bibinfo{person}{Rif~A. Saurous}, \bibinfo{person}{Bryan Seybold},
  \bibinfo{person}{Malcolm Slaney}, \bibinfo{person}{Ron~J. Weiss}, {and}
  \bibinfo{person}{Kevin~W. Wilson}.} \bibinfo{year}{2017}\natexlab{}.
\newblock \showarticletitle{{CNN} architectures for large-scale audio
  classification}. In \bibinfo{booktitle}{\emph{{ICASSP}}}.
  \bibinfo{publisher}{{IEEE}}, \bibinfo{pages}{131--135}.
\newblock


\bibitem[Hu et~al\mbox{.}(2019)]%
        {DBLP:conf/sigir/HuZPL19}
\bibfield{author}{\bibinfo{person}{Peng Hu}, \bibinfo{person}{Liangli Zhen},
  \bibinfo{person}{Dezhong Peng}, {and} \bibinfo{person}{Pei Liu}.}
  \bibinfo{year}{2019}\natexlab{}.
\newblock \showarticletitle{Scalable Deep Multimodal Learning for Cross-Modal
  Retrieval}. In \bibinfo{booktitle}{\emph{{SIGIR}}}.
  \bibinfo{publisher}{{ACM}}, \bibinfo{pages}{635--644}.
\newblock


\bibitem[Hu et~al\mbox{.}(2023)]%
        {DBLP:conf/acl/HuGTKY23}
\bibfield{author}{\bibinfo{person}{Xuming Hu}, \bibinfo{person}{Zhijiang Guo},
  \bibinfo{person}{Zhiyang Teng}, \bibinfo{person}{Irwin King}, {and}
  \bibinfo{person}{Philip~S. Yu}.} \bibinfo{year}{2023}\natexlab{}.
\newblock \showarticletitle{Multimodal Relation Extraction with Cross-Modal
  Retrieval and Synthesis}. In \bibinfo{booktitle}{\emph{{ACL} {(2)}}}.
  \bibinfo{publisher}{Association for Computational Linguistics},
  \bibinfo{pages}{303--311}.
\newblock


\bibitem[Huang et~al\mbox{.}(2022)]%
        {DBLP:conf/cikm/HuangXW0Y22}
\bibfield{author}{\bibinfo{person}{Chao Huang}, \bibinfo{person}{Lianghao Xia},
  \bibinfo{person}{Xiang Wang}, \bibinfo{person}{Xiangnan He}, {and}
  \bibinfo{person}{Dawei Yin}.} \bibinfo{year}{2022}\natexlab{}.
\newblock \showarticletitle{Self-Supervised Learning for Recommendation}. In
  \bibinfo{booktitle}{\emph{{CIKM}}}. \bibinfo{publisher}{{ACM}},
  \bibinfo{pages}{5136--5139}.
\newblock


\bibitem[Jannach et~al\mbox{.}(2015)]%
        {DBLP:journals/umuai/JannachLKJ15}
\bibfield{author}{\bibinfo{person}{Dietmar Jannach}, \bibinfo{person}{Lukas
  Lerche}, \bibinfo{person}{Iman Kamehkhosh}, {and} \bibinfo{person}{Michael
  Jugovac}.} \bibinfo{year}{2015}\natexlab{}.
\newblock \showarticletitle{What recommenders recommend: an analysis of
  recommendation biases and possible countermeasures}.
\newblock \bibinfo{journal}{\emph{User Model. User Adapt. Interact.}}
  \bibinfo{volume}{25}, \bibinfo{number}{5} (\bibinfo{year}{2015}),
  \bibinfo{pages}{427--491}.
\newblock


\bibitem[Jia et~al\mbox{.}(2015)]%
        {DBLP:conf/bigdataconf/JiaWLXXZ15}
\bibfield{author}{\bibinfo{person}{Xiaowei Jia}, \bibinfo{person}{Aosen Wang},
  \bibinfo{person}{Xiaoyi Li}, \bibinfo{person}{Guangxu Xun},
  \bibinfo{person}{Wenyao Xu}, {and} \bibinfo{person}{Aidong Zhang}.}
  \bibinfo{year}{2015}\natexlab{}.
\newblock \showarticletitle{Multi-modal learning for video recommendation based
  on mobile application usage}. In \bibinfo{booktitle}{\emph{{IEEE} BigData}}.
  \bibinfo{publisher}{{IEEE} Computer Society}, \bibinfo{pages}{837--842}.
\newblock


\bibitem[Kawasaki and Seki(2021)]%
        {DBLP:conf/wacv/KawasakiS21}
\bibfield{author}{\bibinfo{person}{Atsushi Kawasaki} {and}
  \bibinfo{person}{Akihito Seki}.} \bibinfo{year}{2021}\natexlab{}.
\newblock \showarticletitle{Multimodal Trajectory Predictions for Autonomous
  Driving without a Detailed Prior Map}. In \bibinfo{booktitle}{\emph{{WACV}}}.
  \bibinfo{publisher}{{IEEE}}, \bibinfo{pages}{3722--3731}.
\newblock


\bibitem[Khosla et~al\mbox{.}(2020)]%
        {DBLP:conf/nips/KhoslaTWSTIMLK20}
\bibfield{author}{\bibinfo{person}{Prannay Khosla}, \bibinfo{person}{Piotr
  Teterwak}, \bibinfo{person}{Chen Wang}, \bibinfo{person}{Aaron Sarna},
  \bibinfo{person}{Yonglong Tian}, \bibinfo{person}{Phillip Isola},
  \bibinfo{person}{Aaron Maschinot}, \bibinfo{person}{Ce Liu}, {and}
  \bibinfo{person}{Dilip Krishnan}.} \bibinfo{year}{2020}\natexlab{}.
\newblock \showarticletitle{Supervised Contrastive Learning}. In
  \bibinfo{booktitle}{\emph{NeurIPS}}.
\newblock


\bibitem[Kim et~al\mbox{.}(2022)]%
        {DBLP:conf/cikm/KimLSK22}
\bibfield{author}{\bibinfo{person}{Taeri Kim}, \bibinfo{person}{Yeon{-}Chang
  Lee}, \bibinfo{person}{Kijung Shin}, {and} \bibinfo{person}{Sang{-}Wook
  Kim}.} \bibinfo{year}{2022}\natexlab{}.
\newblock \showarticletitle{{MARIO:} Modality-Aware Attention and
  Modality-Preserving Decoders for Multimedia Recommendation}. In
  \bibinfo{booktitle}{\emph{{CIKM}}}. \bibinfo{publisher}{{ACM}},
  \bibinfo{pages}{993--1002}.
\newblock


\bibitem[Kim(2014)]%
        {DBLP:conf/emnlp/Kim14}
\bibfield{author}{\bibinfo{person}{Yoon Kim}.} \bibinfo{year}{2014}\natexlab{}.
\newblock \showarticletitle{Convolutional Neural Networks for Sentence
  Classification}. In \bibinfo{booktitle}{\emph{{EMNLP}}}.
  \bibinfo{publisher}{{ACL}}, \bibinfo{pages}{1746--1751}.
\newblock


\bibitem[Kipf and Welling(2017)]%
        {DBLP:conf/iclr/KipfW17}
\bibfield{author}{\bibinfo{person}{Thomas~N. Kipf} {and} \bibinfo{person}{Max
  Welling}.} \bibinfo{year}{2017}\natexlab{}.
\newblock \showarticletitle{Semi-Supervised Classification with Graph
  Convolutional Networks}. In \bibinfo{booktitle}{\emph{{ICLR} (Poster)}}.
  \bibinfo{publisher}{OpenReview.net}.
\newblock


\bibitem[Koren et~al\mbox{.}(2009)]%
        {DBLP:journals/computer/KorenBV09}
\bibfield{author}{\bibinfo{person}{Yehuda Koren}, \bibinfo{person}{Robert~M.
  Bell}, {and} \bibinfo{person}{Chris Volinsky}.}
  \bibinfo{year}{2009}\natexlab{}.
\newblock \showarticletitle{Matrix Factorization Techniques for Recommender
  Systems}.
\newblock \bibinfo{journal}{\emph{Computer}} \bibinfo{volume}{42},
  \bibinfo{number}{8} (\bibinfo{year}{2009}), \bibinfo{pages}{30--37}.
\newblock


\bibitem[Lee et~al\mbox{.}(2023)]%
        {DBLP:conf/cvpr/LeeTCL23}
\bibfield{author}{\bibinfo{person}{Yi{-}Lun Lee}, \bibinfo{person}{Yi{-}Hsuan
  Tsai}, \bibinfo{person}{Wei{-}Chen Chiu}, {and} \bibinfo{person}{Chen{-}Yu
  Lee}.} \bibinfo{year}{2023}\natexlab{}.
\newblock \showarticletitle{Multimodal Prompting with Missing Modalities for
  Visual Recognition}. In \bibinfo{booktitle}{\emph{{CVPR}}}.
  \bibinfo{publisher}{{IEEE}}, \bibinfo{pages}{14943--14952}.
\newblock


\bibitem[Lei et~al\mbox{.}(2023)]%
        {10.1145/3573010}
\bibfield{author}{\bibinfo{person}{Fei Lei}, \bibinfo{person}{Zhongqi Cao},
  \bibinfo{person}{Yuning Yang}, \bibinfo{person}{Yibo Ding}, {and}
  \bibinfo{person}{Cong Zhang}.} \bibinfo{year}{2023}\natexlab{}.
\newblock \showarticletitle{Learning the User’s Deeper Preferences for
  Multi-Modal Recommendation Systems}.
\newblock \bibinfo{journal}{\emph{ACM Trans. Multimedia Comput. Commun. Appl.}}
  \bibinfo{volume}{19}, \bibinfo{number}{3s}, Article \bibinfo{articleno}{138}
  (\bibinfo{date}{feb} \bibinfo{year}{2023}), \bibinfo{numpages}{18}~pages.
\newblock
\showISSN{1551-6857}
\urldef\tempurl%
\url{https://doi.org/10.1145/3573010}
\showDOI{\tempurl}


\bibitem[Lei et~al\mbox{.}(2021)]%
        {DBLP:journals/eswa/LeiHZSZ21}
\bibfield{author}{\bibinfo{person}{Zhenfeng Lei}, \bibinfo{person}{Anwar~Ul
  Haq}, \bibinfo{person}{Adnan Zeb}, \bibinfo{person}{Md Suzauddola}, {and}
  \bibinfo{person}{Defu Zhang}.} \bibinfo{year}{2021}\natexlab{}.
\newblock \showarticletitle{Is the suggested food your desired?: Multi-modal
  recipe recommendation with demand-based knowledge graph}.
\newblock \bibinfo{journal}{\emph{Expert Syst. Appl.}}  \bibinfo{volume}{186}
  (\bibinfo{year}{2021}), \bibinfo{pages}{115708}.
\newblock


\bibitem[Li et~al\mbox{.}(2023)]%
        {DBLP:journals/ijon/LiWLZ23}
\bibfield{author}{\bibinfo{person}{Jiang Li}, \bibinfo{person}{Xiaoping Wang},
  \bibinfo{person}{Guoqing Lv}, {and} \bibinfo{person}{Zhigang Zeng}.}
  \bibinfo{year}{2023}\natexlab{}.
\newblock \showarticletitle{GraphMFT: {A} graph network based multimodal fusion
  technique for emotion recognition in conversation}.
\newblock \bibinfo{journal}{\emph{Neurocomputing}}  \bibinfo{volume}{550}
  (\bibinfo{year}{2023}), \bibinfo{pages}{126427}.
\newblock


\bibitem[Li et~al\mbox{.}(2021)]%
        {DBLP:conf/sigir/Li0YSCZS21}
\bibfield{author}{\bibinfo{person}{Jiao Li}, \bibinfo{person}{Xing Xu},
  \bibinfo{person}{Wei Yu}, \bibinfo{person}{Fumin Shen}, \bibinfo{person}{Zuo
  Cao}, \bibinfo{person}{Kai Zuo}, {and} \bibinfo{person}{Heng~Tao Shen}.}
  \bibinfo{year}{2021}\natexlab{}.
\newblock \showarticletitle{Hybrid Fusion with Intra- and Cross-Modality
  Attention for Image-Recipe Retrieval}. In
  \bibinfo{booktitle}{\emph{{SIGIR}}}. \bibinfo{publisher}{{ACM}},
  \bibinfo{pages}{244--254}.
\newblock


\bibitem[Li et~al\mbox{.}(2015)]%
        {DBLP:journals/mta/LiPGCZ15}
\bibfield{author}{\bibinfo{person}{Zhan Li}, \bibinfo{person}{Jinye Peng},
  \bibinfo{person}{Guohua Geng}, \bibinfo{person}{Xiaojiang Chen}, {and}
  \bibinfo{person}{Pan{-}Pan Zheng}.} \bibinfo{year}{2015}\natexlab{}.
\newblock \showarticletitle{Video recommendation based on multi-modal
  information and multiple kernel}.
\newblock \bibinfo{journal}{\emph{Multim. Tools Appl.}} \bibinfo{volume}{74},
  \bibinfo{number}{13} (\bibinfo{year}{2015}), \bibinfo{pages}{4599--4616}.
\newblock


\bibitem[Liang et~al\mbox{.}(2020)]%
        {DBLP:conf/mm/LiangLCYSC20}
\bibfield{author}{\bibinfo{person}{Hongru Liang}, \bibinfo{person}{Wenqiang
  Lei}, \bibinfo{person}{Paul~Yaozhu Chan}, \bibinfo{person}{Zhenglu Yang},
  \bibinfo{person}{Maosong Sun}, {and} \bibinfo{person}{Tat{-}Seng Chua}.}
  \bibinfo{year}{2020}\natexlab{}.
\newblock \showarticletitle{PiRhDy: Learning Pitch-, Rhythm-, and
  Dynamics-aware Embeddings for Symbolic Music}. In
  \bibinfo{booktitle}{\emph{{ACM} Multimedia}}. \bibinfo{publisher}{{ACM}},
  \bibinfo{pages}{574--582}.
\newblock


\bibitem[Liu et~al\mbox{.}(2023a)]%
        {DBLP:conf/mm/LiuC0NK23}
\bibfield{author}{\bibinfo{person}{Fan Liu}, \bibinfo{person}{Huilin Chen},
  \bibinfo{person}{Zhiyong Cheng}, \bibinfo{person}{Liqiang Nie}, {and}
  \bibinfo{person}{Mohan~S. Kankanhalli}.} \bibinfo{year}{2023}\natexlab{a}.
\newblock \showarticletitle{Semantic-Guided Feature Distillation for Multimodal
  Recommendation}. In \bibinfo{booktitle}{\emph{{ACM} Multimedia}}.
  \bibinfo{publisher}{{ACM}}, \bibinfo{pages}{6567--6575}.
\newblock


\bibitem[Liu et~al\mbox{.}(2019)]%
        {DBLP:conf/mm/LiuCSWNK19}
\bibfield{author}{\bibinfo{person}{Fan Liu}, \bibinfo{person}{Zhiyong Cheng},
  \bibinfo{person}{Changchang Sun}, \bibinfo{person}{Yinglong Wang},
  \bibinfo{person}{Liqiang Nie}, {and} \bibinfo{person}{Mohan~S. Kankanhalli}.}
  \bibinfo{year}{2019}\natexlab{}.
\newblock \showarticletitle{User Diverse Preference Modeling by Multimodal
  Attentive Metric Learning}. In \bibinfo{booktitle}{\emph{{ACM} Multimedia}}.
  \bibinfo{publisher}{{ACM}}, \bibinfo{pages}{1526--1534}.
\newblock


\bibitem[Liu et~al\mbox{.}(2023c)]%
        {10075502}
\bibfield{author}{\bibinfo{person}{Kang Liu}, \bibinfo{person}{Feng Xue},
  \bibinfo{person}{Dan Guo}, \bibinfo{person}{Peijie Sun},
  \bibinfo{person}{Shengsheng Qian}, {and} \bibinfo{person}{Richang Hong}.}
  \bibinfo{year}{2023}\natexlab{c}.
\newblock \showarticletitle{Multimodal Graph Contrastive Learning for
  Multimedia-Based Recommendation}.
\newblock \bibinfo{journal}{\emph{IEEE Transactions on Multimedia}}
  (\bibinfo{year}{2023}), \bibinfo{pages}{1--13}.
\newblock
\urldef\tempurl%
\url{https://doi.org/10.1109/TMM.2023.3251108}
\showDOI{\tempurl}


\bibitem[Liu et~al\mbox{.}(2023d)]%
        {DBLP:journals/tois/LiuXGWLH23}
\bibfield{author}{\bibinfo{person}{Kang Liu}, \bibinfo{person}{Feng Xue},
  \bibinfo{person}{Dan Guo}, \bibinfo{person}{Le Wu}, \bibinfo{person}{Shujie
  Li}, {and} \bibinfo{person}{Richang Hong}.} \bibinfo{year}{2023}\natexlab{d}.
\newblock \showarticletitle{{MEGCF:} Multimodal Entity Graph Collaborative
  Filtering for Personalized Recommendation}.
\newblock \bibinfo{journal}{\emph{{ACM} Trans. Inf. Syst.}}
  \bibinfo{volume}{41}, \bibinfo{number}{2} (\bibinfo{year}{2023}),
  \bibinfo{pages}{30:1--30:27}.
\newblock


\bibitem[Liu et~al\mbox{.}(2023b)]%
        {DBLP:journals/corr/abs-2302-03883}
\bibfield{author}{\bibinfo{person}{Qidong Liu}, \bibinfo{person}{Jiaxi Hu},
  \bibinfo{person}{Yutian Xiao}, \bibinfo{person}{Jingtong Gao}, {and}
  \bibinfo{person}{Xiangyu Zhao}.} \bibinfo{year}{2023}\natexlab{b}.
\newblock \showarticletitle{Multimodal Recommender Systems: {A} Survey}.
\newblock \bibinfo{journal}{\emph{CoRR}}  \bibinfo{volume}{abs/2302.03883}
  (\bibinfo{year}{2023}).
\newblock


\bibitem[Liu et~al\mbox{.}(2021)]%
        {DBLP:conf/mm/LiuYLWTZSM21}
\bibfield{author}{\bibinfo{person}{Yong Liu}, \bibinfo{person}{Susen Yang},
  \bibinfo{person}{Chenyi Lei}, \bibinfo{person}{Guoxin Wang},
  \bibinfo{person}{Haihong Tang}, \bibinfo{person}{Juyong Zhang},
  \bibinfo{person}{Aixin Sun}, {and} \bibinfo{person}{Chunyan Miao}.}
  \bibinfo{year}{2021}\natexlab{}.
\newblock \showarticletitle{Pre-training Graph Transformer with Multimodal Side
  Information for Recommendation}. In \bibinfo{booktitle}{\emph{{ACM}
  Multimedia}}. \bibinfo{publisher}{{ACM}}, \bibinfo{pages}{2853--2861}.
\newblock


\bibitem[Liu et~al\mbox{.}(2022)]%
        {DBLP:conf/mir/LiuMSO022}
\bibfield{author}{\bibinfo{person}{Zhuang Liu}, \bibinfo{person}{Yunpu Ma},
  \bibinfo{person}{Matthias Schubert}, \bibinfo{person}{Yuanxin Ouyang}, {and}
  \bibinfo{person}{Zhang Xiong}.} \bibinfo{year}{2022}\natexlab{}.
\newblock \showarticletitle{Multi-Modal Contrastive Pre-training for
  Recommendation}. In \bibinfo{booktitle}{\emph{{ICMR}}}.
  \bibinfo{publisher}{{ACM}}, \bibinfo{pages}{99--108}.
\newblock


\bibitem[Lv et~al\mbox{.}(2021)]%
        {DBLP:conf/cvpr/LvCHDL21}
\bibfield{author}{\bibinfo{person}{Fengmao Lv}, \bibinfo{person}{Xiang Chen},
  \bibinfo{person}{Yanyong Huang}, \bibinfo{person}{Lixin Duan}, {and}
  \bibinfo{person}{Guosheng Lin}.} \bibinfo{year}{2021}\natexlab{}.
\newblock \showarticletitle{Progressive Modality Reinforcement for Human
  Multimodal Emotion Recognition From Unaligned Multimodal Sequences}. In
  \bibinfo{booktitle}{\emph{{CVPR}}}. \bibinfo{publisher}{Computer Vision
  Foundation / {IEEE}}, \bibinfo{pages}{2554--2562}.
\newblock


\bibitem[Ma et~al\mbox{.}(2022a)]%
        {DBLP:conf/cvpr/0002R0T022}
\bibfield{author}{\bibinfo{person}{Mengmeng Ma}, \bibinfo{person}{Jian Ren},
  \bibinfo{person}{Long Zhao}, \bibinfo{person}{Davide Testuggine}, {and}
  \bibinfo{person}{Xi Peng}.} \bibinfo{year}{2022}\natexlab{a}.
\newblock \showarticletitle{Are Multimodal Transformers Robust to Missing
  Modality?}. In \bibinfo{booktitle}{\emph{{CVPR}}}.
  \bibinfo{publisher}{{IEEE}}, \bibinfo{pages}{18156--18165}.
\newblock


\bibitem[Ma et~al\mbox{.}(2021)]%
        {DBLP:conf/aaai/MaRZTWP21}
\bibfield{author}{\bibinfo{person}{Mengmeng Ma}, \bibinfo{person}{Jian Ren},
  \bibinfo{person}{Long Zhao}, \bibinfo{person}{Sergey Tulyakov},
  \bibinfo{person}{Cathy Wu}, {and} \bibinfo{person}{Xi Peng}.}
  \bibinfo{year}{2021}\natexlab{}.
\newblock \showarticletitle{{SMIL:} Multimodal Learning with Severely Missing
  Modality}. In \bibinfo{booktitle}{\emph{{AAAI}}}. \bibinfo{publisher}{{AAAI}
  Press}, \bibinfo{pages}{2302--2310}.
\newblock


\bibitem[Ma et~al\mbox{.}(2022b)]%
        {DBLP:journals/kbs/MaZWLK22}
\bibfield{author}{\bibinfo{person}{Yao Ma}, \bibinfo{person}{Shilin Zhao},
  \bibinfo{person}{Weixiao Wang}, \bibinfo{person}{Yaoman Li}, {and}
  \bibinfo{person}{Irwin King}.} \bibinfo{year}{2022}\natexlab{b}.
\newblock \showarticletitle{Multimodality in meta-learning: {A} comprehensive
  survey}.
\newblock \bibinfo{journal}{\emph{Knowl. Based Syst.}}  \bibinfo{volume}{250}
  (\bibinfo{year}{2022}), \bibinfo{pages}{108976}.
\newblock


\bibitem[MacKenzie et~al\mbox{.}(2013)]%
        {mackenzie2013retailers}
\bibfield{author}{\bibinfo{person}{Ian MacKenzie}, \bibinfo{person}{Chris
  Meyer}, {and} \bibinfo{person}{Steve Noble}.}
  \bibinfo{year}{2013}\natexlab{}.
\newblock \showarticletitle{How retailers can keep up with consumers}.
\newblock \bibinfo{journal}{\emph{McKinsey \& Company}}  \bibinfo{volume}{18}
  (\bibinfo{year}{2013}).
\newblock


\bibitem[Malitesta et~al\mbox{.}(2023a)]%
        {DBLP:conf/kdd/MalitestaCPDN23}
\bibfield{author}{\bibinfo{person}{Daniele Malitesta},
  \bibinfo{person}{Giandomenico Cornacchia}, \bibinfo{person}{Claudio Pomo},
  {and} \bibinfo{person}{Tommaso {Di Noia}}.} \bibinfo{year}{2023}\natexlab{a}.
\newblock \showarticletitle{Disentangling the Performance Puzzle of
  Multimodal-aware Recommender Systems}. In
  \bibinfo{booktitle}{\emph{EvalRS@KDD}} \emph{(\bibinfo{series}{{CEUR}
  Workshop Proceedings}, Vol.~\bibinfo{volume}{3450})}.
  \bibinfo{publisher}{CEUR-WS.org}.
\newblock


\bibitem[Malitesta et~al\mbox{.}(2023b)]%
        {DBLP:conf/mmir/MalitestaCPN23}
\bibfield{author}{\bibinfo{person}{Daniele Malitesta},
  \bibinfo{person}{Giandomenico Cornacchia}, \bibinfo{person}{Claudio Pomo},
  {and} \bibinfo{person}{Tommaso~Di Noia}.} \bibinfo{year}{2023}\natexlab{b}.
\newblock \showarticletitle{On Popularity Bias of Multimodal-aware Recommender
  Systems: {A} Modalities-driven Analysis}. In
  \bibinfo{booktitle}{\emph{MMIR@MM}}. \bibinfo{publisher}{{ACM}},
  \bibinfo{pages}{59--68}.
\newblock


\bibitem[Malitesta et~al\mbox{.}(2023c)]%
        {DBLP:conf/mm/MalitestaGPN23}
\bibfield{author}{\bibinfo{person}{Daniele Malitesta},
  \bibinfo{person}{Giuseppe Gassi}, \bibinfo{person}{Claudio Pomo}, {and}
  \bibinfo{person}{Tommaso~Di Noia}.} \bibinfo{year}{2023}\natexlab{c}.
\newblock \showarticletitle{Ducho: {A} Unified Framework for the Extraction of
  Multimodal Features in Recommendation}. In \bibinfo{booktitle}{\emph{{ACM}
  Multimedia}}. \bibinfo{publisher}{{ACM}}, \bibinfo{pages}{9668--9671}.
\newblock


\bibitem[Malitesta et~al\mbox{.}(2023d)]%
        {DBLP:journals/corr/abs-2308-10778}
\bibfield{author}{\bibinfo{person}{Daniele Malitesta}, \bibinfo{person}{Claudio
  Pomo}, \bibinfo{person}{Vito~Walter Anelli}, \bibinfo{person}{Alberto
  Carlo~Maria Mancino}, \bibinfo{person}{Eugenio~Di Sciascio}, {and}
  \bibinfo{person}{Tommaso~Di Noia}.} \bibinfo{year}{2023}\natexlab{d}.
\newblock \showarticletitle{A Topology-aware Analysis of Graph Collaborative
  Filtering}.
\newblock \bibinfo{journal}{\emph{CoRR}}  \bibinfo{volume}{abs/2308.10778}
  (\bibinfo{year}{2023}).
\newblock


\bibitem[Malitesta et~al\mbox{.}(2024)]%
        {DBLP:journals/corr/abs-2403-19841}
\bibfield{author}{\bibinfo{person}{Daniele Malitesta},
  \bibinfo{person}{Emanuele Rossi}, \bibinfo{person}{Claudio Pomo},
  \bibinfo{person}{Fragkiskos~D. Malliaros}, {and} \bibinfo{person}{Tommaso~Di
  Noia}.} \bibinfo{year}{2024}\natexlab{}.
\newblock \showarticletitle{Dealing with Missing Modalities in Multimodal
  Recommendation: a Feature Propagation-based Approach}.
\newblock \bibinfo{journal}{\emph{CoRR}}  \bibinfo{volume}{abs/2403.19841}
  (\bibinfo{year}{2024}).
\newblock


\bibitem[McAuley et~al\mbox{.}(2015)]%
        {DBLP:conf/sigir/McAuleyTSH15}
\bibfield{author}{\bibinfo{person}{Julian~J. McAuley},
  \bibinfo{person}{Christopher Targett}, \bibinfo{person}{Qinfeng Shi}, {and}
  \bibinfo{person}{Anton van~den Hengel}.} \bibinfo{year}{2015}\natexlab{}.
\newblock \showarticletitle{Image-Based Recommendations on Styles and
  Substitutes}. In \bibinfo{booktitle}{\emph{{SIGIR}}}.
  \bibinfo{publisher}{{ACM}}, \bibinfo{pages}{43--52}.
\newblock


\bibitem[Mikolov et~al\mbox{.}(2013)]%
        {DBLP:journals/corr/abs-1301-3781}
\bibfield{author}{\bibinfo{person}{Tom{\'{a}}s Mikolov}, \bibinfo{person}{Kai
  Chen}, \bibinfo{person}{Greg Corrado}, {and} \bibinfo{person}{Jeffrey Dean}.}
  \bibinfo{year}{2013}\natexlab{}.
\newblock \showarticletitle{Efficient Estimation of Word Representations in
  Vector Space}. In \bibinfo{booktitle}{\emph{{ICLR} (Workshop Poster)}}.
\newblock


\bibitem[Min et~al\mbox{.}(2020)]%
        {DBLP:journals/tmm/MinJJ20}
\bibfield{author}{\bibinfo{person}{Weiqing Min}, \bibinfo{person}{Shuqiang
  Jiang}, {and} \bibinfo{person}{Ramesh~C. Jain}.}
  \bibinfo{year}{2020}\natexlab{}.
\newblock \showarticletitle{Food Recommendation: Framework, Existing Solutions,
  and Challenges}.
\newblock \bibinfo{journal}{\emph{{IEEE} Trans. Multim.}} \bibinfo{volume}{22},
  \bibinfo{number}{10} (\bibinfo{year}{2020}), \bibinfo{pages}{2659--2671}.
\newblock


\bibitem[Mu et~al\mbox{.}(2022)]%
        {DBLP:conf/mm/MuZT0T22}
\bibfield{author}{\bibinfo{person}{Zongshen Mu}, \bibinfo{person}{Yueting
  Zhuang}, \bibinfo{person}{Jie Tan}, \bibinfo{person}{Jun Xiao}, {and}
  \bibinfo{person}{Siliang Tang}.} \bibinfo{year}{2022}\natexlab{}.
\newblock \showarticletitle{Learning Hybrid Behavior Patterns for Multimedia
  Recommendation}. In \bibinfo{booktitle}{\emph{{ACM} Multimedia}}.
  \bibinfo{publisher}{{ACM}}, \bibinfo{pages}{376--384}.
\newblock


\bibitem[Ngiam et~al\mbox{.}(2011)]%
        {DBLP:conf/icml/NgiamKKNLN11}
\bibfield{author}{\bibinfo{person}{Jiquan Ngiam}, \bibinfo{person}{Aditya
  Khosla}, \bibinfo{person}{Mingyu Kim}, \bibinfo{person}{Juhan Nam},
  \bibinfo{person}{Honglak Lee}, {and} \bibinfo{person}{Andrew~Y. Ng}.}
  \bibinfo{year}{2011}\natexlab{}.
\newblock \showarticletitle{Multimodal Deep Learning}. In
  \bibinfo{booktitle}{\emph{{ICML}}}. \bibinfo{publisher}{Omnipress},
  \bibinfo{pages}{689--696}.
\newblock


\bibitem[Nie et~al\mbox{.}(2016)]%
        {DBLP:journals/mta/NieLZS16}
\bibfield{author}{\bibinfo{person}{Weizhi Nie}, \bibinfo{person}{Anan Liu},
  \bibinfo{person}{Xiaorong Zhu}, {and} \bibinfo{person}{Yuting Su}.}
  \bibinfo{year}{2016}\natexlab{}.
\newblock \showarticletitle{Quality models for venue recommendation in
  location-based social network}.
\newblock \bibinfo{journal}{\emph{Multim. Tools Appl.}} \bibinfo{volume}{75},
  \bibinfo{number}{20} (\bibinfo{year}{2016}), \bibinfo{pages}{12521--12534}.
\newblock


\bibitem[Oramas et~al\mbox{.}(2017)]%
        {DBLP:conf/recsys/OramasNSS17}
\bibfield{author}{\bibinfo{person}{Sergio Oramas}, \bibinfo{person}{Oriol
  Nieto}, \bibinfo{person}{Mohamed Sordo}, {and} \bibinfo{person}{Xavier
  Serra}.} \bibinfo{year}{2017}\natexlab{}.
\newblock \showarticletitle{A Deep Multimodal Approach for Cold-start Music
  Recommendation}. In \bibinfo{booktitle}{\emph{DLRS@RecSys}}.
  \bibinfo{publisher}{{ACM}}, \bibinfo{pages}{32--37}.
\newblock


\bibitem[Pan et~al\mbox{.}(2022)]%
        {DBLP:conf/acl/PanCGZWL22}
\bibfield{author}{\bibinfo{person}{Xichen Pan}, \bibinfo{person}{Peiyu Chen},
  \bibinfo{person}{Yichen Gong}, \bibinfo{person}{Helong Zhou},
  \bibinfo{person}{Xinbing Wang}, {and} \bibinfo{person}{Zhouhan Lin}.}
  \bibinfo{year}{2022}\natexlab{}.
\newblock \showarticletitle{Leveraging Unimodal Self-Supervised Learning for
  Multimodal Audio-Visual Speech Recognition}. In
  \bibinfo{booktitle}{\emph{{ACL} {(1)}}}. \bibinfo{publisher}{Association for
  Computational Linguistics}, \bibinfo{pages}{4491--4503}.
\newblock


\bibitem[Paraskevopoulos et~al\mbox{.}(2020)]%
        {DBLP:conf/acl/Paraskevopoulos20}
\bibfield{author}{\bibinfo{person}{Georgios Paraskevopoulos},
  \bibinfo{person}{Srinivas Parthasarathy}, \bibinfo{person}{Aparna Khare},
  {and} \bibinfo{person}{Shiva Sundaram}.} \bibinfo{year}{2020}\natexlab{}.
\newblock \showarticletitle{Multimodal and Multiresolution Speech Recognition
  with Transformers}. In \bibinfo{booktitle}{\emph{{ACL}}}.
  \bibinfo{publisher}{Association for Computational Linguistics},
  \bibinfo{pages}{2381--2387}.
\newblock


\bibitem[Reimers and Gurevych(2019)]%
        {DBLP:conf/emnlp/ReimersG19}
\bibfield{author}{\bibinfo{person}{Nils Reimers} {and} \bibinfo{person}{Iryna
  Gurevych}.} \bibinfo{year}{2019}\natexlab{}.
\newblock \showarticletitle{Sentence-BERT: Sentence Embeddings using Siamese
  BERT-Networks}. In \bibinfo{booktitle}{\emph{{EMNLP/IJCNLP} {(1)}}}.
  \bibinfo{publisher}{Association for Computational Linguistics},
  \bibinfo{pages}{3980--3990}.
\newblock


\bibitem[Rendle et~al\mbox{.}(2009)]%
        {DBLP:conf/uai/RendleFGS09}
\bibfield{author}{\bibinfo{person}{Steffen Rendle}, \bibinfo{person}{Christoph
  Freudenthaler}, \bibinfo{person}{Zeno Gantner}, {and} \bibinfo{person}{Lars
  Schmidt{-}Thieme}.} \bibinfo{year}{2009}\natexlab{}.
\newblock \showarticletitle{{BPR:} Bayesian Personalized Ranking from Implicit
  Feedback}. In \bibinfo{booktitle}{\emph{{UAI}}}.
\newblock


\bibitem[Rossi et~al\mbox{.}(2022)]%
        {DBLP:conf/log/RossiK0C0B22}
\bibfield{author}{\bibinfo{person}{Emanuele Rossi}, \bibinfo{person}{Henry
  Kenlay}, \bibinfo{person}{Maria~I. Gorinova}, \bibinfo{person}{Benjamin~Paul
  Chamberlain}, \bibinfo{person}{Xiaowen Dong}, {and}
  \bibinfo{person}{Michael~M. Bronstein}.} \bibinfo{year}{2022}\natexlab{}.
\newblock \showarticletitle{On the Unreasonable Effectiveness of Feature
  Propagation in Learning on Graphs With Missing Node Features}. In
  \bibinfo{booktitle}{\emph{LoG}} \emph{(\bibinfo{series}{Proceedings of
  Machine Learning Research}, Vol.~\bibinfo{volume}{198})}.
  \bibinfo{publisher}{{PMLR}}, \bibinfo{pages}{11}.
\newblock


\bibitem[Salah et~al\mbox{.}(2020)]%
        {DBLP:journals/jmlr/SalahTL20}
\bibfield{author}{\bibinfo{person}{Aghiles Salah}, \bibinfo{person}{Quoc{-}Tuan
  Truong}, {and} \bibinfo{person}{Hady~W. Lauw}.}
  \bibinfo{year}{2020}\natexlab{}.
\newblock \showarticletitle{Cornac: {A} Comparative Framework for Multimodal
  Recommender Systems}.
\newblock \bibinfo{journal}{\emph{J. Mach. Learn. Res.}}  \bibinfo{volume}{21}
  (\bibinfo{year}{2020}), \bibinfo{pages}{95:1--95:5}.
\newblock


\bibitem[Sang et~al\mbox{.}(2021)]%
        {DBLP:journals/tmm/SangXQMLW21}
\bibfield{author}{\bibinfo{person}{Lei Sang}, \bibinfo{person}{Min Xu},
  \bibinfo{person}{Shengsheng Qian}, \bibinfo{person}{Matt Martin},
  \bibinfo{person}{Peter Li}, {and} \bibinfo{person}{Xindong Wu}.}
  \bibinfo{year}{2021}\natexlab{}.
\newblock \showarticletitle{Context-Dependent Propagating-Based Video
  Recommendation in Multimodal Heterogeneous Information Networks}.
\newblock \bibinfo{journal}{\emph{{IEEE} Trans. Multim.}}  \bibinfo{volume}{23}
  (\bibinfo{year}{2021}), \bibinfo{pages}{2019--2032}.
\newblock


\bibitem[Shani and Gunawardana(2011)]%
        {DBLP:reference/rsh/ShaniG11}
\bibfield{author}{\bibinfo{person}{Guy Shani} {and} \bibinfo{person}{Asela
  Gunawardana}.} \bibinfo{year}{2011}\natexlab{}.
\newblock \showarticletitle{Evaluating Recommendation Systems}.
\newblock In \bibinfo{booktitle}{\emph{Recommender Systems Handbook}}.
  \bibinfo{publisher}{Springer}, \bibinfo{pages}{257--297}.
\newblock


\bibitem[Shen et~al\mbox{.}(2020)]%
        {DBLP:conf/ijcnn/Shen0LWC20}
\bibfield{author}{\bibinfo{person}{Tiancheng Shen}, \bibinfo{person}{Jia Jia},
  \bibinfo{person}{Yan Li}, \bibinfo{person}{Hanjie Wang}, {and}
  \bibinfo{person}{Bo Chen}.} \bibinfo{year}{2020}\natexlab{}.
\newblock \showarticletitle{Enhancing Music Recommendation with Social Media
  Content: an Attentive Multimodal Autoencoder Approach}. In
  \bibinfo{booktitle}{\emph{{IJCNN}}}. \bibinfo{publisher}{{IEEE}},
  \bibinfo{pages}{1--8}.
\newblock


\bibitem[Simonyan and Zisserman(2015)]%
        {DBLP:journals/corr/SimonyanZ14a}
\bibfield{author}{\bibinfo{person}{Karen Simonyan} {and}
  \bibinfo{person}{Andrew Zisserman}.} \bibinfo{year}{2015}\natexlab{}.
\newblock \showarticletitle{Very Deep Convolutional Networks for Large-Scale
  Image Recognition}. In \bibinfo{booktitle}{\emph{{ICLR}}}.
\newblock


\bibitem[Sun et~al\mbox{.}(2020)]%
        {DBLP:conf/cikm/SunCZWZZWZ20}
\bibfield{author}{\bibinfo{person}{Rui Sun}, \bibinfo{person}{Xuezhi Cao},
  \bibinfo{person}{Yan Zhao}, \bibinfo{person}{Junchen Wan},
  \bibinfo{person}{Kun Zhou}, \bibinfo{person}{Fuzheng Zhang},
  \bibinfo{person}{Zhongyuan Wang}, {and} \bibinfo{person}{Kai Zheng}.}
  \bibinfo{year}{2020}\natexlab{}.
\newblock \showarticletitle{Multi-modal Knowledge Graphs for Recommender
  Systems}. In \bibinfo{booktitle}{\emph{{CIKM}}}. \bibinfo{publisher}{{ACM}},
  \bibinfo{pages}{1405--1414}.
\newblock


\bibitem[Sun et~al\mbox{.}(2019)]%
        {DBLP:conf/www/SunKNS19}
\bibfield{author}{\bibinfo{person}{Wenlong Sun}, \bibinfo{person}{Sami
  Khenissi}, \bibinfo{person}{Olfa Nasraoui}, {and} \bibinfo{person}{Patrick
  Shafto}.} \bibinfo{year}{2019}\natexlab{}.
\newblock \showarticletitle{Debiasing the Human-Recommender System Feedback
  Loop in Collaborative Filtering}. In \bibinfo{booktitle}{\emph{{WWW}
  (Companion Volume)}}. \bibinfo{publisher}{{ACM}}, \bibinfo{pages}{645--651}.
\newblock


\bibitem[Sun et~al\mbox{.}(2021)]%
        {DBLP:conf/icassp/SunMLHN021}
\bibfield{author}{\bibinfo{person}{Wangbin Sun}, \bibinfo{person}{Fei Ma},
  \bibinfo{person}{Yang Li}, \bibinfo{person}{Shao{-}Lun Huang},
  \bibinfo{person}{Shiguang Ni}, {and} \bibinfo{person}{Lin Zhang}.}
  \bibinfo{year}{2021}\natexlab{}.
\newblock \showarticletitle{Semi-Supervised Multimodal Image Translation for
  Missing Modality Imputation}. In \bibinfo{booktitle}{\emph{{ICASSP}}}.
  \bibinfo{publisher}{{IEEE}}, \bibinfo{pages}{4320--4324}.
\newblock


\bibitem[Sundar and Heck(2022)]%
        {DBLP:conf/acl-convai/SundarH22}
\bibfield{author}{\bibinfo{person}{Anirudh Sundar} {and} \bibinfo{person}{Larry
  Heck}.} \bibinfo{year}{2022}\natexlab{}.
\newblock \showarticletitle{Multimodal Conversational {AI:} {A} Survey of
  Datasets and Approaches}. In \bibinfo{booktitle}{\emph{ConvAI@ACL}}.
  \bibinfo{publisher}{Association for Computational Linguistics},
  \bibinfo{pages}{131--147}.
\newblock


\bibitem[Szegedy et~al\mbox{.}(2016)]%
        {DBLP:conf/cvpr/SzegedyVISW16}
\bibfield{author}{\bibinfo{person}{Christian Szegedy}, \bibinfo{person}{Vincent
  Vanhoucke}, \bibinfo{person}{Sergey Ioffe}, \bibinfo{person}{Jonathon
  Shlens}, {and} \bibinfo{person}{Zbigniew Wojna}.}
  \bibinfo{year}{2016}\natexlab{}.
\newblock \showarticletitle{Rethinking the Inception Architecture for Computer
  Vision}. In \bibinfo{booktitle}{\emph{{CVPR}}}. \bibinfo{publisher}{{IEEE}
  Computer Society}, \bibinfo{pages}{2818--2826}.
\newblock


\bibitem[Tan et~al\mbox{.}(2018)]%
        {DBLP:conf/icann/TanSKZYL18}
\bibfield{author}{\bibinfo{person}{Chuanqi Tan}, \bibinfo{person}{Fuchun Sun},
  \bibinfo{person}{Tao Kong}, \bibinfo{person}{Wenchang Zhang},
  \bibinfo{person}{Chao Yang}, {and} \bibinfo{person}{Chunfang Liu}.}
  \bibinfo{year}{2018}\natexlab{}.
\newblock \showarticletitle{A Survey on Deep Transfer Learning}. In
  \bibinfo{booktitle}{\emph{{ICANN} {(3)}}} \emph{(\bibinfo{series}{Lecture
  Notes in Computer Science}, Vol.~\bibinfo{volume}{11141})}.
  \bibinfo{publisher}{Springer}, \bibinfo{pages}{270--279}.
\newblock


\bibitem[Tang et~al\mbox{.}(2022)]%
        {DBLP:journals/tip/TangHLD22}
\bibfield{author}{\bibinfo{person}{Wei Tang}, \bibinfo{person}{Fazhi He},
  \bibinfo{person}{Yu Liu}, {and} \bibinfo{person}{Yansong Duan}.}
  \bibinfo{year}{2022}\natexlab{}.
\newblock \showarticletitle{{MATR:} Multimodal Medical Image Fusion via
  Multiscale Adaptive Transformer}.
\newblock \bibinfo{journal}{\emph{{IEEE} Trans. Image Process.}}
  \bibinfo{volume}{31} (\bibinfo{year}{2022}), \bibinfo{pages}{5134--5149}.
\newblock


\bibitem[Tao et~al\mbox{.}(2020)]%
        {DBLP:journals/ipm/TaoWWHHC20}
\bibfield{author}{\bibinfo{person}{Zhulin Tao}, \bibinfo{person}{Yinwei Wei},
  \bibinfo{person}{Xiang Wang}, \bibinfo{person}{Xiangnan He},
  \bibinfo{person}{Xianglin Huang}, {and} \bibinfo{person}{Tat{-}Seng Chua}.}
  \bibinfo{year}{2020}\natexlab{}.
\newblock \showarticletitle{{MGAT:} Multimodal Graph Attention Network for
  Recommendation}.
\newblock \bibinfo{journal}{\emph{Inf. Process. Manag.}} \bibinfo{volume}{57},
  \bibinfo{number}{5} (\bibinfo{year}{2020}), \bibinfo{pages}{102277}.
\newblock


\bibitem[Tawfik et~al\mbox{.}(2021)]%
        {DBLP:journals/mta/TawfikEFDE21}
\bibfield{author}{\bibinfo{person}{Nahed Tawfik}, \bibinfo{person}{Heba~A.
  Elnemr}, \bibinfo{person}{Mahmoud Fakhr}, \bibinfo{person}{Moawad~I.
  Dessouky}, {and} \bibinfo{person}{Fathi E.~Abd El{-}Samie}.}
  \bibinfo{year}{2021}\natexlab{}.
\newblock \showarticletitle{Survey study of multimodality medical image fusion
  methods}.
\newblock \bibinfo{journal}{\emph{Multim. Tools Appl.}} \bibinfo{volume}{80},
  \bibinfo{number}{4} (\bibinfo{year}{2021}), \bibinfo{pages}{6369--6396}.
\newblock


\bibitem[Truong et~al\mbox{.}(2021)]%
        {DBLP:conf/recsys/TruongSL21}
\bibfield{author}{\bibinfo{person}{Quoc{-}Tuan Truong},
  \bibinfo{person}{Aghiles Salah}, {and} \bibinfo{person}{Hady~W. Lauw}.}
  \bibinfo{year}{2021}\natexlab{}.
\newblock \showarticletitle{Multi-Modal Recommender Systems: Hands-On
  Exploration}. In \bibinfo{booktitle}{\emph{RecSys}}.
  \bibinfo{publisher}{{ACM}}, \bibinfo{pages}{834--837}.
\newblock


\bibitem[Vargas(2014)]%
        {DBLP:conf/sigir/Vargas14}
\bibfield{author}{\bibinfo{person}{Sa{\'{u}}l Vargas}.}
  \bibinfo{year}{2014}\natexlab{}.
\newblock \showarticletitle{Novelty and diversity enhancement and evaluation in
  recommender systems and information retrieval}. In
  \bibinfo{booktitle}{\emph{{SIGIR}}}. \bibinfo{publisher}{{ACM}},
  \bibinfo{pages}{1281}.
\newblock


\bibitem[Vargas and Castells(2011)]%
        {DBLP:conf/recsys/VargasC11}
\bibfield{author}{\bibinfo{person}{Saul Vargas} {and} \bibinfo{person}{Pablo
  Castells}.} \bibinfo{year}{2011}\natexlab{}.
\newblock \showarticletitle{Rank and relevance in novelty and diversity metrics
  for recommender systems}. In \bibinfo{booktitle}{\emph{RecSys}}.
  \bibinfo{publisher}{{ACM}}, \bibinfo{pages}{109--116}.
\newblock


\bibitem[Vaswani et~al\mbox{.}(2021)]%
        {DBLP:conf/bigmm/VaswaniAA21}
\bibfield{author}{\bibinfo{person}{Kunal Vaswani}, \bibinfo{person}{Yudhik
  Agrawal}, {and} \bibinfo{person}{Vinoo Alluri}.}
  \bibinfo{year}{2021}\natexlab{}.
\newblock \showarticletitle{Multimodal Fusion Based Attentive Networks for
  Sequential Music Recommendation}. In \bibinfo{booktitle}{\emph{BigMM}}.
  \bibinfo{publisher}{{IEEE}}, \bibinfo{pages}{25--32}.
\newblock


\bibitem[Wang et~al\mbox{.}(2018)]%
        {DBLP:conf/emnlp/WangNL18}
\bibfield{author}{\bibinfo{person}{Cheng Wang}, \bibinfo{person}{Mathias
  Niepert}, {and} \bibinfo{person}{Hui Li}.} \bibinfo{year}{2018}\natexlab{}.
\newblock \showarticletitle{{LRMM:} Learning to Recommend with Missing
  Modalities}. In \bibinfo{booktitle}{\emph{{EMNLP}}}.
  \bibinfo{publisher}{Association for Computational Linguistics}.
\newblock


\bibitem[Wang et~al\mbox{.}(2023)]%
        {DBLP:journals/tmm/WangWYWSN23}
\bibfield{author}{\bibinfo{person}{Qifan Wang}, \bibinfo{person}{Yinwei Wei},
  \bibinfo{person}{Jianhua Yin}, \bibinfo{person}{Jianlong Wu},
  \bibinfo{person}{Xuemeng Song}, {and} \bibinfo{person}{Liqiang Nie}.}
  \bibinfo{year}{2023}\natexlab{}.
\newblock \showarticletitle{DualGNN: Dual Graph Neural Network for Multimedia
  Recommendation}.
\newblock \bibinfo{journal}{\emph{{IEEE} Trans. Multim.}}  \bibinfo{volume}{25}
  (\bibinfo{year}{2023}), \bibinfo{pages}{1074--1084}.
\newblock


\bibitem[Wang et~al\mbox{.}(2021a)]%
        {DBLP:journals/tomccap/WangDJJSN21}
\bibfield{author}{\bibinfo{person}{Wenjie Wang}, \bibinfo{person}{Ling{-}Yu
  Duan}, \bibinfo{person}{Hao Jiang}, \bibinfo{person}{Peiguang Jing},
  \bibinfo{person}{Xuemeng Song}, {and} \bibinfo{person}{Liqiang Nie}.}
  \bibinfo{year}{2021}\natexlab{a}.
\newblock \showarticletitle{Market2Dish: Health-aware Food Recommendation}.
\newblock \bibinfo{journal}{\emph{{ACM} Trans. Multim. Comput. Commun. Appl.}}
  \bibinfo{volume}{17}, \bibinfo{number}{1} (\bibinfo{year}{2021}),
  \bibinfo{pages}{33:1--33:19}.
\newblock


\bibitem[Wang et~al\mbox{.}(2019)]%
        {DBLP:conf/sigir/Wang0WFC19}
\bibfield{author}{\bibinfo{person}{Xiang Wang}, \bibinfo{person}{Xiangnan He},
  \bibinfo{person}{Meng Wang}, \bibinfo{person}{Fuli Feng}, {and}
  \bibinfo{person}{Tat{-}Seng Chua}.} \bibinfo{year}{2019}\natexlab{}.
\newblock \showarticletitle{Neural Graph Collaborative Filtering}. In
  \bibinfo{booktitle}{\emph{{SIGIR}}}. \bibinfo{publisher}{{ACM}},
  \bibinfo{pages}{165--174}.
\newblock


\bibitem[Wang et~al\mbox{.}(2021b)]%
        {DBLP:conf/www/0012OM21}
\bibfield{author}{\bibinfo{person}{Xi Wang}, \bibinfo{person}{Iadh Ounis},
  {and} \bibinfo{person}{Craig Macdonald}.} \bibinfo{year}{2021}\natexlab{b}.
\newblock \showarticletitle{Leveraging Review Properties for Effective
  Recommendation}. In \bibinfo{booktitle}{\emph{{WWW}}}.
  \bibinfo{publisher}{{ACM} / {IW3C2}}, \bibinfo{pages}{2209--2219}.
\newblock


\bibitem[Wei et~al\mbox{.}(2023)]%
        {DBLP:conf/www/WeiHXZ23}
\bibfield{author}{\bibinfo{person}{Wei Wei}, \bibinfo{person}{Chao Huang},
  \bibinfo{person}{Lianghao Xia}, {and} \bibinfo{person}{Chuxu Zhang}.}
  \bibinfo{year}{2023}\natexlab{}.
\newblock \showarticletitle{Multi-Modal Self-Supervised Learning for
  Recommendation}. In \bibinfo{booktitle}{\emph{{WWW}}}.
  \bibinfo{publisher}{{ACM}}, \bibinfo{pages}{790--800}.
\newblock


\bibitem[Wei et~al\mbox{.}(2022)]%
        {DBLP:journals/tmm/WeiWHNRC22}
\bibfield{author}{\bibinfo{person}{Yinwei Wei}, \bibinfo{person}{Xiang Wang},
  \bibinfo{person}{Xiangnan He}, \bibinfo{person}{Liqiang Nie},
  \bibinfo{person}{Yong Rui}, {and} \bibinfo{person}{Tat{-}Seng Chua}.}
  \bibinfo{year}{2022}\natexlab{}.
\newblock \showarticletitle{Hierarchical User Intent Graph Network for
  Multimedia Recommendation}.
\newblock \bibinfo{journal}{\emph{{IEEE} Trans. Multim.}}  \bibinfo{volume}{24}
  (\bibinfo{year}{2022}), \bibinfo{pages}{2701--2712}.
\newblock


\bibitem[Wei et~al\mbox{.}(2020)]%
        {DBLP:conf/mm/WeiWN0C20}
\bibfield{author}{\bibinfo{person}{Yinwei Wei}, \bibinfo{person}{Xiang Wang},
  \bibinfo{person}{Liqiang Nie}, \bibinfo{person}{Xiangnan He}, {and}
  \bibinfo{person}{Tat{-}Seng Chua}.} \bibinfo{year}{2020}\natexlab{}.
\newblock \showarticletitle{Graph-Refined Convolutional Network for Multimedia
  Recommendation with Implicit Feedback}. In \bibinfo{booktitle}{\emph{{ACM}
  Multimedia}}. \bibinfo{publisher}{{ACM}}, \bibinfo{pages}{3541--3549}.
\newblock


\bibitem[Wei et~al\mbox{.}(2019)]%
        {DBLP:conf/mm/WeiWN0HC19}
\bibfield{author}{\bibinfo{person}{Yinwei Wei}, \bibinfo{person}{Xiang Wang},
  \bibinfo{person}{Liqiang Nie}, \bibinfo{person}{Xiangnan He},
  \bibinfo{person}{Richang Hong}, {and} \bibinfo{person}{Tat{-}Seng Chua}.}
  \bibinfo{year}{2019}\natexlab{}.
\newblock \showarticletitle{{MMGCN:} Multi-modal Graph Convolution Network for
  Personalized Recommendation of Micro-video}. In
  \bibinfo{booktitle}{\emph{{ACM} Multimedia}}. \bibinfo{publisher}{{ACM}},
  \bibinfo{pages}{1437--1445}.
\newblock


\bibitem[Wu et~al\mbox{.}(2022)]%
        {DBLP:conf/sigir/WuWQZHX22}
\bibfield{author}{\bibinfo{person}{Chuhan Wu}, \bibinfo{person}{Fangzhao Wu},
  \bibinfo{person}{Tao Qi}, \bibinfo{person}{Chao Zhang},
  \bibinfo{person}{Yongfeng Huang}, {and} \bibinfo{person}{Tong Xu}.}
  \bibinfo{year}{2022}\natexlab{}.
\newblock \showarticletitle{MM-Rec: Visiolinguistic Model Empowered Multimodal
  News Recommendation}. In \bibinfo{booktitle}{\emph{{SIGIR}}}.
  \bibinfo{publisher}{{ACM}}, \bibinfo{pages}{2560--2564}.
\newblock


\bibitem[Wu et~al\mbox{.}(2021)]%
        {DBLP:conf/sigir/WuWF0CLX21}
\bibfield{author}{\bibinfo{person}{Jiancan Wu}, \bibinfo{person}{Xiang Wang},
  \bibinfo{person}{Fuli Feng}, \bibinfo{person}{Xiangnan He},
  \bibinfo{person}{Liang Chen}, \bibinfo{person}{Jianxun Lian}, {and}
  \bibinfo{person}{Xing Xie}.} \bibinfo{year}{2021}\natexlab{}.
\newblock \showarticletitle{Self-supervised Graph Learning for Recommendation}.
  In \bibinfo{booktitle}{\emph{{SIGIR}}}. \bibinfo{publisher}{{ACM}},
  \bibinfo{pages}{726--735}.
\newblock


\bibitem[Xiao et~al\mbox{.}(2022)]%
        {DBLP:journals/tits/XiaoCGUL22}
\bibfield{author}{\bibinfo{person}{Yi Xiao}, \bibinfo{person}{Felipe
  Codevilla}, \bibinfo{person}{Akhil Gurram}, \bibinfo{person}{Onay
  Urfalioglu}, {and} \bibinfo{person}{Antonio~M. L{\'{o}}pez}.}
  \bibinfo{year}{2022}\natexlab{}.
\newblock \showarticletitle{Multimodal End-to-End Autonomous Driving}.
\newblock \bibinfo{journal}{\emph{{IEEE} Trans. Intell. Transp. Syst.}}
  \bibinfo{volume}{23}, \bibinfo{number}{1} (\bibinfo{year}{2022}),
  \bibinfo{pages}{537--547}.
\newblock


\bibitem[Yang et~al\mbox{.}(2017)]%
        {DBLP:journals/tois/YangHYPDBCE17}
\bibfield{author}{\bibinfo{person}{Longqi Yang}, \bibinfo{person}{Cheng{-}Kang
  Hsieh}, \bibinfo{person}{Hongjian Yang}, \bibinfo{person}{John~P. Pollak},
  \bibinfo{person}{Nicola Dell}, \bibinfo{person}{Serge~J. Belongie},
  \bibinfo{person}{Curtis Cole}, {and} \bibinfo{person}{Deborah Estrin}.}
  \bibinfo{year}{2017}\natexlab{}.
\newblock \showarticletitle{Yum-Me: {A} Personalized Nutrient-Based Meal
  Recommender System}.
\newblock \bibinfo{journal}{\emph{{ACM} Trans. Inf. Syst.}}
  \bibinfo{volume}{36}, \bibinfo{number}{1} (\bibinfo{year}{2017}),
  \bibinfo{pages}{7:1--7:31}.
\newblock


\bibitem[Yang et~al\mbox{.}(2020b)]%
        {DBLP:journals/tcss/YangWLLGDW20}
\bibfield{author}{\bibinfo{person}{Qi Yang}, \bibinfo{person}{Gaosheng Wu},
  \bibinfo{person}{Yuhua Li}, \bibinfo{person}{Ruixuan Li},
  \bibinfo{person}{Xiwu Gu}, \bibinfo{person}{Huicai Deng}, {and}
  \bibinfo{person}{Junzhuang Wu}.} \bibinfo{year}{2020}\natexlab{b}.
\newblock \showarticletitle{{AMNN:} Attention-Based Multimodal Neural Network
  Model for Hashtag Recommendation}.
\newblock \bibinfo{journal}{\emph{{IEEE} Trans. Comput. Soc. Syst.}}
  \bibinfo{volume}{7}, \bibinfo{number}{3} (\bibinfo{year}{2020}),
  \bibinfo{pages}{768--779}.
\newblock


\bibitem[Yang et~al\mbox{.}(2020a)]%
        {DBLP:conf/aaai/YangDW20}
\bibfield{author}{\bibinfo{person}{Xun Yang}, \bibinfo{person}{Xiaoyu Du},
  {and} \bibinfo{person}{Meng Wang}.} \bibinfo{year}{2020}\natexlab{a}.
\newblock \showarticletitle{Learning to Match on Graph for Fashion
  Compatibility Modeling}. In \bibinfo{booktitle}{\emph{{AAAI}}}.
  \bibinfo{publisher}{{AAAI} Press}, \bibinfo{pages}{287--294}.
\newblock


\bibitem[Yi and Chen(2022)]%
        {DBLP:journals/tmm/YiC22}
\bibfield{author}{\bibinfo{person}{Jing Yi} {and} \bibinfo{person}{Zhenzhong
  Chen}.} \bibinfo{year}{2022}\natexlab{}.
\newblock \showarticletitle{Multi-Modal Variational Graph Auto-Encoder for
  Recommendation Systems}.
\newblock \bibinfo{journal}{\emph{{IEEE} Trans. Multim.}}  \bibinfo{volume}{24}
  (\bibinfo{year}{2022}), \bibinfo{pages}{1067--1079}.
\newblock


\bibitem[Yi et~al\mbox{.}(2022)]%
        {DBLP:conf/sigir/Yi0OM22}
\bibfield{author}{\bibinfo{person}{Zixuan Yi}, \bibinfo{person}{Xi Wang},
  \bibinfo{person}{Iadh Ounis}, {and} \bibinfo{person}{Craig MacDonald}.}
  \bibinfo{year}{2022}\natexlab{}.
\newblock \showarticletitle{Multi-modal Graph Contrastive Learning for
  Micro-video Recommendation}. In \bibinfo{booktitle}{\emph{{SIGIR}}}.
  \bibinfo{publisher}{{ACM}}, \bibinfo{pages}{1807--1811}.
\newblock


\bibitem[Yin et~al\mbox{.}(2023)]%
        {DBLP:journals/corr/abs-2306-13549}
\bibfield{author}{\bibinfo{person}{Shukang Yin}, \bibinfo{person}{Chaoyou Fu},
  \bibinfo{person}{Sirui Zhao}, \bibinfo{person}{Ke Li}, \bibinfo{person}{Xing
  Sun}, \bibinfo{person}{Tong Xu}, {and} \bibinfo{person}{Enhong Chen}.}
  \bibinfo{year}{2023}\natexlab{}.
\newblock \showarticletitle{A Survey on Multimodal Large Language Models}.
\newblock \bibinfo{journal}{\emph{CoRR}}  \bibinfo{volume}{abs/2306.13549}
  (\bibinfo{year}{2023}).
\newblock


\bibitem[Ying et~al\mbox{.}(2018)]%
        {DBLP:conf/kdd/YingHCEHL18}
\bibfield{author}{\bibinfo{person}{Rex Ying}, \bibinfo{person}{Ruining He},
  \bibinfo{person}{Kaifeng Chen}, \bibinfo{person}{Pong Eksombatchai},
  \bibinfo{person}{William~L. Hamilton}, {and} \bibinfo{person}{Jure
  Leskovec}.} \bibinfo{year}{2018}\natexlab{}.
\newblock \showarticletitle{Graph Convolutional Neural Networks for Web-Scale
  Recommender Systems}. In \bibinfo{booktitle}{\emph{{KDD}}}.
  \bibinfo{publisher}{{ACM}}, \bibinfo{pages}{974--983}.
\newblock


\bibitem[Yu et~al\mbox{.}(2023)]%
        {DBLP:conf/mm/Yu0LB23}
\bibfield{author}{\bibinfo{person}{Penghang Yu}, \bibinfo{person}{Zhiyi Tan},
  \bibinfo{person}{Guanming Lu}, {and} \bibinfo{person}{Bing{-}Kun Bao}.}
  \bibinfo{year}{2023}\natexlab{}.
\newblock \showarticletitle{Multi-View Graph Convolutional Network for
  Multimedia Recommendation}. In \bibinfo{booktitle}{\emph{{ACM} Multimedia}}.
  \bibinfo{publisher}{{ACM}}, \bibinfo{pages}{6576--6585}.
\newblock


\bibitem[Yu et~al\mbox{.}(2019)]%
        {DBLP:conf/mm/YuSZZJ19}
\bibfield{author}{\bibinfo{person}{Tong Yu}, \bibinfo{person}{Yilin Shen},
  \bibinfo{person}{Ruiyi Zhang}, \bibinfo{person}{Xiangyu Zeng}, {and}
  \bibinfo{person}{Hongxia Jin}.} \bibinfo{year}{2019}\natexlab{}.
\newblock \showarticletitle{Vision-Language Recommendation via Attribute
  Augmented Multimodal Reinforcement Learning}. In
  \bibinfo{booktitle}{\emph{{ACM} Multimedia}}. \bibinfo{publisher}{{ACM}},
  \bibinfo{pages}{39--47}.
\newblock


\bibitem[Zeng et~al\mbox{.}(2022)]%
        {DBLP:conf/sigir/ZengL022}
\bibfield{author}{\bibinfo{person}{Jiandian Zeng}, \bibinfo{person}{Tianyi
  Liu}, {and} \bibinfo{person}{Jiantao Zhou}.} \bibinfo{year}{2022}\natexlab{}.
\newblock \showarticletitle{Tag-assisted Multimodal Sentiment Analysis under
  Uncertain Missing Modalities}. In \bibinfo{booktitle}{\emph{{SIGIR}}}.
  \bibinfo{publisher}{{ACM}}, \bibinfo{pages}{1545--1554}.
\newblock


\bibitem[Zhan et~al\mbox{.}(2022)]%
        {DBLP:journals/tmm/ZhanLASDK22}
\bibfield{author}{\bibinfo{person}{Huijing Zhan}, \bibinfo{person}{Jie Lin},
  \bibinfo{person}{Kenan~Emir Ak}, \bibinfo{person}{Boxin Shi},
  \bibinfo{person}{Ling{-}Yu Duan}, {and} \bibinfo{person}{Alex~C. Kot}.}
  \bibinfo{year}{2022}\natexlab{}.
\newblock \showarticletitle{{\textdollar}A{\^{}}3{\textdollar}-FKG: Attentive
  Attribute-Aware Fashion Knowledge Graph for Outfit Preference Prediction}.
\newblock \bibinfo{journal}{\emph{{IEEE} Trans. Multim.}}  \bibinfo{volume}{24}
  (\bibinfo{year}{2022}), \bibinfo{pages}{819--831}.
\newblock


\bibitem[Zhang et~al\mbox{.}(2022a)]%
        {DBLP:conf/kdd/ZhangCMZWWZ22}
\bibfield{author}{\bibinfo{person}{Chaohe Zhang}, \bibinfo{person}{Xu Chu},
  \bibinfo{person}{Liantao Ma}, \bibinfo{person}{Yinghao Zhu},
  \bibinfo{person}{Yasha Wang}, \bibinfo{person}{Jiangtao Wang}, {and}
  \bibinfo{person}{Junfeng Zhao}.} \bibinfo{year}{2022}\natexlab{a}.
\newblock \showarticletitle{M3Care: Learning with Missing Modalities in
  Multimodal Healthcare Data}. In \bibinfo{booktitle}{\emph{{KDD}}}.
  \bibinfo{publisher}{{ACM}}, \bibinfo{pages}{2418--2428}.
\newblock


\bibitem[Zhang et~al\mbox{.}(2021a)]%
        {DBLP:conf/mm/Zhang00WWW21}
\bibfield{author}{\bibinfo{person}{Jinghao Zhang}, \bibinfo{person}{Yanqiao
  Zhu}, \bibinfo{person}{Qiang Liu}, \bibinfo{person}{Shu Wu},
  \bibinfo{person}{Shuhui Wang}, {and} \bibinfo{person}{Liang Wang}.}
  \bibinfo{year}{2021}\natexlab{a}.
\newblock \showarticletitle{Mining Latent Structures for Multimedia
  Recommendation}. In \bibinfo{booktitle}{\emph{{ACM} Multimedia}}.
  \bibinfo{publisher}{{ACM}}, \bibinfo{pages}{3872--3880}.
\newblock


\bibitem[Zhang et~al\mbox{.}(2021b)]%
        {DBLP:journals/corr/abs-2111-00678}
\bibfield{author}{\bibinfo{person}{Jinghao Zhang}, \bibinfo{person}{Yanqiao
  Zhu}, \bibinfo{person}{Qiang Liu}, \bibinfo{person}{Mengqi Zhang},
  \bibinfo{person}{Shu Wu}, {and} \bibinfo{person}{Liang Wang}.}
  \bibinfo{year}{2021}\natexlab{b}.
\newblock \showarticletitle{Latent Structures Mining with Contrastive Modality
  Fusion for Multimedia Recommendation}.
\newblock \bibinfo{journal}{\emph{CoRR}}  \bibinfo{volume}{abs/2111.00678}
  (\bibinfo{year}{2021}).
\newblock


\bibitem[Zhang et~al\mbox{.}(2017)]%
        {DBLP:conf/ijcai/ZhangWHHG17}
\bibfield{author}{\bibinfo{person}{Qi Zhang}, \bibinfo{person}{Jiawen Wang},
  \bibinfo{person}{Haoran Huang}, \bibinfo{person}{Xuanjing Huang}, {and}
  \bibinfo{person}{Yeyun Gong}.} \bibinfo{year}{2017}\natexlab{}.
\newblock \showarticletitle{Hashtag Recommendation for Multimodal Microblog
  Using Co-Attention Network}. In \bibinfo{booktitle}{\emph{{IJCAI}}}.
  \bibinfo{publisher}{ijcai.org}, \bibinfo{pages}{3420--3426}.
\newblock


\bibitem[Zhang(2017)]%
        {DBLP:journals/corr/abs-1708-06409}
\bibfield{author}{\bibinfo{person}{Yongfeng Zhang}.}
  \bibinfo{year}{2017}\natexlab{}.
\newblock \showarticletitle{Explainable Recommendation: Theory and
  Applications}.
\newblock \bibinfo{journal}{\emph{CoRR}}  \bibinfo{volume}{abs/1708.06409}
  (\bibinfo{year}{2017}).
\newblock


\bibitem[Zhang and Chen(2020)]%
        {DBLP:journals/ftir/ZhangC20}
\bibfield{author}{\bibinfo{person}{Yongfeng Zhang} {and} \bibinfo{person}{Xu
  Chen}.} \bibinfo{year}{2020}\natexlab{}.
\newblock \showarticletitle{Explainable Recommendation: {A} Survey and New
  Perspectives}.
\newblock \bibinfo{journal}{\emph{Found. Trends Inf. Retr.}}
  \bibinfo{volume}{14}, \bibinfo{number}{1} (\bibinfo{year}{2020}),
  \bibinfo{pages}{1--101}.
\newblock
\urldef\tempurl%
\url{https://doi.org/10.1561/1500000066}
\showDOI{\tempurl}


\bibitem[Zhang et~al\mbox{.}(2022b)]%
        {DBLP:journals/ijon/ZhangLWPD22}
\bibfield{author}{\bibinfo{person}{Ziqi Zhang}, \bibinfo{person}{Zeyu Li},
  \bibinfo{person}{Kun Wei}, \bibinfo{person}{Siduo Pan}, {and}
  \bibinfo{person}{Cheng Deng}.} \bibinfo{year}{2022}\natexlab{b}.
\newblock \showarticletitle{A survey on multimodal-guided visual content
  synthesis}.
\newblock \bibinfo{journal}{\emph{Neurocomputing}}  \bibinfo{volume}{497}
  (\bibinfo{year}{2022}), \bibinfo{pages}{110--128}.
\newblock


\bibitem[Zhao et~al\mbox{.}(2022)]%
        {DBLP:conf/cikm/ZhaoHPYZLZBTSCX22}
\bibfield{author}{\bibinfo{person}{Wayne~Xin Zhao}, \bibinfo{person}{Yupeng
  Hou}, \bibinfo{person}{Xingyu Pan}, \bibinfo{person}{Chen Yang},
  \bibinfo{person}{Zeyu Zhang}, \bibinfo{person}{Zihan Lin},
  \bibinfo{person}{Jingsen Zhang}, \bibinfo{person}{Shuqing Bian},
  \bibinfo{person}{Jiakai Tang}, \bibinfo{person}{Wenqi Sun},
  \bibinfo{person}{Yushuo Chen}, \bibinfo{person}{Lanling Xu},
  \bibinfo{person}{Gaowei Zhang}, \bibinfo{person}{Zhen Tian},
  \bibinfo{person}{Changxin Tian}, \bibinfo{person}{Shanlei Mu},
  \bibinfo{person}{Xinyan Fan}, \bibinfo{person}{Xu Chen}, {and}
  \bibinfo{person}{Ji{-}Rong Wen}.} \bibinfo{year}{2022}\natexlab{}.
\newblock \showarticletitle{RecBole 2.0: Towards a More Up-to-Date
  Recommendation Library}. In \bibinfo{booktitle}{\emph{{CIKM}}}.
  \bibinfo{publisher}{{ACM}}, \bibinfo{pages}{4722--4726}.
\newblock


\bibitem[Zheng et~al\mbox{.}(2017)]%
        {DBLP:conf/wsdm/ZhengNY17}
\bibfield{author}{\bibinfo{person}{Lei Zheng}, \bibinfo{person}{Vahid Noroozi},
  {and} \bibinfo{person}{Philip~S. Yu}.} \bibinfo{year}{2017}\natexlab{}.
\newblock \showarticletitle{Joint Deep Modeling of Users and Items Using
  Reviews for Recommendation}. In \bibinfo{booktitle}{\emph{{WSDM}}}.
  \bibinfo{publisher}{{ACM}}, \bibinfo{pages}{425--434}.
\newblock


\bibitem[Zheng et~al\mbox{.}(2023)]%
        {DBLP:conf/mobicom/ZhengLCW023}
\bibfield{author}{\bibinfo{person}{Tianyue Zheng}, \bibinfo{person}{Ang Li},
  \bibinfo{person}{Zhe Chen}, \bibinfo{person}{Hongbo Wang}, {and}
  \bibinfo{person}{Jun Luo}.} \bibinfo{year}{2023}\natexlab{}.
\newblock \showarticletitle{AutoFed: Heterogeneity-Aware Federated Multimodal
  Learning for Robust Autonomous Driving}. In
  \bibinfo{booktitle}{\emph{MobiCom}}. \bibinfo{publisher}{{ACM}},
  \bibinfo{pages}{15:1--15:15}.
\newblock


\bibitem[Zhou et~al\mbox{.}(2023b)]%
        {DBLP:journals/corr/abs-2302-04473}
\bibfield{author}{\bibinfo{person}{Hongyu Zhou}, \bibinfo{person}{Xin Zhou},
  \bibinfo{person}{Zhiwei Zeng}, \bibinfo{person}{Lingzi Zhang}, {and}
  \bibinfo{person}{Zhiqi Shen}.} \bibinfo{year}{2023}\natexlab{b}.
\newblock \showarticletitle{A Comprehensive Survey on Multimodal Recommender
  Systems: Taxonomy, Evaluation, and Future Directions}.
\newblock \bibinfo{journal}{\emph{CoRR}}  \bibinfo{volume}{abs/2302.04473}
  (\bibinfo{year}{2023}).
\newblock


\bibitem[Zhou and Shen(2023)]%
        {DBLP:conf/mm/ZhouS23}
\bibfield{author}{\bibinfo{person}{Xin Zhou} {and} \bibinfo{person}{Zhiqi
  Shen}.} \bibinfo{year}{2023}\natexlab{}.
\newblock \showarticletitle{A Tale of Two Graphs: Freezing and Denoising Graph
  Structures for Multimodal Recommendation}. In \bibinfo{booktitle}{\emph{{ACM}
  Multimedia}}. \bibinfo{publisher}{{ACM}}, \bibinfo{pages}{935--943}.
\newblock


\bibitem[Zhou et~al\mbox{.}(2023a)]%
        {DBLP:conf/www/ZhouZLZMWYJ23}
\bibfield{author}{\bibinfo{person}{Xin Zhou}, \bibinfo{person}{Hongyu Zhou},
  \bibinfo{person}{Yong Liu}, \bibinfo{person}{Zhiwei Zeng},
  \bibinfo{person}{Chunyan Miao}, \bibinfo{person}{Pengwei Wang},
  \bibinfo{person}{Yuan You}, {and} \bibinfo{person}{Feijun Jiang}.}
  \bibinfo{year}{2023}\natexlab{a}.
\newblock \showarticletitle{Bootstrap Latent Representations for Multi-modal
  Recommendation}. In \bibinfo{booktitle}{\emph{{WWW}}}.
  \bibinfo{publisher}{{ACM}}, \bibinfo{pages}{845--854}.
\newblock


\end{thebibliography}

\end{document}